\documentclass[a4paper,12pt]{article}
\pdfoutput=1
\usepackage{epsfig}
\usepackage{amssymb,subfigure}
\usepackage{amsfonts}
\usepackage{amsmath}
\usepackage{euscript}
\usepackage{verbatim}
\usepackage{latexsym}
\usepackage{graphicx,color}
\usepackage{caption}
\usepackage{hyperref}
\usepackage{float}
\usepackage{slashed}
\usepackage{graphicx}
\graphicspath{ {images/} }
\usepackage{ytableau}
\usepackage{cite}
\usepackage{wrapfig}
	\usepackage[T1]{fontenc}
	\usepackage{tikz}
	\usetikzlibrary{decorations.pathmorphing}
\usepackage{tikz}
\usetikzlibrary{shapes.geometric, arrows,patterns,snakes}
\tikzstyle{ellip} = [ellipse, minimum width=3cm, minimum height=1cm,text centered, draw=black]

\usepackage{relsize}
\usetikzlibrary{calc}
\usepackage{relsize}
\tikzset{fontscale/.style = {font=\relsize{#1}}}
\usetikzlibrary{arrows}
\usetikzlibrary{shapes,arrows,shapes.multipart}

\usepackage{pgfplots}
\usepackage{tikz-feynhand}
\usepackage{slashed}

\newskip\humongous \humongous=0pt plus 1000pt minus 1000pt

\newif\ifdtup

\allowdisplaybreaks[1]

\catcode`\@=11

\@addtoreset{equation}{section}

\def\@normalsize{\@setsize\normalsize{15pt}\xiipt\@xiipt
\abovedisplayskip 14pt plus3pt minus3pt%
\belowdisplayskip \abovedisplayskip
\abovedisplayshortskip \z@ plus3pt%
\belowdisplayshortskip 7pt plus3.5pt minus0pt}

\def\small{\@setsize\small{13.6pt}\xipt\@xipt
\abovedisplayskip 13pt plus3pt minus3pt%
\belowdisplayskip \abovedisplayskip
\abovedisplayshortskip \z@ plus3pt%
\belowdisplayshortskip 7pt plus3.5pt minus0pt
\def\@listi{\parsep 4.5pt plus 2pt minus 1pt
     \itemsep \parsep
     \topsep 9pt plus 3pt minus 3pt}}

\relax

\catcode`@=12

\catcode`\@=11

\def\section{\@startsection{section}{1}{\z@}{3.5ex plus 1ex minus
   .2ex}{2.3ex plus .2ex}{\large\bf}}

\def\SymBoxes#1#2#3#4{\newdimen\un@t \un@t#3%
\raisebox{#1}{\rule{#2\un@t}{#4}\hskip-#2\un@t
\@tempdimb\un@t \advance\@tempdimb by-#4\@tempcntb#2\relax%
\@whilenum{\@tempcntb>0}\do{
\rule{#4}{\un@t}\hskip\@tempdimb \advance\@tempcntb by\m@ne}%
\hskip-#2\un@t \rule[\un@t]{#2\un@t}{#4}%
\rule[\un@t]{#4}{#4}\hskip-#4
\rule{#4}{\un@t}}\hskip-#4}                

\begin{document}


\newcommand{\beq}{\begin{equation}}
\newcommand{\eeq}{\end{equation}}
\newcommand{\bea}{\begin{eqnarray}}
\newcommand{\eea}{\end{eqnarray}}
\newcommand{\beas}{\begin{eqnarray*}}
\newcommand{\eeas}{\end{eqnarray*}}
\newcommand{\defi}{\stackrel{\rm def}{=}}
\newcommand{\non}{\nonumber}
\newcommand{\bquo}{\begin{quote}}
\newcommand{\enqu}{\end{quote}}
\renewcommand{\(}{\begin{equation}}
\renewcommand{\)}{\end{equation}}

\def \eqn#1#2{\begin{equation}#2\label{#1}\end{equation}}
\def\IZ{{\mathbb Z}}
\def\IR{{\mathbb R}}
\def\IC{{\mathbb C}}
\def\IQ{{\mathbb Q}}
\def\de{\partial}
\def\Tr{ \hbox{\rm Tr}}
\def\H{ \hbox{\rm H}}
\def\HE{ \hbox{$\rm H^{even}$}}
\def\HO{ \hbox{$\rm H^{odd}$}}
\def\K{ \hbox{\rm K}}
\def\Im{ \hbox{\rm Im}}
\def\Ker{ \hbox{\rm Ker}}
\def\const{\hbox {\rm const.}}
\def\o{\over}
\def\im{\hbox{\rm Im}}
\def\re{\hbox{\rm Re}}
\def\bra{\langle}\def\ket{\rangle}
\def\Arg{\hbox {\rm Arg}}
\def\Re{\hbox {\rm Re}}
\def\Im{\hbox {\rm Im}}
\def\exo{\hbox {\rm exp}}
\def\diag{\hbox{\rm diag}}
\def\longvert{{\rule[-2mm]{0.1mm}{7mm}}\,}
\def\a{\alpha}
\def\dag{{}^{\dagger}}
\def\tq{{\widetilde q}}
\def\p{{}^{\prime}}
\def\W{W}
\def\N{{\cal N}}
\def\hsp{,\hspace{.7cm}}

\def\br{\nonumber\\}
\def\IZ{{\mathbb Z}}
\def\IR{{\mathbb R}}
\def\IC{{\mathbb C}}
\def\IQ{{\mathbb Q}}
\def\IP{{\mathbb P}}
\def \eqn#1#2{\begin{equation}#2\label{#1}\end{equation}}

\newcommand{\sgm}[1]{\sigma_{#1}}
\newcommand{\idd}{\mathbf{1}}

\newcommand{\C}{\ensuremath{\mathbb C}}
\newcommand{\Z}{\ensuremath{\mathbb Z}}
\newcommand{\R}{\ensuremath{\mathbb R}}
\newcommand{\rp}{\ensuremath{\mathbb {RP}}}
\newcommand{\cp}{\ensuremath{\mathbb {CP}}}
\newcommand{\vac}{\ensuremath{|0\rangle}}
\newcommand{\vact}{\ensuremath{|00\rangle}                    }
\newcommand{\oc}{\ensuremath{\overline{c}}}
\begin{titlepage}
\bigskip
\def\thefootnote{\fnsymbol{footnote}}

\title{\textbf{\huge{The Semiclassical Approximation: \\ Its Application to Holography and the Information Paradox}}}
\author{\large{ Rifath Khan \footnote{rifathkhantheo@gmail.com}}}
 \date{\today}
\maketitle

 \begin{center}
 {DAMTP, University of Cambridge}
\end{center}

\renewcommand{\thefootnote}{\arabic{footnote}}

\noindent
\begin{center} {\bf Abstract} \end{center}
	  In this research, we explore the semiclassical approximation to canonical quantum gravity and how a classical background emerges from the Wheeler-DeWitt (WDW) states. By employing the Wigner functional analysis, we derive the backreacted Einstein-Hamilton-Jacobi equation as an approximation to the WDW equation, along with the requisite validity conditions. We then apply this understanding to both AdS/CFT and dS/CFT correspondences in conjunction with Cauchy slice holography, to explain how the bulk is encoded in the correlation functions of the dual field theory. 
      
      We then explain an appropriate description for scenarios in which gravity behaves quantum mechanically in certain regions of spacetime and explain its relation to subregion holography. We derive the validity conditions for gravity to be semiclassical near any co-dimension 1 time-like surface and employ these conditions to explore the black hole information paradox. Our analysis suggests that for evaporating black holes, there might be a violation of semiclassical gravity in the near-horizon region close to the Page time, although this is contingent upon certain assumptions. This also provides insights into the fate of information trapped within evaporating black holes. We then explore this issue from the perspectives of both external and infalling observers. We then explain how to employ the framework of Cauchy slice holography to study the retrieval of information from evaporating black holes, presenting a comprehensive approach to tackle this complex issue in quantum gravity.

\vspace{1.6 cm}
\vfill

\end{titlepage}

\setcounter{page}{2}
\tableofcontents

\setcounter{footnote}{0}


\section{Introduction}

In physics, the concept of emergence is integral to bridging the chasm between different scales and realms of understanding. For instance, classical physics is understood to emerge from quantum mechanics as a viable approximation, which facilitates our comprehension of the macroscopic world from the microscopic perspective. 

In a similar vein, one might expect that classical gravity would emerge from quantum gravity through a comparable approximation. However, the situation is complicated by the fact that our very platform for doing physics - the spacetime background - is structured by the spacetime metric. This spacetime metric is an object within classical gravity and therefore is also emergent as classical gravity emerges. This leads to a complicated interplay where the foundational canvas for our physical investigations is a byproduct of the emergent phenomena we are probing. Consequently, it is crucial that a theory of quantum gravity should not depend on a pre-existing spacetime background. Instead, it needs to be formulated in a background-independent manner, allowing for the emergence of spacetime to be a byproduct of the theory, not a pre-condition.

While there are several approaches to quantum gravity, one of the most prominent is the holographic principle \cite{tHooft:1993dmi,Susskind:1994vu}. This theoretical assertion posits that all the information contained within a given volume of space can be fully described by information on its boundary. In essence, it suggests that the behavior of a system can be entirely encapsulated by a lower-dimensional representation, thereby implying a form of dimensional reduction or ``holography'' in nature. The Anti-deSitter/Conformal Field Theory (AdS/CFT) correspondence is a concrete realization of the holographic principle within the framework of string theory \cite{Maldacena:1997re,Witten:1998qj}. It establishes a duality between a $d+1$ dimensional theory of quantum gravity in asymptotically AdS space-time and a $d$-dimensional CFT residing on its boundary. Essentially, it posits that every phenomenon occurring within the AdS space-time, including gravitational interactions, can be precisely described by the dynamics of the CFT on its boundary. This correspondence has provided vital insights into the nature of quantum gravity and has been an instrumental tool in advancing our understanding of the structure of space-time at the quantum level.

The holographic CFT encompasses all quantum gravitational information of the bulk without needing to reference a bulk background, thereby rendering it bulk background-independent. To decipher this information about the bulk from the CFT data, a dictionary is necessary that provides a mapping between bulk and boundary concepts. This dictionary is well-established in certain aspects of the theory. For instance, the ground state of the CFT is dual to an empty AdS space-time, and low-energy excitation on the CFT ground state translates to quanta propagating within this global AdS background. Similarly, a Thermofield Double (TFD) state on the boundary is dual to the maximally extended Schwarzschild AdS spacetime in the bulk \cite{Maldacena:2001kr}. In principle, any bulk geometry should have a dual state in the CFT. Additionally, as one can create superpositions of these CFT states, it is crucial to understand the implications of superpositions involving vastly different spacetime geometries in the bulk.

In the canonical theory of quantum gravity \cite{DeWitt:1967yk}, the state is characterized by a superposition of spatial metrics and matter fields, denoted as $\Psi[g,\phi]$, which satisfies the Hamiltonian (also known as the Wheeler-DeWitt equation) and momentum constraints. This wavefunction, $\Psi[g,\phi]$, is also referred to as a Wheeler-DeWitt (WDW) state. To derive classical gravity under suitable conditions, one must apply semiclassical approximations to this WDW state. Previous work on this can be found in references \cite{Kiefer:1993fg,Kiefer:1991ef,Kiefer:1993yg}. Building on these previous studies, we aim to derive the back-reacted Einstein-Hamilton-Jacobi function, thereby revealing how semiclassical gravity emerges from the WDW state.

To decode the CFT information into a bulk description that can also accommodate the superposition of bulk geometries, a dictionary is required to map arbitrary CFT states to a corresponding bulk WDW state. This mapping is precisely what is provided by the framework of Cauchy Slice Holography \cite{GRW}. Hereafter we refer to this map as the CSH dictionary. The question of how the bulk background emerges from the boundary data then essentially reduces to the question of how the bulk background emerges from the associated WDW state. This is exactly the purpose served by the semiclassical approximation. Hence, studying the semiclassical approximation of WDW states has profound implications in the understanding of AdS/CFT correspondence, particularly in terms of how the boundary encodes the bulk.

Another motivation for understanding how semiclassical gravity emerges from a WDW state is the black hole information paradox. Hawking originally formulated this paradox within the semiclassical regime \cite{Hawking:1976ra}. According to the AdS/CFT correspondence, the dual theory evolves unitarily, suggesting that information isn't lost on the boundary side. However, on the bulk side, when operating within the semiclassical regime, it is clear that information is lost due to the non-globally hyperbolic spacetime depicted by the Penrose diagram of collapsing and evaporating black holes. This raises a question: why should we trust the semiclassical Penrose diagram? The reasoning generally follows that if one assumes quantum gravity effects only become significant at the Planck scale or near the singularity, then it becomes necessary to accept the validity of semiclassical gravity until the Cauchy slices approach the singularity. The paradox arises even before this point, at the so-called Page time. By understanding how semiclassical gravity emerges from the WDW states, it is possible to derive the conditions under which this approximation is valid, enabling an assessment of whether the semiclassical gravity can be trusted close to the Page time. 

On the other hand, in quantum cosmology \cite{Hartle:1983ai,Halliwell:1984eu}, specifically for closed universes (those for which the Cauchy slices have no boundaries), the wavefunction of the universe (e.g., the Hartle-Hawking wavefunction) does not have an external time parameter. This scenario gives rise to the ``problem of time'' in canonical quantum gravity. DeWitt elucidates in his research that time is also an emergent degree of freedom, and the arguments of the WDW state establish a clock through the semiclassical approximation \cite{DeWitt:1967yk}. Moreover, the wavefunction of the universe encompasses everything, including the observers within the universe. Consequently, it becomes necessary to find a split between the observers and the systems they observe. This results in a partition of arguments within the WDW state. Depending on the nature of this division, the WDW state gives rise to multiple emergent descriptions of the universe's subsystems as seen by other subsystems. Furthermore, since the entirety of the universe is described by one state, it prompts a crucial question: how is it possible for observers within this universe to perceive subsystems of the rest of the universe in different states? We will explore these intriguing questions in the course of this paper.

In the context of dS/CFT, a significant challenge arises due to the absence of an asymptotic time-like boundary where a dual field theoretical description could be situated. Instead, in some proposals, the CFT is conceived to exist on the asymptotic space-like boundary \cite{Strominger:2001pn}. Under the framework of Cauchy slice holography, the dual field theory's partition function supplies the wave function of the universe. Once we understand the semiclassical approximation, we can correctly identify the correlation functions of the dual theory that must be computed in order to obtain information about bulk concepts within the semiclassical regime.

Finally, one can question the situation when the metric is quantum mechanically fluctuating within a finite region of spacetime but is classical outside of it. What would observers describe this region as? Studying such a description would require applying a partial semiclassical approximation to the WDW equation on subregions, a process we refer to as ``subregion classicalisation''. This process is linked to the notion of emergent time-like boundaries hosting induced WDW states, which we will refer to as WDW screens. We believe this process is either the same as, or related to, subregion holography applicable to either AdS or dS. At the level of the WDW states, one could only understand the framework of such emergent subregion holography. However, the details of the specific theory will depend on the choice of WDW states and the choice of non-perturbative quantum gravity. As it was understood in a previous study \cite{GRW}, canonical quantum gravity functions as an ``effective'' theory of quantum gravity. \footnote{A part of this work was presented at the Cambridge/LMU 2022 conference and is available on youtube: \cite{youtube}.}

\subsection*{Plan of Paper}

In Section \ref{semiclassicalparticles}, we review the WKB approximation and explain the emergence of Hamilton-Jacobi theory from the Schrödinger equation. Following that, we review DeWitt's approach of applying the WKB approximation to a subset of degrees of freedom, thereby deriving a hybrid classical-quantum dynamics from the comprehensive quantum theory \cite{DeWitt:1967yk}.

In Section \ref{wkbinqg}, we review the work of Kiefer, Padmanabhan, and Singh \cite{Kiefer:1993fg,Kiefer:1991ef,Kiefer:1993yg,Padmanabhan:1990bk} towards applying the semiclassical approximation to canonical quantum gravity. This approximation gives the Einstein-Hamilton-Jacobi theory from a Wheeler-DeWitt state. Subsequently, we explain how a classical background can be constructed from the Einstein-Hamilton's principal function. We then extend the Wigner function analysis conducted by Halliwell, Padmanabhan, and Singh \cite{Halliwell:1987eu,Padmanabhan:1990fn} to obtain the backreacted Einstein-Hamilton-Jacobi equations. This derivation enables us to identify certain conditions that the emergent QFT state must meet for semiclassical gravity to remain valid.

In Section \ref{AAdSCFT}, we discuss how the aforementioned semiclassical analysis can be applied to reconstruct the bulk from the boundary in the AdS/CFT correspondence. We formulate criteria for determining the existence of semiclassical duals for a generic holographic CFT state and outline a procedure to derive them in terms of the correlation functions of the dual theory residing on Cauchy slices. This methodology employs the framework of Cauchy slice holography.

In Section \ref{dSCFTclosed}, we delve into the exploration of a minisuperspace model to explicitly showcase the ideas discussed earlier. We also introduce the concepts of WKB branch splitting and merging. The approach to interpreting dS/CFT in light of the semiclassical approximation is detailed, followed by an explanation of how the correlation functions of the dual theory can be related to bulk objects. We proceed to define operational observables, which closely align with realistic observables, and explain how to interpret the wave function of the universe. Finally, we detail the phenomenon where observers in certain WKB branches can observe numerous states for a subsystem, even when the wave function of the entire universe is a single state.

In Section \ref{subregionclassicalisationscreens}, we delve into the approach of treating gravity quantum mechanically within a finite region, while concurrently treating it classically outside this defined area. Through this process, we elaborate on the emergence of a time-like boundary, referred to as a WDW screen. The WDW states that exist on this emergent boundary would then encode the quantum gravitational information of the interior. Following this, we establish the validity conditions necessary for gravity to be classical near a WDW screen. Utilizing these insights, we then enhance our understanding of the framework of finite region holography within AdS and dS spacetimes.

Finally in Section \ref{BHinfoparadox}, we will apply the concepts we've developed thus far to better understand the black hole information paradox. This analysis involves examining when and where violations of the validity conditions for semiclassical gravity occur in the context of evaporating black holes, particularly post-Page time and within the near-horizon region. We find that the validity conditions suggest a breakdown of semiclassical gravity in the near-horizon region close to the Page time. Subsequently, we outline a framework for calculating the future semiclassical state, provided such a state exists, using the data from the past semiclassical state. From this perspective, we then discuss the experiences of both external observers and those falling into the black hole.


\subsection*{Cauchy slices and generalized Cauchy surfaces}

The term ``Cauchy surface" is well-defined in standard mathematical GR literature. However, the term ``Cauchy slice" is not as established. Even so, it is sometimes used interchangeably with Cauchy surfaces. For the purposes of holography, we will now define a Cauchy slice specifically for asymptotically-AdS and asymptotically-dS spacetimes, differentiating it from Cauchy surfaces.\footnote{Should the reader find themselves unacquainted with the foundational definitions in mathematical General Relativity, we recommend the comprehensive lecture notes on black holes by Prof. Harvey Reall, which serve as an exemplary reference. These notes can be accessed in the reference \cite{HarveyReall}.}

A \textbf{Cauchy slice} in a $d+1$ dimensional asymptotically-AdS or asymptotically-dS Lorentzian spacetime $(\mathcal{M},\mathbf{g})$ is defined as a $d$ dimensional submanifold $\Sigma$ such that no two points on $\Sigma$ are connected by a causal curve in $\mathcal{M}$, with $\partial \Sigma$ being a Cauchy surface of $\partial \mathcal{M}$ if $\mathcal{M}$ is asymptotically-AdS, and $\partial \Sigma = \emptyset$ if $\mathcal{M}$ is asymptotically-dS. We do not define a notion of a Cauchy slice for asymptotically flat spacetime in this paper. Also, note that while this definition applies only to spacetimes with one asymptotic end, it can be straightforwardly generalized to spacetimes with 
$n$ asymptotic AdS ends. However, it needs an extension for $n$ asymptotic dS ends, and we will not delve into such notions in this paper.

In a globally hyperbolic asymptotically-dS spacetime, we conjecture that a Cauchy slice becomes a Cauchy surface. However, if an asymptotically-dS spacetime is not globally hyperbolic, it still contains Cauchy slices but not Cauchy surfaces. For instance, due to its naked singularity at the final evaporation point, the spacetime of an asymptotically dS evaporating black hole will be non-globally 
hyperbolic, so it contains Cauchy slices but not Cauchy surfaces.

On the other hand, a global AdS spacetime is not globally hyperbolic in the strictest sense. This is because there exist inextendible causal curves that reach the boundary of this spacetime without cutting through the Cauchy slices. Even so, it is commonly said in the holography community that the global AdS spacetime becomes globally hyperbolic after imposing reflective boundary conditions. What is meant here is that the causal curves are allowed to be reflected on the boundary. Let us formalize this notion now.

In an asymptotically-AdS spacetime $\mathcal{B}$, an \textbf{inextendible reflective causal curve} is an inextendible causal curve which, upon reaching the boundary of $\mathcal{B}$, reflects off the boundary and continues as an inextendible causal curve within $\mathcal{B}$. The \textbf{generalized future domain of dependence} of a partial Cauchy surface in $\mathcal{B}$ is defined as the set of $p \in \mathcal{B}$, such that every past inextendible reflective causal curve through $p$ intersects this partial Cauchy surface. The generalized past domain of dependence is defined similarly. The \textbf{generalized domain of dependence} of a partial Cauchy surface in $\mathcal{B}$ is defined to be the union of the generalized future domain of dependence and the generalized past domain of dependence of the partial Cauchy surface. A \textbf{generalized Cauchy surface} in $\mathcal{B}$ is defined as a partial Cauchy surface whose generalized domain of dependence is $\mathcal{B}$. Finally, $\mathcal{B}$ is said to be \textbf{globally generalized hyperbolic} if it admits a generalized Cauchy surface. In a globally generalized hyperbolic asymptotically AdS spacetime, we conjecture that a Cauchy slice becomes a generalized Cauchy surface. However, this is not true when $\mathcal{B}$ contains evaporating black holes, as it is not globally generalized hyperbolic due to the naked singularity at the final evaporation point.

\section{Semiclassical approximation in quantum physics of point particles} \label{semiclassicalparticles}

\subsection{A non-relativistic particle in flat space}

Consider a particle with mass $m$ in a 2-dimensional Minkowski spacetime, which is non-relativistic and subject to an external potential $V(x)$. In quantum mechanics, the state of the particle is described by the wavefunction $\psi(x,t)$ which satisfies the Schr\"{o}dinger equation:
\begin{equation} \label{SE}
    i \hbar \frac{\partial}{\partial t} \psi(x,t) = \Big\{-\frac{\hbar^2}{2 m} \frac{\partial^2}{\partial x^2} + V(x) \Big\} \psi(x,t).
\end{equation}

We will now see how we recover classical mechanics by taking the limit $\hbar \to 0$.\footnote{Often in the literature, the limit $\hbar \to 0$ is taken to recover classical physics from quantum theory. However, $\hbar$ is a fundamental physical constant of nature with dimensions $\text{M} \text{L}^2 \text{T}^{-1}$. So, what does it mean to change $\hbar$, which is supposed to be a constant and a dimensionful quantity? The answer is that we are not actually changing $\hbar$, but instead taking the limit of a dimensionless quantity like $S/\hbar \to \infty$, where $S$ could be the phase of the wave function or the action of a theory, depending on the context. Alternatively, we could consider this as changing the unit system such that the numerical value of $\hbar$ in that system changes and can be made arbitrarily small while not changing the numerical value of $S$. The same applies to the gravitational constant (with dimensions $\text{M}^{-1} \text{L}^{3} \text{T}^{-2}$) and the speed of light (with dimensions $\text{L} \text{T}^{-1}$). This implies that the physical $S$ will change, as $S$ has dimensions. An alternative perspective is that there must exist some dimensionless parameter used for approximation.

} Write the wavefunction as  
\begin{equation}
    \psi(x,t) = \text{N} \ e^{\frac{i}{\hbar} S(x,t)},
\end{equation}
where $\text{N}$ is the normalization constant and $S(x,t)$ is a complex-valued function (so both the phase and amplitude of $\psi$ is expressed by $S$). Now expand $S(x,t)$ in a $\hbar$ series:
\begin{equation}
    S(x,t) = \sum_{n=0}^{\infty} \hbar^{n} S_n(x,t).
\end{equation}

The Schr\"{o}dinger equation \eqref{SE} becomes:
\begin{eqnarray}
    - \sum_{n=0}^{\infty} \hbar^{n} \frac{\partial S_n(x,t)}{\partial t} &=& -\frac{i \hbar}{2 m} \sum_{n=0}^{\infty} \hbar^{n} \frac{\partial^2 S_n(x,t)}{\partial x^2}+ \frac{1}{2 m} \left( \sum_{n=0}^{\infty} \hbar^{n} \frac{\partial S_n(x,t)}{\partial x}\right)^2 + V(x). \label{hbarexpansion}
\end{eqnarray}


To the leading order term in $\hbar$ ($O(\hbar^0)$), the above equation becomes:
\begin{equation}
    - \frac{\partial S_0(x,t)}{\partial t} =  \frac{1}{2 m} \left( \frac{\partial S_0(x,t)}{\partial x}\right)^2 + V(x), \label{HJqm}
\end{equation}
which is the Hamilton-Jacobi equation of classical mechanics, with $S_0(x,t)$ being the Hamilton's principal function but remember that $S_0(x,t)$ is still complex. Let's split $S(x,t)$ into its real and imaginary parts as $S(x,t) = S^R(x,t) + i \ S^I(x,t)$ and write the wavefunction as
\begin{equation}
    \psi(x,t)= \text{N} \ e^{\frac{i}{\hbar} S^R(x,t)} e^{-\frac{1}{\hbar} S^I(x,t)}.
\end{equation}

From this point forward, if $S_0^I$ possesses a unique global minimum at a specific point in space and if $\frac{\partial S_0^R}{\partial x}$ does not become zero at that location, then let us say that the state exhibits a \textbf{Type-1 property}. For such states, the particle localizes at a single point in the phase space, as demonstrated below.

When working in the leading order of $\hbar$, the higher order terms in $S^I(x,t)$ can be thrown out:
\begin{equation}
    \psi(x,t)= \text{N} \ e^{\frac{i}{\hbar} S^R(x,t)} e^{-\frac{1}{\hbar} S_0^I(x,t)}. \label{leadingS_0}
\end{equation}
At a time $t$, let $x_{\mathrm{cl}}(t)$ be the point in space where $S^I_0(x,t)$ attains its global minimum. 
In the limit $\hbar \to 0$, the probability distribution function of the position becomes a Dirac delta function at $x_{\mathrm{cl}}(t)$:
\bea
\lim_{\hbar \to 0} \psi^*(x,t) \psi(x,t) &=& \lim_{\hbar \to 0}\text{N}^2 \ e^{-\frac{2}{\hbar} S_0^I(x,t)}, \\
&=& \lim_{\hbar \to 0} \frac{1}{ \int dx   \frac{1}{\sqrt{\pi \hbar}} \ e^{-\frac{2}{\hbar} S_0^I(x,t)} } \frac{1}{\sqrt{\pi \hbar}} e^{-\frac{2}{\hbar} S_0^I(x,t)}, \\
 &=&  \delta(x-x_{\mathrm{cl}}(t)).
\eea
So when measured, one will always find the particle to be very localized in space at $x_{\mathrm{cl}}(t)$ and we say the position of the classical particle is $x_{\mathrm{cl}}(t)$.

Next express the wavefunction in momentum space: 
\begin{eqnarray}
    \Tilde{\psi}(p,t) &=& \frac{1}{\sqrt{2 \pi \hbar} }   \int_{-\infty}^{\infty} dx \ e^{-i p x/\hbar} \ \psi(x,t),
    \\
    &=& \frac{\text{N}}{\sqrt{2 \pi \hbar}}  \int_{-\infty}^{\infty}  dx \ e^{-i p x/\hbar} \ e^{\frac{i}{\hbar} S^R(x,t)} \ e^{-\frac{1}{\hbar} S_0^I(x,t)}.
\end{eqnarray}
This integral gets its contribution only in the neighbourhood of $x_{\mathrm{cl}}(t)$ because the integrand peaks there and so we Taylor expand $S^R(x,t)$ and $S^I_0(x,t)$ around $x_{\mathrm{cl}}(t)$ and keep only up to the linear term ($S_0^{I\prime}(x_{\mathrm{cl}}(t),t)=0$ because it is a minima):
\begin{eqnarray}
    \Tilde{\psi}(p,t) &=& \frac{\text{N}}{\sqrt{2 \pi \hbar}}  \int  dx \ e^{- \frac{i}{\hbar} \left( p x - S^R(x,t)\right)}  \ e^{-\frac{1}{\hbar} S_0^I(x,t)},
    \\
    &=& \frac{\text{N}}{\sqrt{2 \pi \hbar}}  \int  dx \ e^{- \frac{i}{\hbar} \left( p x - S^R(x_{\mathrm{cl}}(t),t) - S^{R\prime}(x_{\mathrm{cl}}(t),t) (x-x_{\mathrm{cl}}(t)) \right)}  \ e^{-\frac{1}{\hbar} \left( S_0^I(x_{\mathrm{cl}}(t),t) \right) },
    \\
    &=& \frac{\text{N}}{\sqrt{2 \pi \hbar}}   e^{ \frac{i}{\hbar} \left( S^R(x_{\mathrm{cl}}(t),t) -  S^{R\prime}(x_{\mathrm{cl}}(t),t) x_{\mathrm{cl}}(t) + i S_0^I(x_{\mathrm{cl}}(t),t) \right)} 
    \int  dx \ e^{- \frac{i}{\hbar} \left( p   - S^{R\prime}(x_{\mathrm{cl}}(t),t) \right) x}.
\end{eqnarray}
In $\hbar \to 0$ limit, the probability distribution function of the momentum will become a Dirac delta function at $S_0^{R\prime}(x_{\mathrm{cl}}(t),t)$:
\bea
\lim_{\hbar \to 0} \Tilde{\psi}^*(p,t)\Tilde{\psi}(p,t) &=& \lim_{\hbar \to 0} \frac{\text{N}^2 e^{ -\frac{2}{\hbar} S_0^I(x_{\mathrm{cl}})}}{2 \pi \hbar}   \int dx dy \ e^{ \frac{-i}{\hbar} \left( p   - S^{R\prime}(x_{\mathrm{cl}}(t),t) \right) \left(x-y\right)}, \\
&=& \delta\left( p   - S_0^{R\prime}(x_{\mathrm{cl}}(t),t) \right).
\eea
So when measured, one will always find the particle to have the momentum:
\begin{equation}
    p_{\mathrm{cl}}(t)  = \frac{\partial S_0}{\partial x}(x_{\mathrm{cl}}(t),t), \label{classicalmomentum}
\end{equation}
where I have added back the imaginary part because its derivative at $x_{\mathrm{cl}}(t)$ is anyways zero.
Notice $S_0^I(x_{\mathrm{cl}}(t),t)$ is independent of time as: 
\bea
\dot{S}_0^I(x_{\mathrm{cl}}(t),t) &=& \frac{\partial S_0^I}{\partial t}(x_{\mathrm{cl}}(t),t) + \frac{\partial S_0^I}{\partial x}(x_{\mathrm{cl}}(t),t) \dot{x}_{\mathrm{cl}}(t), \\
&=&\frac{\partial S_0^I}{\partial x}(x_{\mathrm{cl}}(t),t) \left(-\frac{1}{m} \frac{\partial S_0^R}{\partial x}(x_{\mathrm{cl}}(t),t) + \dot{x}_{\mathrm{cl}}(t)\right), \\
&=& 0,
\eea
where we have used \eqref{HJqm} and $\frac{\partial S_0^I}{\partial x}(x_{\mathrm{cl}}(t),t) =0$ and the dot denotes time derivative. From this we get:
\bea
0 &=& \ddot{S}_0^I(x_{\mathrm{cl}}(t),t), \\
&=& \frac{\partial^2 S_0^I}{\partial t^2}(x_{\mathrm{cl}}(t),t) + 2 \frac{\partial^2 S_0^I}{\partial x \partial t}(x_{\mathrm{cl}}(t),t) \dot{x}_{\mathrm{cl}}(t) +\frac{\partial^2 S_0^I}{\partial x^2}(x_{\mathrm{cl}}(t),t) \dot{x}^2_{\mathrm{cl}}(t), \\
&=& \frac{1}{m^2}\frac{\partial^2 S_0^I}{\partial x^2}(x_{\mathrm{cl}}(t),t) \left(p_{\mathrm{cl}}(t) - m  \dot{x}_{\mathrm{cl}}(t)\right)^2.
\eea
So the classical momentum is related to the classical velocity:
\bea
p_{\mathrm{cl}}(t) = m  \dot{x}_{\mathrm{cl}}(t).
\eea
As $S_0(x,t)$ satisfies the Hamilton-Jacobi equations: $x_{\mathrm{cl}}(t)$ and $p_{\mathrm{cl}}(t)$ traces a path that satisfies the classical equations of motion.\footnote{An alternative perspective on the aforementioned calculation involves maintaining a fixed unit system (so $\hbar$ remains constant) and altering $S(x,t)$ such that $S(x,t)/\hbar \gg 1$. This means that the wavefunction begins to oscillate rapidly, allowing us to apply the same WKB approximation and obtain identical results. Both of these perspectives are equivalent. In the second approach, it is also possible to cause the wavefunction to oscillate rapidly only in specific spatial regions, while remaining non-oscillatory in others. Consequently, the WKB approximation is only valid in regions where the wavefunction exhibits rapid oscillation. The regions with non-oscillatory behavior represent the classically forbidden areas.}

Next, if $S^I_0$ exhibits a global minimum not at a single point but across an interval, the particle will not localize to a specific point; instead, it will be distributed over the interval as $\hbar$ approaches zero. It is always possible to shift the value of $S^I_0$ within this interval to zero, which may introduce a constant factor multiplying the wave function. However, this factor can be absorbed into the normalization constant. From this point forward, if $\frac{\partial S_0^R}{\partial x}$ does not become zero at any location within this interval, then let us say that the state exhibits a $\textbf{Type-2 property}$. 

Consider a state that possesses a Type-2 property at every point in space at a given time. Such a state will evolve into one where the Type-2 property holds at every point in space for all time. This can be inferred from the imaginary part of the Hamilton-Jacobi equation \eqref{HJqm}:
\bea
-\frac{\partial S_0^I}{\partial t} = \frac{1}{m} \frac{\partial S_0^R}{\partial x} \frac{\partial S_0^I}{\partial x},
\eea
which leads to the conclusion that $S_0^I(x,t)=0$ for all time $t$. In the limit as $\hbar \to 0$, the probability distribution function of position becomes $\rho(x,t) := \text{N}^2 e^{-2 S_1^I(x,t)}$. From the O($\hbar$) terms in \eqref{hbarexpansion}, we obtain:
\bea
-\frac{\partial S_1^I}{\partial t} &=& -\frac{1}{2m} \frac{\partial^2 S_0^R}{\partial x^2} + \frac{1}{m} \frac{\partial S_0^R}{\partial x} \frac{\partial S_1^I}{\partial x}.
\eea
Using this equation, we can solve for $S_1^I(x,t)$ given the initial condition $S_1^I(x,t_0)$ at some time $t=t_0$ and a Hamilton's principal function $S_0^R(x,t)$. It follows that the probability distribution function of position satisfies:
\bea
 \frac{\partial \rho}{\partial t} &=& - \frac{\partial }{\partial x} \left( \rho \frac{1}{m} \frac{\partial S_0^R}{\partial x} \right) =- \frac{\partial j}{\partial x},
\eea
which is simply the continuity equation for the conservation of probability, with the current $j:=\left( \rho \frac{1}{m} \frac{\partial S_0^R}{\partial x} \right)$.\footnote{This can be interpreted as the particle being described by a probability distribution of position, with the particle's velocity given by $\frac{1}{m} \frac{\partial S_0^R}{\partial x}$ when the particle is at position $x$.}

Consider a probability distribution on the phase space defined as:
\bea
\Tilde{\rho}(x,p,t) := \rho(x,t)  \ \delta\left(p-\frac{\partial S_0^R}{\partial x} \right) = \rho(x,t) \int \frac{d \lambda}{2\pi} \ e^{i\left(p-\frac{\partial S_0^R}{\partial x}\right)\lambda}  .
\eea
It can be shown that this satisfies the Liouville equation: 
\bea
\frac{\partial \Tilde{\rho}(x,p,t)}{\partial t} + \left\{ \Tilde{\rho}(x,p,t),H\right(x,p)\} &=& 0,
\eea
where the curly brackets denote the Poisson brackets. The probability distribution of momentum now will be:
\begin{align}
\Tilde{\psi}^*(p,t)\Tilde{\psi}(p,t) &= \frac{\text{N}^2}{2 \pi \hbar}  \int  dx dy \ e^{- \frac{i}{\hbar} \left(px-py -S^R(x,t)+S^R(y,t)\right)}  \ e^{- \left(S_1^I(x,t)+S_1^I(y,t)\right)}, \\
&= \frac{\text{N}^2}{2 \pi}  \int  dx d\eta \ e^{- \frac{i}{\hbar} \left(-p\hbar \eta -S^R(x,t)+S^R(x+\hbar \eta,t)\right)}  \ e^{- \left(S_1^I(x,t)+S_1^I(x+\hbar \eta,t)\right)},
\end{align}
and by Taylor expanding about $x$ and throwing away the higher order terms in $\hbar$ we get:
\bea
\Tilde{\psi}^*(p,t)\Tilde{\psi}(p,t) &=& \frac{1}{2 \pi}  \int dx \ \text{N}^2  e^{-2 S_1^I(x,t)} \int d\eta \ e^{i \left(p-S^{R\prime}(x,t)\right)\eta}  , \\
&=&   \int dx \ \Tilde{\rho}(x,p,t),
\eea
which is the momentum marginal of $\Tilde{\rho}$. So in this case, we get the probability distribution of position and probability distribution of momentum of the wavefunction to be the marginals of $\Tilde{\rho}$ (which satisfies the Liouville equation) for all time. So the wavefunction of the particle reduces to a phase space probability distribution of a classical particle evolving according to the Liouville equations.


\subsection{Classical-Quantum split} \label{CQsplitinQM}

Consider a two-particle quantum system with the Hamiltonian:
\begin{equation}
H = -\frac{\hbar^2}{2} \frac{\epsilon}{M}\frac{\partial^2}{\partial Q^2} - \frac{\hbar^2}{2m} \frac{\partial^2}{\partial q^2} + \frac{1}{\epsilon} V_{\mathrm{cl}}(Q) + V(q,Q),
\end{equation}
where $M/\epsilon$ and $m$ are the masses of the particles with positions $Q$ and $q$, respectively. The wavefunction of the system, $\Psi(q,Q,t)$, satisfies the Schr\"{o}dinger equation:
\begin{equation}
i \hbar \frac{\partial}{\partial t} \Psi = H \Psi.
\end{equation}
Express the wavefunction as:
\begin{equation}
\Psi(q,Q,t) = e^{\frac{i}{\hbar} S(q,Q,t)},
\end{equation}
and expand $S(q,Q,t)$ as a power series in $\epsilon$ (which is a dimensionless quantity):
\begin{equation}
S(q,Q,t) = \sum_{n=0}^{\infty} \epsilon^{n-1} S_n(q,Q,t).
\end{equation}

Now, we will analyze the small $\epsilon$ limit. This occurs when one particle has a mass $M/\epsilon$ that is much larger than the other particle's mass $m$. At order $O(\epsilon^{-2})$, we obtain the following from the Schr\"{o}dinger equation:
\begin{equation}
\frac{\partial}{\partial q} S_0(q,Q,t) = 0,
\end{equation}
so $S_0(q,Q,t) = S_0(Q,t)$ is independent of $q$. At order $O(\epsilon^{-1})$, we get from the Schr\"{o}dinger equation:
\begin{equation}
-\frac{\partial S_0}{\partial t} = \frac{1}{2 M} \left(\frac{\partial S_0}{\partial Q} \right)^2 +  V_{\mathrm{cl}},\label{splitHJtdep}
\end{equation}
which is the Hamilton-Jacobi equation for $Q$. So, in the $\epsilon \to 0$ limit, the heavy particle behaves classically with $S_0$ as its Hamilton's principal function.

At order $O(\epsilon^0)$, we get from the Schr\"{o}dinger equation:
\begin{equation}
i \hbar \frac{\partial}{\partial t} e^{\frac{i}{\hbar} S_1(q,Q,t)} = H_{quant} \ e^{\frac{i}{\hbar} S_1(q,Q,t)} -\frac{i\hbar}{2 M} \left( \frac{\partial^2 S_0}{\partial Q^2} + 2 \frac{\partial S_0}{\partial Q} \frac{\partial}{\partial Q} \right)e^{\frac{i}{\hbar} S_1(q,Q,t)}, \label{1ptclqsplt}
\end{equation}
where
\bea
H_{quant} := \left\{ -\frac{\hbar^2}{2m} \frac{\partial^2}{\partial q^2} + V(q,Q) \right\}.
\eea
Define an effective wavefunction $\psi(q,Q,t)$ for particle $q$ in a classical background $Q$ at time $t$ as:
\begin{equation}
\psi(q,Q,t) := N(Q,t) \ e^{\frac{i}{\hbar} S_1(q,Q,t)},
\end{equation}
where $N(Q,t)$ is the normalization factor defined as:
\begin{equation}
N^{-2}(Q,t):= \int dq \left|e^{\frac{i}{\hbar} S_1} \right|^2.
\end{equation}
We can rearrange Equation \eqref{1ptclqsplt} to get:
\bea
i \hbar \left( \frac{\partial}{\partial t} +\frac{1}{M} \frac{\partial S_0}{\partial Q} \frac{\partial }{\partial Q} \right) \psi = \left\{ H_{\mathrm{quant}} + \frac{\hbar}{2 M} \frac{\partial^2 S_0^I}{\partial Q^2} \right\}\ \psi  . \label{rearranged}
\eea
If the state has the Type-2 property, then one can define a parameter $\tau$ as:
\begin{equation}
\frac{\partial}{\partial \tau} :=\frac{1}{M} \frac{\partial S_0}{\partial Q} \frac{\partial }{\partial Q},
\end{equation}
and by rewriting Equation \eqref{rearranged}, one gets:
\begin{equation}
i \hbar \left(\frac{\partial}{\partial t} + \frac{\partial}{\partial \tau} \right) \psi(q,\tau,t) = H_{quant} \ \psi .
\end{equation}
This equation describes the evolution of the state of a quantum particle $\psi(q,\tau,t)$ which is coupled to a classical particle.

Now let us consider a stationary state $\Psi(q,Q)$ (independent of $t$) with an energy eigenvalue $E$ which is also expanded in a power series in $\epsilon$:
\begin{equation}\label{stationaryqmstate}
H \ \Psi(q,Q) = E \ \Psi(q,Q) = \left(\sum_{n=0}^{\infty} \epsilon^{n-1} E_n \right) e^{\frac{i}{\hbar} \sum_{n=0}^{\infty} \epsilon^{n-1} S_n(q,Q)}.
\end{equation}
Following the same analysis, we get $S_0$ to be independent of $q$ and satisfying the Hamilton-Jacobi equation:
\begin{equation}
\frac{1}{2 M} \left(\frac{\partial S_0}{\partial Q} \right)^2 + V_{\mathrm{cl}}(Q) = E_0, \label{HJforsplit}
\end{equation}
and the effective wavefunction of $q$ satisfying:
\bea\label{emergentparticleschrodingerequation}
i \hbar \frac{\partial}{\partial \tau} \psi(q,\tau) = \left\{ H_{quant} - E_1 \right\} \psi(q,\tau).
\eea
By identifying $\tau(Q)$ as the time parameter, the above equation is the emergent Schr\"{o}dinger equation describing the quantum particle $q$ with the clock variable $\tau(Q)$.

\subsection{Backreaction of the quantum particle on the classical particle from Wigner function analysis} \label{wignerbackreaction}

In Section \ref{CQsplitinQM}, we studied the case of a heavy particle with mass $M$ and position $Q$ coupled to a light particle with mass $m$ and position $q$. Using the WKB approximation, we found that at the leading order in $\epsilon$ (i.e., $O(\epsilon^{-1})$), the heavy particle behaves classically, while at the next order in $\epsilon$ (i.e., $O(\epsilon^0)$), the light particle exhibits quantum mechanical behavior, with the classical heavy particle providing a background potential. At order $O(\epsilon^{-1})$, the quantum particle (light particle) has no effect on the classical particle (heavy particle). However, at order $O(\epsilon^0)$, the quantum particle can backreact on the classical particle, modifying its classical equations of motion. In this section, we review the calculation of backreaction using the Wigner function, as discussed in \cite{Halliwell:1987eu,Padmanabhan:1990fn}.

The Wigner function corresponding to the quantum state $\Psi(q,Q,t)$ is given by:
\begin{align}
W(q,p,Q,P,t) = \frac{1}{(\pi \hbar)^2} \int dx \ dy \ \Psi^*(q+x,Q+y,t) \ \Psi(q-x,Q-y,t) \ e^{2i (px+Py)/\hbar},
\end{align}
where $p$ and $P$ are the canonical momenta conjugate to $q$ and $Q$, respectively.
The Wigner function is a quasi-probability distribution that becomes a true probability distribution in the $\hbar \to 0$ limit. In \cite{Halliwell:1987eu}, Halliwell elucidates the form the Wigner function must take in order for a particle to exhibit classical behavior. A quantum particle $Q$ is said to behave classically (i.e., classical physics emerging from quantum physics) in the classical limit (using the WKB approximation) if and only if its Wigner function reduces to the classical Wigner function:
\bea\label{firstHalliwellcriterion}
W^{classical}(Q,P) := \mathcal{P}(Q) \ \delta\left(P-f(Q)\right),
\eea
where $\mathcal{P}(Q)$ represents the probability distribution of $Q$ and $f(Q)$ is any function of $Q$. This implies that classical mechanics emerges from a quantum theory when there are strong correlations between the position $Q$ and its conjugate momentum $P$. We will henceforth refer to this condition—that the Wigner function takes the above form in Equation \eqref{firstHalliwellcriterion}—as the \textbf{Halliwell criterion}.

As we are analyzing the $\epsilon \to 0$ limit, let's rewrite the Wigner function accordingly:
\begin{equation}
W(q,p,Q,P,t) =  \int \frac{\epsilon \ dx \ dy}{(\pi \hbar)^2 M m} \ \Psi^*(q+\frac{x}{m},Q+\frac{\epsilon \ y}{M},t) \ \Psi(q-\frac{x}{m},Q-\frac{\epsilon \ y}{M},t) e^{\frac{2i}{\hbar} (\frac{px}{m}+\frac{\epsilon Py}{M})}.    
\end{equation}
We can obtain the reduced Wigner function for the heavy particle by integrating over the position $q$ and momentum $p$ of the light particle:
\begin{align}
W(Q,P,t) &=  \int dq \ dp \ W(q,p,Q,P,t), \\
&=\frac{\epsilon}{2 \pi^2 \hbar M } \int dy \ dq \ \Psi^*(q,Q+\frac{\epsilon \ y}{M},t) \ \Psi(q,Q-\frac{\epsilon \ y}{M},t) e^{\frac{2i \epsilon Py}{M \hbar}}, \\
&=  \frac{\epsilon }{2 \pi^2 \hbar M } \int   dy \ dq
  \  e^{\frac{i}{\hbar}\left(S^R(q,Q-\frac{\epsilon  y}{M},t)- S^R(q,Q+\frac{\epsilon  y}{M},t) + \frac{2 \epsilon Py}{M}\right)} \\ 
&\;    e^{-\frac{1}{\hbar} \left(S^I(q,Q-\frac{\epsilon y}{M},t) +S^I(q,Q+\frac{\epsilon y}{M},t) \right)}. \label{Wig}
\end{align}
Next, we expand $S$ in a power series in $\epsilon$ and perform a Taylor series expansion of the integrand around $y=0$:
\bea
W(Q,P,t) &=&  \frac{\epsilon }{2 \pi^2 \hbar M } \int dy  dq \  e^{-\frac{2}{\hbar} \left( \epsilon^{-1} S_0^I(Q,t) + S_1^I(q,Q,t) \right) + O(\epsilon^{1})} \\
&\;& \ e^{\frac{2 i \epsilon}{M \hbar}\left(P - \frac{1}{\epsilon} \frac{\partial S_0^R}{\partial Q}(Q,t) -\frac{\partial S_1^R}{\partial Q}(q,Q,t) \right)y + O(\epsilon^{2})},
\eea
and by throwing away the higher-order terms in $\epsilon$, and also neglecting $\frac{\partial S_1^R}{\partial Q}$ compared to $\frac{1}{\epsilon} \frac{\partial S_0^R}{\partial Q}$ (but not neglecting $S_1^I$ to include the case of $S_0^I$ being zero or not), the $q$ and $y$ integrals separate:
\begin{align}
W(Q,P,t) &=  \frac{  \epsilon \ e^{-\frac{2}{\hbar \epsilon}  S_0^I(Q,t)}}{2 \pi^2 \hbar M } \left( \int dq   e^{-\frac{2}{\hbar} S_1^I(q,Q,t) } \right)  \left( \int dy  e^{\frac{2 i \epsilon}{M \hbar}\left(P - \frac{1}{\epsilon} \frac{\partial S_0^R}{\partial Q}(Q,t) \right)y} \right), \\
&=  \frac{  e^{-\frac{2}{\hbar \epsilon}  S_0^I(Q,t)}}{4 \pi^2 } \left( \int dq   e^{-\frac{2}{\hbar} S_1^I(q,Q,t) } \right)  \left( \int dy  e^{i\left(P - \frac{1}{\epsilon} \frac{\partial S_0^R}{\partial Q}(Q,t) \right)y} \right), \\
&=  \frac{  e^{-\frac{2}{\hbar \epsilon}  S_0^I(Q,t)}}{2 \pi } \left( \int dq   e^{-\frac{2}{\hbar} S_1^I(q,Q,t) } \right)  \delta\left(P - \frac{1}{\epsilon} \frac{\partial S_0^R}{\partial Q}(Q,t) \right).
\end{align}
When the state possesses the Type-1 property, the reduced Wigner function takes the following form:
\bea
W^{(1)}(Q,P,t) &=&  \delta \left(Q-Q_{\mathrm{cl}}(t) \right)  \delta\left(P - \frac{1}{\epsilon} \frac{\partial S_0^R}{\partial Q}(Q_{\mathrm{cl}}(t),t) \right),
\eea
where $Q_{\mathrm{cl}}(t)$ denotes the point in space where $S_0^I$ attains its global minimum at time $t$. In the presence of the Type-1 property, the reduced Wigner function takes the form of a probability distribution function represented by a Dirac delta function in the classical phase space path.

On the other hand, for states with the Type-2 property, the reduced Wigner function takes a different form:
\bea
W^{(2)}(Q,P,t) &=&  \left(  \int dq  \ e^{-\frac{2}{\hbar} S_1^I(q,Q,t) } \right)  \delta\left(P - \frac{1}{\epsilon} \frac{\partial S_0^R}{\partial Q}(Q,t) \right), 
\eea
which is a probability distribution in phase space ($Q,P$) and is subject to the Liouville equation. It is worth noting that at this stage of analysis, there is no backreaction from the quantum particle on the classical particle. However, by considering the subsequent terms in $\epsilon$ in Equation \eqref{Wig} and disregarding the higher-order terms, we can account for the backreaction and its effects:
\bea
W^{(2)}(Q,P,t) &=&  \frac{N^2 \epsilon}{2 \pi^2 \hbar M } \int dy \  dq \  e^{-\frac{2}{\hbar} S_1^I(q,Q,t) }  \ e^{\frac{2 i \epsilon}{M \hbar}\left(P - \frac{1}{\epsilon} \frac{\partial S_0^R}{\partial Q}(Q,t) -\frac{\partial S_1^R}{\partial Q}(q,Q,t) \right)y}, \\
&=&  \frac{N^2 }{4 \pi^2  } \int dy \ e^{i\left(P - \frac{1}{\epsilon} \frac{\partial S_0^R}{\partial Q}(Q,t) \right)y} \int dq  \  e^{-\frac{2}{\hbar} S_1^I(q,Q,t)} \  e^{-i\left(\frac{\partial S_1^R}{\partial Q}(q,Q,t) \right)y} , \\
&=&   \mathcal{P}(Q,t) \int \frac{dy}{2\pi} \ e^{i\left(P - \frac{1}{\epsilon} \frac{\partial S_0^R}{\partial Q}(Q,t) \right)y} \int dq  \  \mathcal{P}(q/Q,t) \  e^{-i\left(\frac{\partial S_1^R}{\partial Q}(q,Q,t) \right)y} ,\\
&=&  \mathcal{P}(Q,t) \int \frac{dy}{2\pi} \ e^{i\left(P - \frac{1}{\epsilon} \frac{\partial S_0^R}{\partial Q}(Q,t) \right)y} \left\langle  e^{-i y \frac{\partial S_1^R}{\partial Q}(q,Q,t) }\right\rangle_{(Q,t)},\label{semiclassicalisingwignerfunctionforparticle}
\eea
where $\mathcal{P}(Q,t):=\frac{1}{2\pi}\int dq \ e^{-\frac{2}{\hbar} S_1^I(q,Q,t)}$ represents the probability distribution of $Q$ at time $t$. Similarly, $\mathcal{P}(q/Q,t):=\frac{1}{\mathcal{P}(Q,t)} e^{-\frac{2}{\hbar} S_1^I(q,Q,t)}$ denotes the conditional probability distribution of $q$ at time $t$ given $Q$, while $\langle A \rangle_{(Q,t)}:=\int dq \ \mathcal{P}(q/Q,t) \ A$ stands for the conditional expectation value of $A$ given $Q$ at time $t$.

If the following unrealistically strong condition is met:
\bea
\left\langle  e^{-i y \frac{\partial S_1^R}{\partial Q}(q,Q,t) }\right\rangle_{(Q,t)} = e^{-i y\left\langle   \frac{\partial S_1^R}{\partial Q}(q,Q,t) \right\rangle_{(Q,t)}}, \label{semiclassicalcondition}
\eea
the reduced Wigner function in \eqref{semiclassicalisingwignerfunctionforparticle} becomes:
\bea
W^{(2)}(Q,P,t) &=&  \mathcal{P}(Q,t) \int \frac{dy}{2\pi} \ e^{i\left(P - \frac{1}{\epsilon} \frac{\partial S_0^R}{\partial Q}(Q,t) - \left\langle   \frac{\partial S_1^R}{\partial Q}(q,Q,t) \right\rangle_{(Q,t)} \right)y}, \\
&=&  \mathcal{P}(Q,t) \ \delta \left(P - \frac{1}{\epsilon} \frac{\partial S_0^R}{\partial Q}(Q,t) - \left\langle   \frac{\partial S_1^R}{\partial Q} \right\rangle_{(Q,t)} \right), \label{Halliwellcriterionparticle}
\eea
which satisfies the Halliwell criterion. This implies that the momentum of the heavy classical particle at $Q$ will be measured as:
\bea
P = \frac{1}{\epsilon} \frac{\partial S_0^R}{\partial Q}(Q,t) + \left\langle   \frac{\partial S_1^R}{\partial Q} \right\rangle_{(Q,t)}. 
\eea
By squaring the above equation and neglecting the term of order $O(\epsilon^0)$, we get:
\bea
  \left(\frac{\partial S_0^R}{\partial Q}(Q,t) \right)^2 &=& \epsilon^2 P^2 - 2 \epsilon \frac{\partial S_0^R}{\partial Q}(Q,t)  \left\langle   \frac{\partial S_1^R}{\partial Q} \right\rangle_{(Q,t)}, \\
  &=& \epsilon^2 P^2 - 2 \epsilon M   \left\langle   \frac{\partial S_1^R}{\partial \tau} \right\rangle_{(Q,t)}, 
\eea
and by substituting it in the Hamilton-Jacobi equation \eqref{splitHJtdep}, we get:
\bea
\frac{\epsilon P^2}{2 M}  + \frac{1}{\epsilon} V_{\mathrm{cl}}(Q) - \left\langle   \frac{\partial S_1^R}{\partial \tau} \right\rangle_{(Q,t)}&=& - \frac{1}{\epsilon} \frac{\partial S^R_0}{\partial t},
\eea
which is the backreacted Hamilton-Jacobi equation for the classical heavy particle with mass $M/\epsilon$ subjected to the potential $\frac{1}{\epsilon} V_{\mathrm{cl}}(Q)$, and the backreaction $\left\langle\frac{\partial S_1^R}{\partial \tau} \right\rangle$.

In case the wave function $\Psi(q,p,Q,P,t)$ is a stationary state as given in Equation \ref{stationaryqmstate} (keeping in mind that we are still considering states with the Type-2 property), the above equation can be simplified to:
\begin{equation}
\frac{\epsilon P^2}{2 M} + \frac{1}{\epsilon} V_{\mathrm{cl}}(Q) + \left\langle H_{quant} \right\rangle_{\tau} = \frac{1}{\epsilon} E_0 + E_1,
\end{equation}
where we have utilized the emergent Schr\"{o}dinger equation \ref{emergentparticleschrodingerequation} to relate the expectation value of $\frac{\partial S_1^R}{\partial \tau}$ to the expectation value of the Hamiltonian of the $q$ particle.

The validity conditions \ref{semiclassicalcondition} are very strong and restrictive, but this is the cost of requiring the $Q$ particle to behave perfectly classically. Let us partially relax this requirement and allow the $Q$ particle to have a Gaussian width in its Wigner function, of the form:
\bea \label{quasiclassicalwignerfunction}
W^{quasiclassical}(Q,P) := &\;& \mathcal{P}(Q) \frac{1}{\sqrt{\pi (\alpha(Q))^2}} e^{-\frac{1}{(\alpha(Q))^2} \left(P-p(Q)\right)^2}, \\
p(Q) := &\;&\frac{1}{\epsilon} \frac{\partial S_0^R}{\partial Q}(Q) + \left\langle   \frac{\partial S_1^R}{\partial Q} \right\rangle_{(Q)},
\eea
where $\alpha(Q)$ is a function of $Q$ with dimensions of $(\text{momentum})^2$, which characterizes the deviation of the state from a classical state at $Q$ along the Gaussian direction. Such a state behaves similar to a classical state and is referred to as a quasi-classical state. 

By imposing the requirement that the reduced Wigner function given in Equation \ref{semiclassicalisingwignerfunctionforparticle} becomes a quasi-classical Wigner function, we arrive at a broader validity condition for the emergent wave function of the $q$ particle:
\bea
\left\langle  e^{-i y \frac{\partial S_1^R}{\partial Q}(q,Q) }\right\rangle_{(Q)} = e^{-\frac{(\alpha(Q))^2}{4} y^2 -i y\left\langle   \frac{\partial S_1^R}{\partial Q}(q,Q) \right\rangle_{(Q)}}. \label{quasisemiclassicalcondition}
\eea

\section{Semiclassical approximation in Canonical Quantum Gravity} \label{wkbinqg}

In the canonical theory of $d+1$-dimensional Quantum Gravity \cite{DeWitt:1967yk}, the state is described by a superposition of metrics and matter fields, denoted by $\Psi[g,\Phi]$. This state satisfies the Wheeler-DeWitt (WDW) equation, which is also referred to as the Hamiltonian constraint, as well as the momentum constraints (in the units $\hbar=c=1$):
\begin{align}
\mathcal{H}(x) & \ \Psi[g,\Phi]:=\Bigg\{\frac{16\pi G_N}{\sqrt{g}}:\Big(\Pi_{ab}\Pi^{ab}-\frac{1}{d-1}\Pi^2\Big):-\frac{\sqrt{g}}{16\pi G_N}(R-2\Lambda) \nonumber \\
&+\frac{1}{2}\left(\frac{1}{\sqrt{g}}
\!:\!\Pi_\Phi^2\!:+\sqrt{g}\left(g^{ab}\nabla_a\Phi\nabla_b\Phi+m^2\Phi^2+V(\Phi,g)\right)\right)\Bigg\}\Psi[g,\Phi]=0, \label{matterHC2}
\end{align}
\begin{equation}
    \mathcal{D}^a(x)\Psi[g,\Phi]:=\left\{-2(\nabla_b\Pi^{ab}) + (\nabla^a\Phi) \Pi_{\Phi} \right\}\Psi[g,\Phi]=0,
\end{equation}
where $g_{ab}(x)$ and $\Phi(x)$ represent the metric and matter field on a $d$-dimensional Cauchy slice $\Sigma$, while $\Pi^{ab}(x)$ and $\Pi_\Phi(x)$ denote the canonical momenta conjugate to $g_{ab}(x)$ and $\Phi(x)$, respectively:
\bea
    \Pi^{ab}(x)&=&-i  \frac{\delta}{\delta g_{ab}(x)}, \\
     \Pi_{\Phi}(x)&=&-i  \frac{\delta}{\delta \Phi(x)},
\eea
and $V(\Phi,g)$ is an arbitrary matter potential. These states are called WDW-states (Wheeler DeWitt states).

Cauchy slices in asymptotically dS spacetimes are closed, and so $\Sigma$ has no boundaries. We refer to this case as a \textbf{closed universe} or \textbf{closed spacetime}. On the other hand, Cauchy slices in asymptotically AdS or asymptotically flat spacetimes are open, and so $\Sigma$ has a boundary $\partial\Sigma$. We refer to this case as an \textbf{open universe} or \textbf{open spacetime}. In the case of a closed universe, the universe exists in a single state known as the wave function of the universe.\footnote{By this, we do not mean that there is a unique wave function of the universe. Instead, we mean that among all possible states, once a wave function of the universe is chosen, it describes the entire contents of the universe, including its past and future. Any notion of time evolution and multiple states of subsystems will only be emergent concepts, which will be explained later.}

For an asymptotically AdS spacetime, the WDW-state on $\Sigma$ depends on where $\Sigma$ is anchored on the boundary (i.e., $\partial\Sigma$). Since $\partial\Sigma$ can be time evolved, the WDW-state depends on the boundary time and satisfies a boundary Schrödinger equation:
\bea
i  \frac{\delta}{\delta T(x)} \Psi[g,\Phi,T] = \mathcal{H}_{\text{ADM}}(x) \Psi[g,\Phi,T],
\eea
where  $T$ represents the boundary time, such that $T(x)=\mathrm{const}$ defines a boundary Cauchy slice $\partial\Sigma_T$, and $\mathcal{H}_{\text{ADM}}(x)$ is the ADM Hamiltonian density at a boundary point (i.e., $x \in \partial\Sigma_t$). For a minimally coupled scalar field, the ADM Hamiltonian is given as follows, according to Hayward and Wong \cite{Hayward:1992ix}:
\bea \label{HADM}
\mathcal{H}_{\text{ADM}} = \frac{1}{\sqrt{1-\left(V\right)^2}} \left(\frac{\sqrt{\sigma}}{8\pi G_N} \Gamma + 2 V M \Pi^{rr}   \right) ,
\eea
where $\Gamma$ is the trace of the extrinsic $(d-1)$-curvature of $\partial\Sigma$ as embedded in the $d$-geometry of $\Sigma$, $\sigma$ is the determinant of the induced metric on $\partial\Sigma$, and $r$ is the radial coordinate (with $\partial\Sigma$ being a constant $r$ surface in $\Sigma$). Additionally, $V:=\frac{N^rM}{N}$, where $N$ is the lapse function, $N^r$ is the $r$-component of the shift vector, and $M$ is the radial lapse associated with the proper distance between constant $r$ slices on $\Sigma$.

\subsection{Emergent spacetime from a WDW state}

Starting from the WDW equation, we apply the semiclassical approximation based on the work of Kiefer, Padmanabhan, and Singh \cite{Kiefer:1993fg,Kiefer:1991ef,Kiefer:1993yg} for both $\Lambda>0$ and $\Lambda<0$ cases. In units where the cosmological constant is set to:
\bea
\Lambda = \pm \frac{d(d-1)}{2},
\eea
(where the plus and minus signs correspond to dS and AdS cases, respectively), $G_N$ becomes dimensionless. We perform the analysis in the $G_N \to 0$ limit, which makes the metric classical while keeping the matter field quantum. By expanding the WDW-state in a power series of $G_N$:
\bea
\Psi[g,\Phi] &=& e^{i S[g,\Phi]}, \\
S[g,\Phi] &=& \sum_{n=0}^\infty (32 \pi G_N)^{n-1} S_n[g,\Phi],
\eea
and inserting it into the Hamiltonian constraint, we get at order $O(G_N^{-2})$:
\bea
\frac{\delta}{\delta \Phi(x)} S_0[g,\Phi] =0,
\eea
and thus $S_0=S_0[g]$ depends only on the metric. At the next order $O(G_N^{-1})$, we obtain the Hamilton-Jacobi-Einstein equations from the Hamiltonian constraint. It is worth noting that the Hamilton-Jacobi-Einstein equation is equivalent to all ten Einstein field equations \cite{Gerlach:1969ph}.
\bea \label{HJEeqn}
\frac{1}{4}G_{abcd} \left(\frac{\delta S_0}{\delta g_{ab}}\right)\left(\frac{\delta S_0}{\delta g_{cd}} \right)-\sqrt{g}(R-2\Lambda)=0,
\eea
where $G_{abcd}$ is the metric on the superspace $\mathcal{S}$ (the set of all metrics on $\Sigma$):
\bea
G_{abcd} := \frac{1}{2 \sqrt{g}} \left(g_{ac}g_{bd}+g_{ad}g_{bc}-\frac{2}{d-1}g_{ab}g_{cd} \right).
\eea
For the asymptotically AdS case, at order $O(G_N^{-1})$ we obtain another equation from the boundary Schr\"{o}dinger equation:
\bea
-\frac{1}{4}  \frac{\delta S_0}{\delta T(x)}  = \frac{1}{\sqrt{1-\left(V\right)^2}} \left(\sqrt{\sigma} \ \Gamma +  V M \frac{\delta S_0}{\delta g_{rr}}\right),
\eea
which is the boundary Hamilton-Jacobi-Einstein equation.

For $d$-geometries that are far from the minima of $S_0^I$, the state $\Psi$ is heavily suppressed in the $G_N \to 0$ limit. Therefore, one needs to constrain the superspace to a subspace in which the imaginary part of the Hamilton's principal function attains its minimum value, i.e., $S_0^I[g]=0$. Note that one could always shift the minimum value of $S_0^I$ to zero by adjusting an overall factor, which can be absorbed into the normalization constant. We refer to this subspace as the semiclassical superspace $\mathcal{S}_{\text{semiclassical}}$:
\bea
\mathcal{S}_{\text{semiclassical}} := \left\{g \in \mathcal{S} \Bigg| S_0^I[g] = 0  \right\}.\label{semiclassicalsuperspace}
\eea
The classical momentum conjugate to the metric can now be obtained from the Hamilton's principal function:
\bea
\Pi^{ab}_{\mathrm{cl}}(x) = \frac{1}{32 \pi G_N} \frac{\delta S_0}{\delta g_{ab}(x)} \Bigg|_{g \in \mathcal{S}_{\text{semiclassical}}}. \label{classicalmetricmomentum}
\eea
The metric $g \in \mathcal{S}_{\text{semiclassical}}$, together with its canonical conjugate momentum $\Pi^{ab}_{\mathrm{cl}}$, provides the initial conditions on a Cauchy slice, which can be used to obtain a $d+1$-dimensional classical spacetime that satisfies the Einstein equations. In this way, a classical spacetime emerges from a WDW state.

\subsection{Constructing a $d+1$ dimensional classical geometry from Einstein-Hamilton principal function}\label{constructingaclassicalspacetime}

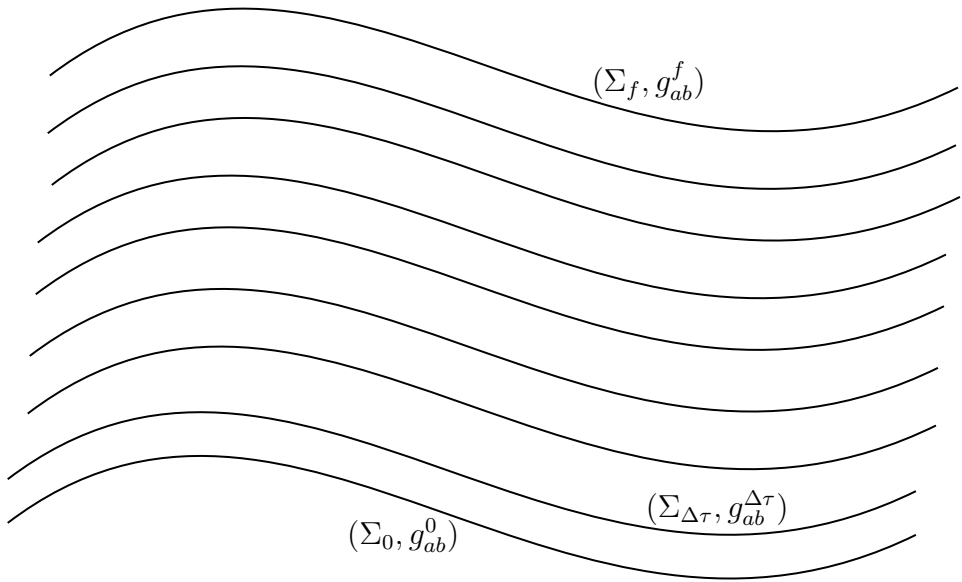
\begin{figure}[h]
\centering

\tikzset{every picture/.style={line width=0.75pt}} 

\begin{tikzpicture}[x=0.75pt,y=0.75pt,yscale=-1,xscale=1]

\draw    (87,344) .. controls (230.33,234) and (366.33,436) .. (540.33,350) ;
\draw    (87,322) .. controls (230.33,212) and (366.33,414) .. (540.33,328) ;
\draw    (97,289) .. controls (240.33,179) and (376.33,381) .. (550.33,295) ;
\draw    (98,260) .. controls (241.33,150) and (377.33,352) .. (551.33,266) ;
\draw    (101,229) .. controls (244.33,119) and (380.33,321) .. (554.33,235) ;
\draw    (102,203) .. controls (245.33,93) and (381.33,295) .. (555.33,209) ;
\draw    (109,174) .. controls (252.33,64) and (388.33,266) .. (562.33,180) ;
\draw    (107,148) .. controls (250.33,38) and (386.33,240) .. (560.33,154) ;
\draw    (108,119) .. controls (251.33,9) and (387.33,211) .. (561.33,125) ;

\draw (255,340.4) node [anchor=north west][inner sep=0.75pt]    {$( \Sigma_0 ,g^{0}_{ab})$};
\draw (404,326.4) node [anchor=north west][inner sep=0.75pt]    {$( \Sigma_{\Delta \tau} ,g^{\Delta \tau}_{ab})$};
\draw (377,110.4) node [anchor=north west][inner sep=0.75pt]    {$( \Sigma_{f} ,g^{f}_{ab})$};

\end{tikzpicture}

\caption{\small Construction of a classical spacetime from Einstein Hamilton principal function. One starts with an initial slice with some metric and finds the extrinsic curvature from the Einstein Hamilton principal function. Then by choosing a gauge for the lapse and shift, one finds the metric on the next slice and repeats this to stitch a spacetime. This stitched spacetime geometry would then satisfy the Einstein equations.}
\label{constructingspacetime}
\end{figure}

To explicitly obtain a $(d+1)$-dimensional spacetime from the Einstein-Hamilton principal function $S_0[g]$, one must perform the following steps. Consider an \emph{abstract} $d$ dimensional Riemannian manifold, $\Sigma_0$, with a metric $g^{0}_{ab}$, such that $g^{0}_{ab} \in \mathcal{S}_{\text{semiclassical}}$ (see Fig. \ref{constructingspacetime}). The extrinsic curvature $K^{ab}$ can be obtained by
\bea\label{classicalK}
K_{ab} = 16 \pi G_N G_{abcd} \Pi^{cd}_{\mathrm{cl}} = \frac{1}{2} G_{abcd}  \frac{\delta S_0}{\delta g_{ab}(x)} \Bigg|_{g \in \mathcal{S}_{\text{semiclassical}}},
\eea
If one had been given a classical spacetime to begin with, then the extrinsic curvature would be related to the lapse $N$, shifts $N_a$, and the derivative of the spatial metric $\partial_t g_{ab}$ with respect to a time coordinate $t$ by:
\bea
K_{ab} = \frac{1}{2N} \left(\partial_t g_{ab} -\nabla_a N_b -\nabla_b N_a \right),
\eea
where $\nabla$ is the connection compatible with the spatial metric $g$. The lapse and the shifts are gauge degrees of freedom and can be picked to be anything. For simplicity, let us pick the shifts to be zero. In this gauge, the extrinsic curvature is related to the derivative of the spatial metric with respect to the proper time $\tau$:
where $\nabla$ is the connection compatible with the spatial metric $g$. The lapse and the shifts are gauge degrees of freedom and can be chosen arbitrarily. For simplicity, let us set the shifts to zero. In this gauge, the extrinsic curvature is related to the derivative of the spatial metric with respect to the proper time $\tau$:
\bea
K_{ab} = \frac{1}{2N} \partial_t g_{ab} = \frac{1}{2} \partial_\tau g_{ab}.
\eea

As the spacetime emerges from the WDW state, the notion of $\tau$ on the right-hand side of the above equation is inherited from $K_{ab}$, which in turn inherits its notion from $\Pi^{ab}_{\mathrm{cl}}$, which derives its notion from the Einstein Hamilton principal function. Now that the derivative of the spatial metric is known, one must consider a ``next" slice, $\Sigma_{\Delta \tau}$, with a new metric $g^{\Delta \tau}_{ab}$ such that:
\bea
2 K_{ab} = \frac{g^{\Delta \tau}_{ab}-g^{0}_{ab}}{\Delta \tau},
\eea
where $\Delta \tau$ is an infinitesimal proper time. Next, one repeats this entire process with the updated metric. As long as the updated metric belongs to the semiclassical superspace, one can proceed with the construction of a $d+1$ dimensional spacetime from $S_0$. Finally, since $S_0$ satisfies the Einstein-Hamilton-Jacobi equation, the constructed spacetime will satisfy the Einstein equations.

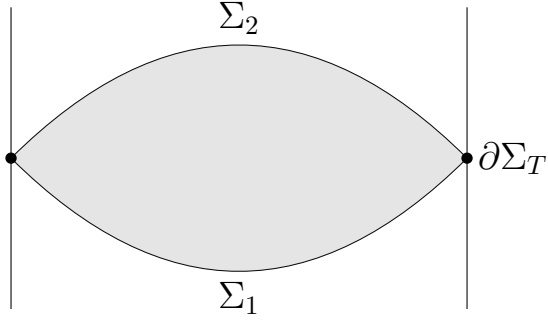
\begin{figure}[h]
\centering
\begin{tikzpicture}[font=\sffamily]
  \def\a{3};
  \def\b{2};
  \draw (-\a,-\b) -- (-\a,\b); 
  \draw (\a,-\b) -- (\a,\b) node[midway,right][fontscale=1]{$\partial\Sigma_{T}$};
  \draw (-\a,0) ..controls (-\a/3,\b) and (\a/3,\b).. (\a,0) node[midway,above] [fontscale=1]{$\Sigma_2$};
  \draw (-\a,0) ..controls (-\a/3,-\b) and (\a/3,-\b).. (\a,0) node[midway,below] [fontscale=1]{$\Sigma_1$};
  \fill [gray,opacity=0.2] (-\a,0) ..controls (-\a/3,\b) and (\a/3,\b).. (\a,0);
  \fill [gray,opacity=0.2] (-\a,0) ..controls (-\a/3,-\b) and (\a/3,-\b).. (\a,0);
  \node at (\a,0)[circle,fill,inner sep=1.5pt]{};
  \node at (-\a,0)[circle,fill,inner sep=1.5pt]{};
  \end{tikzpicture}
\caption{\small When the lapse vanishes at $\partial\Sigma_T$, we get the emergent classical spacetime by WKB approximation between $\Sigma_1$ and $\Sigma_2$ both of which end on $\partial\Sigma_T$.}
\label{asymptoticallyvanishinglapse}
\end{figure}

For closed universes, the above construction is the end of the story. However, for asymptotically AdS spacetimes, one must also take the boundary into account. The choice of lapse at the boundary is not a gauge degree of freedom and it corresponds to the proper time evolution at the boundary. If one chooses a lapse function that vanishes at $\partial\Sigma_T$, then one would obtain a $(d+1)$-dimensional metric on a manifold $\mathcal{M}$ which interpolates between two $d$-dimensional manifolds, $\Sigma_1$ and $\Sigma_2$, both ending on $\partial\Sigma_T$ (i.e., $\partial\Sigma_1=\partial\Sigma_2=\partial\Sigma_T$), as shown in Fig. $\ref{asymptoticallyvanishinglapse}$.

\begin{figure}[h]
\centering
\begin{tikzpicture}[font=\sffamily]
  \def\a{3};
  \def\b{2};
  \draw (-\a,-\b) -- (-\a,\b); 
  \draw (\a,-\b) -- (\a,\b);
  \draw (-\a,\b/2) ..controls (-\a/3,\b) and (\a/3,\b).. (\a,\b/2)node[midway,above] [fontscale=1]{$\Sigma_{T_2}$};
  \draw (-\a,-\b/2) ..controls (-\a/3,-\b) and (\a/3,-\b).. (\a,-\b/2) node[midway,below] [fontscale=1]{$\Sigma_{T_1}$};
  \fill [gray,opacity=0.2] (-\a,\b/2) ..controls (-\a/3,\b) and (\a/3,\b).. (\a,\b/2) --  (\a,-\b/2)..controls (\a/3,-\b) and (-\a/3,-\b) .. (-\a,-\b/2);
  \node at (\a+0.7,\b/2) {$\partial\Sigma_{T_2}$};
  \node at (\a+0.7,-\b/2) {$\partial\Sigma_{T_1}$};
  \node at (\a,\b/2)[circle,fill,inner sep=1.5pt]{};
  \node at (\a,-\b/2)[circle,fill,inner sep=1.5pt]{};
  \node at (-\a,\b/2)[circle,fill,inner sep=1.5pt]{};
  \node at (-\a,-\b/2)[circle,fill,inner sep=1.5pt]{};
  \end{tikzpicture}
\caption{\small When the lapse does not vanish at the boundary of the Cauchy slice, we get the emergent classical spacetime by WKB approximation between $\Sigma_{T_1}$ and $\Sigma_{T_2}$.}
\label{asymptoticallynonvanishinglapse}
\end{figure}
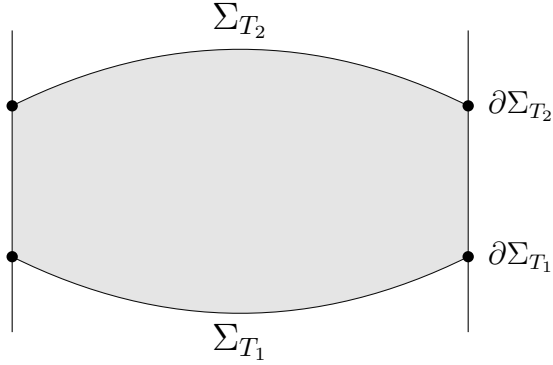

If a non-vanishing lapse is chosen at the boundary, it would correspond to pushing the boundary $\partial\Sigma_T$ along the boundary time. From the semiclassical approximation, one would obtain the $(d+1)$-dimensional metric on a manifold that interpolates between $\Sigma_{T_1}$ (which ends on $\partial\Sigma_{T_1}$) and $\Sigma_{T_2}$ (which ends on $\partial\Sigma_{T_2}$), as shown in Fig. \ref{asymptoticallynonvanishinglapse}.

\subsection{Emergent QFT in curved spacetime from a WDW state}

Proceeding with the WKB approximation, at order $O(G_N^0)$, the Hamiltonian constraint gives us:
\bea
i \left(G_{abcd} \frac{\delta S_0}{\delta g_{ab}}\frac{\delta}{\delta g_{cd}} \right) e^{i S_1} = \mathcal{H}_{\mathrm{matter}} \ e^{i S_1} -\frac{i}{2} G_{abcd} \frac{\delta^2 S_0}{\delta g_{ab}\delta g_{cd}} e^{i S_1}. \label{emergentqftprototype}
\eea
By defining the effective wavefunctional for the field theory (in a fixed classical background) as:
\bea \label{emergentQFTstate}
\psi[\Phi,g]:= \mathcal{N}[g] \ e^{i S_1[g,\Phi]},
\eea
where $\mathcal{N}[g]$ is the normalization factor defined as:
\bea
\mathcal{N}^{-2}[g]:= \int D\phi \left|e^{i S_1[\phi,g]} \right|^2.
\eea
By rearranging equation \eqref{emergentqftprototype}, one gets:
\bea 
i \left(G_{abcd} \frac{\delta S_0}{\delta g_{ab}}\frac{\delta}{\delta g_{cd}} \right) \psi = \left\{ \mathcal{H}_{\mathrm{matter}} + \frac{1}{2} G_{abcd} \frac{\delta^2 S_0^I}{\delta g_{ab}\delta g_{cd}}  \right\} \ \psi.
\eea
As we have restricted to the semiclassical superspace $\mathcal{S}_{\text{semiclassical}}$, the above equation becomes:
\bea
i \left(32 \pi G_N G_{abcd} \Pi^{ab}_{\mathrm{cl}}\frac{\delta}{\delta g_{cd}} \right) \psi = \left\{ \mathcal{H}_{\mathrm{matter}} + \frac{1}{2} G_{abcd} \frac{\delta^2 S_0^I}{\delta g_{ab}\delta g_{cd}}  \right\}  \ \psi. \label{emergentqftprototype2}
\eea
The classical momentum $\Pi^{ab}_{\mathrm{cl}}$ is related to the $d$-dimensional extrinsic curvature $K_{ab}$ of the Cauchy slices embedded in a classical $(d+1)$-dimensional background metric, as described by Equation \ref{classicalK}. Furthermore, the extrinsic curvature $K_{ab}$ can be expressed in terms of the time derivative of the metric on the Cauchy slices, as well as the lapse function and shift vector:
\bea
K_{ab} = \frac{1}{2N}\left(\partial_t g_{ab} - \nabla_{a}N_{b}-\nabla_{b}N_{a}\right).
\eea
Equation \eqref{emergentqftprototype2} now becomes:
\bea
i \left(2 K_{ab}\frac{\delta}{\delta g_{ab}} \right) \psi &=& \left\{ \mathcal{H}_{\mathrm{matter}} + \frac{1}{2} G_{abcd} \frac{\delta^2 S_0^I}{\delta g_{ab}\delta g_{cd}}  \right\}  \ \psi,\\
i \left(\frac{1}{N}\left(\partial_t g_{ab} - \nabla_{a}N_{b}-\nabla_{b}N_{a}\right)  \right)\frac{\delta}{\delta g_{ab}} \psi &=& \left\{ \mathcal{H}_{\mathrm{matter}} + \frac{1}{2} G_{abcd} \frac{\delta^2 S_0^I}{\delta g_{ab}\delta g_{cd}}  \right\}  \ \psi.
\eea
The $t$ derivative divided by the lapse is just the derivative with respect to the proper time $\tau$. By fixing the shifts to zero, the above equation becomes:
\bea
i\frac{\delta}{\delta \tau(x)} \psi =  \left\{ \mathcal{H}_{\mathrm{matter}} + \frac{1}{2} G_{abcd} \frac{\delta^2 S_0^I}{\delta g_{ab}\delta g_{cd}}  \right\}  \ \psi,\label{preemergentTSequation}
\eea
where the functional derivative with respect to $\tau(x)$ is:
\bea
\frac{\delta}{\delta \tau(x)} := \frac{1}{N(x)}\partial_t g_{ab(x)} \frac{\delta}{\delta g_{ab}(x)}.
\eea
For states with the Type-1 property, the second term in the RHS of Equation \eqref{preemergentTSequation} can be non-zero and must be taken as a contribution to the QFT Hamiltonian. However, for states with the Type-2 property, it vanishes, and we get:
\bea
i\frac{\delta}{\delta \tau(x)} \psi =   \mathcal{H}_{\mathrm{matter}}   \ \psi,\label{emergentTSequation}
\eea
which is the Tomonaga-Schwinger equation \cite{Tomonaga:1946zz,Koba:1947rzy,Schwinger:1948yk} for a quantum field theory in curved spacetime.

\subsection{Semiclassical Einstein equation from a WDW state} \label{semiclassicaleinsteinequationfromWDW}

The semiclassical Einstein equation is:
\bea \label{SEeqn}
\textbf{R}_{\mu \nu}-\frac{1}{2}\textbf{R} \textbf{g}_{\mu \nu} + \Lambda \textbf{g}_{\mu \nu} = \frac{8 \pi G_N}{c^4} \langle T_{\mu \nu} \rangle_{\psi},
\eea
where $\textbf{g}_{\mu \nu}$ is the $d+1$-dimensional metric and its Ricci tensor is $\textbf{R}_{\mu\nu}$. We will derive this with an analogous Wigner functional analysis for quantum gravity. The Wigner functional for gravity coupled to a scalar field is\footnote{We are unaware of any formulation of Wigner functional for quantum gravity. Therefore, we extrapolated the form of the Wigner function in quantum mechanics to quantum gravity. Ideally, one must formulate a QG Wigner functional in the same spirit as how it is formulated in QM and repeat the whole upcoming analysis if it is different from the one used in this paper.}
\bea\label{WDWWignerfunctional}
W[g,\Pi,\phi,\Pi_\phi] = \int Dh D\tilde{\phi} \Psi^*[g+h,\phi+\tilde{\phi}]\Psi[g-h,\phi-\tilde{\phi}] e^{\left(2 i \int d^dx \left(h_{ab} \Pi^{ab} + \tilde{\phi} \Pi_\phi \right) \right)},
\eea
where the integration is done over all possible perturbations $h_{ab}$ of the metric (such that $g_{ab}+h_{ab}$ spans the entire superspace). By rescaling $h_{ab}$, the Wigner functional becomes:
\bea
W[g,\Pi,\phi,\Pi_\phi] &=& \int D(32\pi G_N h) D\tilde{\phi} \  e^{\left(2 i \int d^dx \left((32\pi G_N)h_{ab} \Pi^{ab} + \tilde{\phi} \Pi_\phi \right) \right)} \\
&\;& \Psi^*[g+(32\pi G_N) h,\phi+\tilde{\phi}]\Psi[g-(32\pi G_N) h,\phi-\tilde{\phi}].
\eea
The reduced Wigner functional of gravity is obtained by integrating out $\phi$ and $\Pi_\phi$:
\bea
W[g,\Pi] &=& \int D\phi D\Pi_\phi \ W[g,\Pi,\phi,\Pi_\phi],\\
&=& \int D\phi  \int D(32\pi G_N h) \ e^{\left(2 i (32\pi G_N) \int d^dx h_{ab} \Pi^{ab}  \right)} \\
&\;& \Psi^*[g+(32\pi G_N) h,\phi]\Psi[g-(32\pi G_N) h,\phi].
\eea
For $g \in \mathcal{S}_{\mathrm{semiclassical}}$, the Wigner functional takes the following form after a power series expansion in $G_N$ and Taylor expansion of the integrand about $h_{ab}=0$:\footnote{For $g \notin \mathcal{S}_{\mathrm{semiclassical}}$, the Wigner functional gets heavily suppressed in the $G_N \to 0$ limit.}
\begin{align}
W[g,\Pi] &=  \int   D(32\pi G_N h) \ \exp\left(64 i \pi G_N \int d^dx \left(\Pi^{ab}-\frac{1}{32\pi G_N}\frac{\delta S_0^R}{\delta g_{ab}}\right) h_{ab}   \right) \\
&\; \int D\phi \ e^{-2 S_1^I[g,\phi]} \ \exp\left(- 64 i \pi G_N \int d^dx \frac{\delta S_1^R}{\delta g_{ab}} h_{ab} \right), \\
&= \mathcal{P}[g] \int   Dh \ \exp\left(i \int d^dx \left(\Pi^{ab}-\frac{1}{32\pi G_N}\frac{\delta S_0^R}{\delta g_{ab}}\right) h_{ab}   \right) \\
&\; \int D\phi \ \mathcal{P}[\phi/g] \ \exp\left(-i  \int d^dx \frac{\delta S_1^R}{\delta g_{ab}} h_{ab} \right),
\end{align}
where $\mathcal{P}[g]:=\int D\phi \ e^{-2 S_1^I[g,\phi]}$ and $\mathcal{P}[\phi/g]:=\frac{1}{\mathcal{P}[g]}   e^{-2 S_1^I[g,\phi]}$.
If the following condition is satisfied (which we will take as a necessary condition for the subsequent analysis to hold):
\bea \label{WDWscvalidity}
\left\langle e^{- i  \int d^dx \frac{\delta S_1^R}{\delta g_{ab}} h_{ab}} \right\rangle =  e^{- i  \int d^dx \left\langle \frac{\delta S_1^R}{\delta g_{ab}} \right\rangle h_{ab}}  ,
\eea
where $\langle A \rangle:=\int D\phi \ \mathcal{P}[\phi/g] \ A$, the Wigner functional becomes:
\begin{align}
W[g,\Pi] &= \mathcal{P}[g] \int   Dh \ \exp\left(i  \int d^dx \left[\Pi^{ab}-\frac{1}{32\pi G_N}\frac{\delta S_0^R}{\delta g_{ab}} - \left\langle \frac{\delta S_1^R}{\delta g_{ab}} \right\rangle\right] h_{ab}   \right), \\
&= \mathcal{P}[g] \ \delta^\infty\left[\frac{1}{\sqrt{g}}\left(\Pi^{ab}-\frac{1}{32\pi G_N}\frac{\delta S_0^R}{\delta g_{ab}} - \left\langle \frac{\delta S_1^R}{\delta g_{ab}} \right\rangle\right)\right]. \label{DiracdeltaWignergravity}
\end{align}
This Wigner functional is analogous to Equation \ref{Halliwellcriterionparticle}. We will also refer to the condition requiring the Wigner functional to take this form as the Halliwell criterion, but for the case of gravity. The distinction should be clear from the context. Unlike the earlier case, for gravity, the Halliwell criterion becomes a local criterion that needs to be satisfied everywhere on $\Sigma$ for gravity to be treated classically everywhere on $\Sigma$. It is conceivable that the Halliwell criterion is satisfied in some regions of $\Sigma$ while it is not in others. We will discuss this scenario later in section \ref{subregionclassicalisationscreens}. This section is dedicated to the case where the Halliwell criterion is satisfied everywhere on $\Sigma$.

As a result, the Wigner functional dictates that the conjugate momentum to the metric, denoted by $\Pi^{ab}$, should take the form:
\bea\label{backreactedmomentum}
\Pi^{ab}=\frac{1}{32\pi G_N}\frac{\delta S_0^R}{\delta g_{ab}} + \left\langle \frac{\delta S_1^R}{\delta g_{ab}} \right\rangle.
\eea
From the above equation, we obtain (ignoring the $O(G_N)$ term):
\bea
G_{abcd} \left(\frac{\delta S^R_0}{\delta g_{ab}}\right)\left(\frac{\delta S^R_0}{\delta g_{cd}} \right)&=&(32\pi G_N)^2 G_{abcd} \Pi^{ab}\Pi^{cd} - 2 (32\pi G_N) \left\langle \frac{\delta S_1^R}{\delta \tau} \right\rangle,
\eea
By substituting this in \eqref{HJEeqn}, we obtain the back-reacted Hamilton-Jacobi-Einstein equation:
\bea \label{brHJEeqn}
(16\pi G_N) G_{abcd} \Pi^{ab}\Pi^{cd} -\frac{\sqrt{g}}{16\pi G_N}(R-2\Lambda) +  \left\langle \mathcal{H}_{\mathrm{matter}} \right\rangle =0,
\eea
where we used $-\left\langle \frac{\delta S_1^R}{\delta \tau(x)} \right\rangle = \left\langle \mathcal{H}_{\mathrm{matter}}(x) \right\rangle$, which follows from \eqref{emergentTSequation}.
The back-reacted Hamilton-Jacobi-Einstein equation given in \eqref{brHJEeqn} implies that the $d+1$-dimensional spacetime satisfies the semiclassical Einstein equation \eqref{SEeqn}.

\textbf{Validity conditions:} For the analysis conducted in this subsection to be valid, we assumed the condition in equation \ref{WDWscvalidity} to be satisfied. These are conditions that states in the emergent QFT must meet. Realistically, only a few QFT states would fulfill this condition. This is the strictest form of the condition, implying that gravity can be treated as perfectly classical. An approximation is said to hold when the error is less than a certain tolerance level. This tolerance level is arbitrarily chosen based on the sensitivity of the measurement apparatus and the precision required for individual experiments, but it must certainly be finite. 

Therefore, as long as the Wigner functional for gravity is peaked in a way similar to equation \ref{DiracdeltaWignergravity} (for example, a Gaussian with the same peak), and the variance (for example, the width of the Gaussian) is less than the tolerance level, we can say that the Halliwell criterion is sufficiently satisfied and gravity behaves classically. However, one thing we cannot allow is for the variance to blow up; in such a case, we cannot consider semiclassical gravity to be valid under any circumstances. This is the weakest form of the validity condition and realistically, an intermediate state between the strictest and weakest conditions would be ideal for experiments.

\section{Applications
in AdS/CFT} \label{AAdSCFT}

In the AdS/CFT correspondence \cite{tHooft:1973alw,Maldacena:1997re,Witten:1998qj,Aharony:1999ti}, the correlation functions of the boundary CFT match the correlation functions of its dual bulk quantum gravity. These are typically computed in the large $N$ limit and around a fixed bulk background. For instance, the $n$-point functions of some low-energy operators sandwiched between the CFT ground state, $\langle \Omega|\mathcal{O}_1 \dots \mathcal{O}_n | \Omega \rangle$, are equal to certain correlation functions in the bulk on a globally AdS spacetime. One might expect the bulk spacetime geometry (on which such correlation functions are computed) to depend on the operators $\mathcal{O}_i$. However, in the large $N$ limit, the backreaction of these operator insertions is negligible, as long as they are low-energy operators, meaning their backreaction does not grow strongly with some power of $N$.

The CFT ground state is dual to the empty global AdS bulk spacetime, and other states in the CFT could be dual to different bulk spacetimes in the large $N$ limit. However, this is not generally true. For example, if $\psi_1$ and $\psi_2$ are CFT states dual to two vastly different bulk spacetime geometries $\mathbf{g}_1$ and $\mathbf{g}_2$, respectively, then the CFT state $\psi_3 = \alpha \ \psi_1 + \beta \ \psi_2$ would not necessarily have a classical geometry dual. Consequently, if one computes correlation functions on the CFT side of the form $\langle \psi_3|\mathcal{O}_1 \dots \mathcal{O}_n | \psi_3 \rangle$, they will not necessarily match the bulk correlation functions on a single background geometry, even in the large $N$ limit. However, by expanding
\bea
\langle \psi_3|\mathcal{O}_1 \dots \mathcal{O}_n | \psi_3 \rangle &= |\alpha|^2 \langle \psi_1|\mathcal{O}_1 \dots \mathcal{O}_n | \psi_1 \rangle + |\beta|^2 \langle \psi_2|\mathcal{O}_1 \dots \mathcal{O}_n | \psi_2 \rangle \\ 
&+ \alpha \beta^* \langle \psi_2|\mathcal{O}_1 \dots \mathcal{O}_n | \psi_1 \rangle + \alpha^* \beta \langle \psi_1|\mathcal{O}_1 \dots \mathcal{O}_n | \psi_2 \rangle,
\eea
one can expect to match $\langle \psi_1|\mathcal{O}_1 \dots \mathcal{O}_n | \psi_1 \rangle$ to bulk correlation functions on the background $\mathbf{g}_1$, and match $\langle \psi_2|\mathcal{O}_1 \dots \mathcal{O}_n | \psi_2 \rangle$ to bulk correlation functions on the background $\mathbf{g}_2$. However, it remains unclear which bulk background to choose in order for the bulk correlation functions to match $\langle \psi_2|\mathcal{O}_1 \dots \mathcal{O}_n | \psi_1 \rangle$. If one chooses to re-express $| \psi_i \rangle$ as some heavy operators acting on $| \Omega \rangle$, then on the bulk side, it could correspond to quanta that heavily backreact on the background geometry, and it is unclear to what extent perturbation theory could be valid.

The situation becomes worse in physical processes where the future large spacetime geometry is heavily influenced by tiny changes in the initial QFT state on some fixed initial spacetime geometry. For example, consider a process in which a massive bowling ball could be split in half and shot in opposite directions, but the direction depends on the spin of an electron. Changing the state of the electron could lead to different spacetimes in the future, and if the electron were in a superposition of the spin states, then the future geometry would also be forced into a superposition.

All of these conceptual points can be resolved when one starts with a dictionary that maps arbitrary CFT states to bulk states by encoding the superposition of bulk geometries. We should allow the bulk QG state to determine the appropriate background in the bulk, if any, on which to compute the correlation functions. Furthermore, the bulk QG state would provide us with precise conditions to determine when a perturbative sector is valid. Such a dictionary is the CSH dictionary formulated in \cite{GRW}, which we briefly describe now.

\subsection{The CSH dictionary}

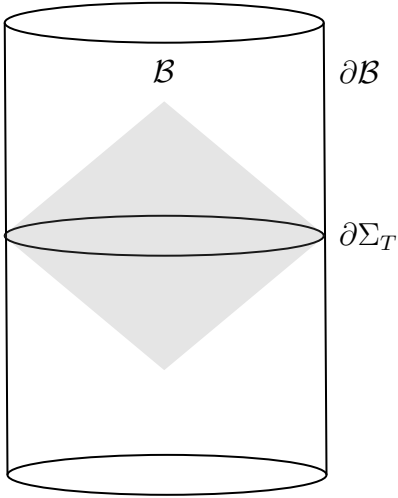
\begin{figure}[h]
    \centering
\tikzset{every picture/.style={line width=0.75pt}} 
\begin{tikzpicture}[x=0.75pt,y=0.75pt,yscale=-1,xscale=1]

\draw    (242.5,22.54) -- (243.88,249.96) ;
\draw    (401.73,22.54) -- (403.12,249.96) ;
\draw   (242.5,129.69) .. controls (242.5,124.14) and (278.15,119.64) .. (322.12,119.64) .. controls (366.09,119.64) and (401.73,124.14) .. (401.73,129.69) .. controls (401.73,135.24) and (366.09,139.73) .. (322.12,139.73) .. controls (278.15,139.73) and (242.5,135.24) .. (242.5,129.69) -- cycle ;
\filldraw[gray,opacity=0.2]  (322.12,62.72) -- (401.73,129.69) -- (322.12,196.65) -- (242.5,129.69) -- cycle ;
\draw   (243.88,249.96) .. controls (243.88,244.41) and (279.53,239.91) .. (323.5,239.91) .. controls (367.47,239.91) and (403.12,244.41) .. (403.12,249.96) .. controls (403.12,255.5) and (367.47,260) .. (323.5,260) .. controls (279.53,260) and (243.88,255.5) .. (243.88,249.96) -- cycle ;
\draw   (242.5,22.54) .. controls (242.5,17) and (278.15,12.5) .. (322.12,12.5) .. controls (366.09,12.5) and (401.73,17) .. (401.73,22.54) .. controls (401.73,28.09) and (366.09,32.59) .. (322.12,32.59) .. controls (278.15,32.59) and (242.5,28.09) .. (242.5,22.54) -- cycle ;

\draw (408,121.86) node [anchor=north west][inner sep=0.75pt]   [align=left] {$\partial\Sigma_T$};
\draw (408,40) node [anchor=north west][inner sep=0.75pt]   [align=left] {$\partial\mathcal{B}$};
\draw (315,40) node [anchor=north west][inner sep=0.75pt]   [align=left] {$\mathcal{B}$};

\end{tikzpicture}
    \caption{\small A Cauchy slice $\partial\Sigma_T$ of the boundary spacetime $\partial\mathcal{B}$ is depicted. The field theory state $\psi_{\mathrm{bdy}}[\{\chi\},T]$ exists on $\partial\Sigma_T$. The WDW-state $\Psi[g,\Phi,T]$, dual to $\psi_{\mathrm{bdy}}[\{\chi\},T]$, resides on $\Sigma_T$, which is a \emph{abstract} $d$-dimensional manifold with $\partial\Sigma_T$ as its boundary. The shaded region illustrates the WDW patch of $\partial\Sigma_T$, which emerges from the WDW state $\Psi[g,\Phi,T]$ when it is a semiclassical state.}
    \label{AdStincan2}
\end{figure}

Consider a $(d+1)$-dimensional quantum gravity in an asymptotically-AdS bulk $\mathcal{B}$, which is dual to a $d$-dimensional holographic CFT living on the boundary $\partial\mathcal{B}$. We choose a \emph{complete commuting set of operators} $\{\chi\}$ acting on the CFT Hilbert space. This selection defines a basis of states labeled by the set of eigenvalues of $\{\chi\}$\footnote{We will be using the same symbol $\{\chi\}$ to denote both the complete set of operators and the set of eigenvalues used as boundary conditions for the partition function. The context should make this distinction clear.}. A general state on $\partial\Sigma_T$ (a Cauchy slice of the boundary) will be a superposition thereof: $\psi_{\mathrm{bdy}}[\{\chi\},T]$ (see Figure \ref{AdStincan2}). The CSH dictionary in \cite{GRW} maps this CFT state to a bulk WDW state as follows:
\bea
\Psi[g,\Phi,T] = Z_{T^2}[g,\phi;\psi_{\mathrm{bdy}}] = \int d\{\chi\}\,  Z_{T^2}[g,\phi;\{\chi\}] \, \psi_{\mathrm{bdy}}[\{\chi\},T],\label{dualWDWstate}
\eea
where, $Z_{T^2}[g,\phi;\{\chi\}]$ is the partition function of the $T^2$ deformed field theory (which satisfies the Hamiltonian constraint) on $\Sigma_T$ (an \emph{abstract} $d$-dimensional manifold whose boundary is $\partial\Sigma_T$) with the metric $g$ and matter field $\phi$ as the background sources.\footnote{
The field theory source $\phi$ is not precisely the same as the bulk matter field $\Phi$. Instead, they are related to each other by a scaling factor involving the AdS length scale $L_{\mathrm{AdS}}$, which, in the units we are working with, is set to one.} Since $\Sigma_T$ has a boundary, $\partial\Sigma_T$, the partition function will also depend on the boundary conditions, and the integration is performed over these boundary conditions, weighted with the CFT state $\psi_{\mathrm{bdy}}[\{\chi\},T]$.

It is important to note that $\Sigma_T$ is not necessarily a Cauchy slice embedded in a $d+1$-dimensional background, because in quantum gravity, a background is not given a priori. However, when the WDW state is a semiclassical state, a $d+1$-dimensional spacetime emerges from $\Psi$ that constructs the WDW patch (illustrated as the shaded region in Figure \ref{AAdSCFT}).\footnote{The WDW patch of a boundary Cauchy slice $\partial\Sigma$, is defined as the set of all points $x \in \mathcal{B}$ for which there exists a spacelike curve connecting $x$ to some point on $\partial\Sigma$. In other words, the WDW patch contains all points in the bulk spacetime that can be reached by a bulk Cauchy slice anchored on $\partial\Sigma$.} In this case, the physics on a Cauchy slice in this WDW patch, with an induced metric $g$, will be encoded in the behavior of $\Psi$ in the vicinity of $g$ on $\Sigma_T$. It is in this sense that one can attribute meaning to $\Sigma_T$ as a Cauchy slice in the emergent spacetime.

The emergence of a $(d+1)$-dimensional bulk spacetime from the boundary CFT state is a central question in the gauge/gravity correspondence. The CSH dictionary maps a boundary CFT state to a bulk WDW state, establishing a connection between the boundary and bulk theories. Consequently, the question of bulk emergence from the boundary is now about understanding how the $(d+1)$-dimensional spacetime emerges from the WDW state, which is precisely the purpose of the semiclassical approximation. We delve into this topic further in the following subsections.

\subsection{Dual bulk geometries}
\label{bulkclassicalduals}

The CSH dictionary maps all states of the boundary CFT to bulk WDW states. In general, these WDW states encode various superpositions of bulk spacetimes. Not all of them correspond to a single bulk spacetime; only the semiclassical WDW states give rise to a classical bulk background and bulk QFT states living on it. Therefore, for a holographic state to have a classical dual, certain conditions must be satisfied. In this subsection, we outline these conditions and describe the procedure for finding the geometric duals.

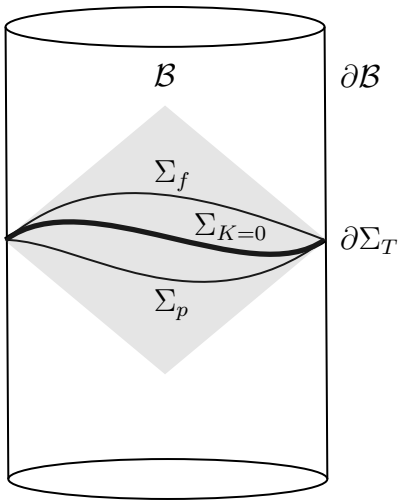
\begin{figure}[h]
    \centering
\tikzset{every picture/.style={line width=0.75pt}} 
\begin{tikzpicture}[x=0.75pt,y=0.75pt,yscale=-1,xscale=1]

\draw    (242.5,22.54) -- (243.88,249.96) ;
\draw    (401.73,22.54) -- (403.12,249.96) ;
\draw   (243.88,249.96) .. controls (243.88,244.41) and (279.53,239.91) .. (323.5,239.91) .. controls (367.47,239.91) and (403.12,244.41) .. (403.12,249.96) .. controls (403.12,255.5) and (367.47,260) .. (323.5,260) .. controls (279.53,260) and (243.88,255.5) .. (243.88,249.96) -- cycle ;
\draw   (242.5,22.54) .. controls (242.5,17) and (278.15,12.5) .. (322.12,12.5) .. controls (366.09,12.5) and (401.73,17) .. (401.73,22.54) .. controls (401.73,28.09) and (366.09,32.59) .. (322.12,32.59) .. controls (278.15,32.59) and (242.5,28.09) .. (242.5,22.54) -- cycle ;
\draw[line width=2pt]    (242.5,129.69) .. controls (282.5,99.69) and (365,156) .. (401.73,129.69) ;
\draw    (242.5,129.69) .. controls (282.5,95.69) and (329,101) .. (401.73,129.69) ;
\draw    (242.5,129.69) .. controls (280,130) and (339,177) .. (401.73,129.69) ;

\filldraw[gray,opacity=0.2]  (322.12,62.72) -- (401.73,129.69) -- (322.12,196.65) -- (242.5,129.69) -- cycle ;
\draw (408,121.86) node [anchor=north west][inner sep=0.75pt]   [align=left] {$\partial\Sigma_T$};
\draw (408,40) node [anchor=north west][inner sep=0.75pt]   [align=left] {$\partial\mathcal{B}$};
\draw (315,40) node [anchor=north west][inner sep=0.75pt]   [align=left] {$\mathcal{B}$};
\draw (315,88) node [anchor=north west][inner sep=0.75pt]   [align=left] {$\Sigma_f$};
\draw (315,153) node [anchor=north west][inner sep=0.75pt]   [align=left] {$\Sigma_p$};
\draw (335,115) node [anchor=north west][inner sep=0.75pt]   [align=left] {$\Sigma_{K=0}$};

\end{tikzpicture}
    \caption{\small Consider a CFT state living on a boundary Cauchy slice $\partial\Sigma_T$, that has a dual bulk geometry $\mathbf{g}$. The WDW patch of $\partial\Sigma_T$ is depicted by the shaded region. The unique maximal volume slice, denoted by $\Sigma_{K=0}$, is represented by the thick curve. For some slice $\Sigma_f$ in the future of $\Sigma_{K=0}$, it might be possible to find a slice $\Sigma_p$ in the past of $\Sigma_{K=0}$, such that the induced metric on these two slices is the same.}
    \label{AdSgeometricdual}
\end{figure}

To analyze the structure of the WDW state when a bulk semiclassical dual exists, let's consider a CFT state $\psi_{\mathrm{bdy}}$ with a bulk dual geometry $\mathbf{g}$. One might naively expect the dual WDW state $\Psi$ to be a WKB state; however, this might not be the case. This is because, given a $d$-dimensional metric $g$, it might be possible to find more than one Cauchy slice in the WDW patch of $\partial\Sigma_T$ such that the metric $\mathbf{g}$ when induced on each of them is $g$. For example, consider two Cauchy slices $\Sigma_f$ and $\Sigma_p$ depicted in Figure \ref{AdSgeometricdual}, both of which have the same induced metric $g$ on them (assuming there is not a third Cauchy slice with the same metric). The physics on these two slices are encoded in the behavior of the WDW state in the neighborhood of the point $g$ in the superspace. Therefore, in this neighborhood, the WDW state would be a sum of two WKB branches, $\Psi_f$ and $\Psi_p$, corresponding to $\Sigma_f$ and $\Sigma_p$, respectively. And, if there exist multiple Cauchy slices in the WDW patch with the same induced metric $g'$, then the WDW state would be a sum of multiple WKB branches in the neighborhood of $g'$ in the superspace. Each of these WKB branches would correspond to a specific Cauchy slice with the induced metric $g'$, reflecting the different ways in which the bulk spacetime can be sliced while preserving the same induced metric. Thus, to study the physics on a specific bulk Cauchy slice, one must isolate the corresponding WKB branch from the WDW state. Currently, we do not know how to isolate the WKB branch corresponding to a specific bulk Cauchy slice, and we leave it as an open problem.\footnote{One way to circumvent this problem is to work in an alternative quantum gravity basis $(K, \gamma_{ab}, \phi)$, where $\gamma_{ab} = g^{-1/d} g_{ab}$ represents the conformal part of the metric, and $K$ is the trace of the extrinsic curvature (see \cite{Hartle:1983ai} for a reference on this point). For a WDW patch in a classical spacetime, there exists a unique foliation such that for any given slice, $K$ is constant everywhere on that slice, and it changes monotonically with the slices. In this case, a CFT state with a dual spacetime geometry would be dual to a semiclassical WDW state $\Psi[K, \gamma, \phi]$. The behavior of the WDW state in the neighborhood of a point corresponding to a particular value of $K$ would encode the physics of the unique Cauchy slice with that value of $K$.}

However, as explained in \cite{GRW}, the maximal volume slice, denoted as $\Sigma_{K=0}$, is unique. It is defined as the Cauchy slice on which the trace of the extrinsic curvature vanishes everywhere (i.e., $K=0$). Here, the extrinsic curvature is calculated by embedding the slice in the a priori given spacetime geometry $\mathbf{g}$. Let the induced metric on $\Sigma_{K=0}$ be denoted as $g_{K=0}$. Regrettably, this is the point in superspace where the semiclassical approximation breaks down, as $K=0$ corresponds to a turning point. Nevertheless, the semiclassical approximation remains valid on points that are arbitrarily close to $g_{K=0}$ in the $G_N \to 0$ limit, provided they are not strictly equal to $g_{K=0}$. Consider two slices $\Sigma^\uparrow$ and $\Sigma^\downarrow$ with the same induced metric, which is very close to $g_{K=0}$, assuming there aren't any other slices with the same induced metric.\footnote{If there are more than two slices with the same induced metric and close to $\Sigma_{K=0}$ (but not equal to), then the same argument will follow but with multiple WKB branches that all merge to a single branch.} The WDW state can be represented as a sum of two WKB branches, each corresponding to one of the slices, $\Sigma^\uparrow$ and $\Sigma^\downarrow$:
\bea
\Psi[g,\Phi] = \Psi^\uparrow[g,\Phi] + \Psi^\downarrow[g,\Phi] = e^{\frac{i}{32 \pi G_N}S^\uparrow_0[g]} e^{i S^\uparrow_1[g,\phi]} + e^{\frac{i}{32 \pi G_N}S^\downarrow_0[g]} e^{i S^\downarrow_1[g,\phi]} .
\eea
The extrinsic curvature and the QFT state on both of these slices, $\Sigma^\uparrow$ and $\Sigma^\downarrow$, would agree when they converge at $\Sigma_{K=0}$. This agreement ensures that, as the slices approach $\Sigma_{K=0}$, their extrinsic curvatures and QFT states become identical, allowing for a consistent description of the physics on and around the maximal volume slice. Therefore, when one takes the limit where $g \to g_{K=0}$ (after having taken the limit $G_N \to 0$), the two branches will merge into a single WKB branch. \emph{In this context, any quantity that we compute with $g=g_{K=0}$ should be interpreted as being evaluated in the limit $g \to g_{K=0}$, after having taken the $G_N \to 0$ limit.}

The extrinsic curvature of $\Sigma_{K=0}$ (as embedded in a spacetime with the $\mathbf{g}$) will be related to the $\Pi^{ab}_{\mathrm{cl}}$ obtained from the WDW state using equation \ref{classicalmetricmomentum}. Since the CSH dictionary maps the CFT state to a bulk WDW state as a $T^2$ partition function $Z_{T^2}[g,\phi;\psi_{\mathrm{bdy}}]$, the functional derivative of the WDW state with respect to the metric is now equivalent to the expectation value of the stress-energy tensor of the $T^2$ theory:
\bea
 -i \frac{\delta}{\delta g_{ab}} \ln \Psi = -i \frac{\delta}{\delta g_{ab}} \ln Z_{T^2} = i\frac{\sqrt{g}}{2} T^{ab} \ln Z_{T^2} = i\frac{\sqrt{g}}{2} \langle T^{ab} \rangle_{\psi_{\mathrm{bdy}}},
\eea
where $\langle T^{ab} \rangle_{\psi_{\mathrm{bdy}}}$ represents the expectation value of the stress-energy tensor $T^{ab}$ in the $T^2$ theory, with background sources $(g, \phi)$ and boundary conditions provided by the CFT state $\psi_{\mathrm{bdy}}$. The classical momentum $\Pi^{ab}_{\mathrm{cl}}$ is the $O(G_N^{-1})$ term in the above equation. From the holographic correspondence, the $(d+1)$-dimensional gravitational constant $G_N$ is related to the number of colors in the dual CFT, denoted by $N$, as follows:
\bea
(8 \pi G_N)^{-1} = \mathcal{V} N^{2},
\eea
where $\mathcal{V}$ is related to the volume of the compactified dimensions. Thus, the $d$-dimensional extrinsic curvature $K_{ab}$ can be obtained from the $T^2$ theory by employing Equation \ref{classicalK}:
\bea \label{KrelationtoT1ptfun}
K_{ab} = \lim_{N \to \infty}  \frac{i \sqrt{g}}{\mathcal{V} N^2} G_{abcd}\langle T^{cd} \rangle_{\psi_{\mathrm{bdy}}}.
\eea

The analysis up to now has been for the case when the CFT state is dual to a bulk in which the metric is classical, and the matter field is quantum mechanical. There could be CFT states whose duals also have classical matter fields. We do not treat that case in this paper, and we expect it would be straightforward to address.\footnote{The low energy effective action in string theory has a factor of $1/G_N$ in front of the whole action, including matter fields. One then canonically normalizes the matter fields, and as a result, $1/G_N$ appears only in front of the gravity action and not the matter action. It is from this action that the WDW equation in \ref{matterHC2} is derived. This action also gives rise to the semiclassical Einstein equations \ref{SEeqn}. In order to treat the case where the matter fields are classical in the bulk, one must repeat the entire $T^2$ deformation corresponding to the WDW equation derived from the initial action, which had $1/G_N$ in front of both gravity and matter parts. And then follow a similar expansion in $G_N$, which would be the same as the $\hbar$ expansion, would correspond to the full WKB approximation.}

In summary, for an arbitrary CFT state to have a semiclassical bulk dual, the following criterion is necessary:

\textbf{Holographic Semiclassical Criterion:} There must exist a metric $g_{K=0}$ for which the $O(N^2)$ term of $\langle T^{ab} \rangle_{\psi_{\mathrm{bdy}}}$ is purely imaginary and traceless ($g_{K=0}^{ab}\langle T_{ab} \rangle_{\psi_{\mathrm{bdy}}}=0$) everywhere on $\Sigma$.

This criterion is necessary but not sufficient because Equation \ref{WDWscvalidity} was used in demonstrating the emergence of semiclassical gravity from the WDW equation. At least, a weaker version of that equation must hold for semiclassical gravity to be reliable. We will explore more on these validity equations in later sections. 
Additionally, we have only explained why this criterion must hold when the bulk is semiclassical, but we have not established the converse—that is, whether the fulfillment of this criterion necessarily implies a semiclassical bulk. However, even if semiclassical gravity is reliable close to the maximal volume slice, it does not necessarily imply that it will hold later. This is because two different semiclassical conditions, differing in the choice of the QFT state on that Cauchy slice, could lead the metric to evolve into two distinct spacetime metrics at a later time, forcing the metric to be in a superposition at that later time. Finally, to determine the entire bulk spacetime dual to a CFT state (which corresponds to a global classical geometry), one must evolve the CFT state through the boundary time using the CFT Hamiltonian and repeatedly find $g_{K=0}$, calculate $K_{ab}$ for each time-evolved state, and construct the spacetime piece by piece, following the procedure outlined earlier.

\subsection{Dual bulk QFT states} \label{bulkQFTduals}

Once the dual bulk classical geometry has been determined (at least in a neighborhood of $\Sigma_{K=0}$), we will now explain how to obtain the dual bulk QFT state living on $\Sigma_{K=0}$ (with the metric $g_{K=0}$) in terms of the correlation functions of the $T^2$ theory. From Equation \ref{emergentQFTstate}, we find the emergent QFT state on $\Sigma_{K=0}$ to be:
\bea
\psi_{\mathrm{bulk}}[\Phi,g_{K=0}]= \mathcal{N}[g_{K=0}]  \ e^{i S_1[g_{K=0},\Phi]}.
\eea
It is challenging to obtain $S_1$ in its complete form from $\psi_{\mathrm{bdy}}$ and $g_{K=0}$. However, one can specify a QFT state by fully specifying the expectation values of all elements of the algebra of observables. We will now elucidate the connection between these expectation values and the correlation functions of the $T^2$ theory.

To begin, let's calculate the expectation value of a product of the field operators, which, according to the Born rule, is given by:
\bea
\left\langle \Phi(x_1)...\Phi(x_m) \right\rangle_{\psi_{\mathrm{bulk}}} &=& \int D\Phi \  \psi^*_{\mathrm{bulk}}[\Phi,g_{K=0}] \ \psi_{\mathrm{bulk}}[\Phi,g_{K=0}] \ \Phi(x_1)...\Phi(x_m), \\
&=& \mathcal{N}^{2}[g_{K=0}] \int D\Phi \ e^{- 2 S_1^I[\Phi,g_{K=0}]} \ \Phi(x_1)...\Phi(x_m),
\eea
where $S^I_1$ represents the imaginary part of $S_1$ (keeping in mind that $S_1$ is complex). The path integral is performed over all field configurations on $\Sigma_{K=0}$ that satisfy the Dirichlet boundary condition on $\partial\Sigma_{K=0}$. The partition function of the holographic CFT living on $\partial\mathcal{B}$ also depends on the background metric and matter field source on $\partial\mathcal{B}$. For simplicity, let us set the matter source on $\partial\mathcal{B}$ to be zero, effectively imposing zero Dirichlet boundary conditions for the matter fields. 

Let us multiply and divide the above integrand by the exponential of some auxiliary functional $S_{\mathrm{aux}}[\Phi,g_{\mathrm{aux}}]$, where $g_{\mathrm{aux}}$ is some auxiliary $d$-metric on $\Sigma_{K=0}$. We will choose $S_{\mathrm{aux}}$ and $g_{\mathrm{aux}}$ according to our convenience. The reason for this will become clear soon.
\bea
\left\langle \Phi(x_1)...\Phi(x_m) \right\rangle_{\psi_{\mathrm{bulk}}} &=&  \mathcal{N}^{2} \int D\Phi \ e^{-S_{\mathrm{aux}}[\Phi]} \ e^{- 2 S_1^I[\Phi]+S_{\mathrm{aux}}[\Phi]} \ \Phi(x_1)...\Phi(x_m),
\eea
where we have notationally suppressed the dependence of $\mathcal{N}$ and $S_1^I$ on $g_{K=0}$, along with the dependence of $S_{\mathrm{aux}}$ on $g_{\mathrm{aux}}$. 

Next, we perform a Taylor expansion of $e^{- 2 S_1^I+S_{\mathrm{aux}}}$ about $\Phi=0$ to obtain:
\bea\label{onepointexpectationvalue}
\left\langle \Phi(x_1)...\Phi(x_m) \right\rangle_{\psi_{\mathrm{bulk}}} &=& \mathcal{N}^{2} \sum_{n=0}^{\infty} \frac{1}{n!}  \int d^dy_1 ... d^dy_n \ \mathcal{G}_n(\mathbf{y_n}) \ \mathcal{P}_{m+n}(\mathbf{x_m},\mathbf{y_n}),
\eea
where $\mathcal{G}$ and $\mathcal{P}$ are defined as:
\bea
\mathcal{G}_n(\mathbf{y_n}) &:=& \left[\frac{\delta^n e^{- 2 S_1^I[\Phi]+S_{\mathrm{aux}}[\Phi]}}{\delta \Phi(y_1)...\delta \Phi(y_n)} \right]_{\Phi=0},\\
\mathcal{P}_{m+n}(\mathbf{x_m},\mathbf{y_n}) &:=& \int D\Phi \ e^{-S_{\mathrm{aux}}[\Phi,g_{\mathrm{aux}}]} \ \Phi(x_1) \dots \Phi(x_m) \Phi(y_1) \dots \Phi(y_n).\label{auxcorrelation}
\eea
Notice that $\mathcal{P}_{m+n}$ can be interpreted as an $(n+m)$-point correlation function of $\Phi$'s in an auxiliary Euclidean QFT with the action $S_{\mathrm{aux}}$, background metric $g_{\mathrm{aux}}$, and boundary conditions $\Phi|_{\partial\Sigma_{K=0}} = 0$.
As $S_{\mathrm{aux}}$ and $g_{\mathrm{aux}}$ can be chosen according to our convenience, it is wise to choose them in such a way that $\mathcal{P}_{m+n}$ can be computed easily. For example, if one chooses the auxiliary theory to be a massless free scalar theory on flat space, as follows:
\bea
S_{\mathrm{aux}}[\Phi,g_{\mathrm{aux}}]=\frac{1}{2}\int d^dx  \ \delta^{ab} \  \partial_a \Phi \partial_b \Phi, 
\eea
then $\mathcal{P}_{m+n}$ is fully known.

As $\mathcal{G}_n$ contains functional derivatives of an exponential, one can expand it as a sum of products of functional derivatives of the exponent. The first functional derivative of $S_{\mathrm{aux}}$ would not contribute because:
\bea
\frac{\delta S_{\mathrm{aux}}}{\delta \Phi(y)} \Bigg|_{\Phi=0} = -   \partial^2 \Phi(y)  \Bigg|_{\Phi=0} =0.
\eea
However, the second functional derivative does contribute:
\bea
\frac{\delta^2 S_{\mathrm{aux}}}{\delta \Phi(y_1) \delta \Phi(y_2)} \Bigg|_{\Phi=0} = -   \partial^2 \delta^d(y_1-y_2),
\eea
whereas the third and higher functional derivatives of $S_{\mathrm{aux}}$ are zero. 

By the CSH dictionary $\Psi=Z_{T^2}$, the functional derivatives of $S^I=-\mathfrak{Re} \{\ln \Psi\}$ are related to the correlation functions of the $T^2$ theory as:
\bea
 \frac{\delta^n S^I}{\delta \Phi(y_1)...\delta \Phi(y_n)}\Bigg|_{\Phi=0} &=& -\mathfrak{Re} \left\{\frac{\delta^n \ln \Psi}{\delta \Phi(y_1)...\delta \Phi(y_n)}\right\}_{\Phi=0} ,\\
 &=& -\mathfrak{Re} \left\{\frac{\delta^n \ln Z_{T^2}}{\delta \Phi(y_1)...\delta \Phi(y_n)}\right\}_{\Phi=0} ,\\
  &=& - \sqrt{g_{K=0}(y_1)}...\sqrt{g_{K=0}(y_n)} \ \mathfrak{Re} \left\langle \mathcal{O}(y_1)...\mathcal{O}(y_n)\right\rangle_{\mathrm{c}}^{T^2},
\eea
where $\left\langle \mathcal{O}(y_1)...\mathcal{O}(y_n)\right\rangle_{\mathrm{c}}^{T^2}$ represents the connected correlation functions of operators in the $T^2$ theory on the background $(g_{K=0},\phi=0)$. Additionally, $\mathcal{O}(x)$ is the operator in the $T^2$ theory that is sourced by $\phi(x)$, analogous to the way the stress-tensor is sourced by the metric. Since $S_0$ is independent of $\Phi$, the leading term in the above equation is of order $O(G_N^0)$, which corresponds to the functional derivatives of $S_1^I$:
\bea\label{realpart}
 \frac{\delta^n S^I_1}{\delta \Phi(y_1)...\delta \Phi(y_n)}\Bigg|_{\Phi=0} = - \sqrt{g_{K=0}(y_1)}...\sqrt{g_{K=0}(y_n)} \lim_{N \to \infty} \ \mathfrak{Re} \left\langle \mathcal{O}(y_1)...\mathcal{O}(y_n)\right\rangle_{\mathrm{c}}^{T^2}.
\eea
Therefore, $\mathcal{G}_n$ can be obtained by computing the real part of the connected correlation functions of dual operators $\mathcal{O}$'s in the $T^2$ theory at large $N$. This, in turn, allows us to obtain the expectation values of a product of field operators in the bulk QFT state $\psi_{\mathrm{bulk}}$, residing on $\Sigma_{K=0}$.

Following the same spirit as before, let us turn to the expectation value of momenta $\Pi_{\Phi}$ conjugate to bulk matter fields, which is once again given by the Born rule:
\bea
\left\langle \Pi_{\Phi}(x) \right\rangle_{\psi_{\mathrm{bulk}}} &=& \int D\Phi \  \psi^*_{\mathrm{bulk}}[\Phi,g_{K=0}] \ \Pi_{\Phi}(x) \ \psi_{\mathrm{bulk}}[\Phi,g_{K=0}], \\
&=& \mathcal{N}^{2}[g_{K=0}] \int D\Phi \ e^{- 2 S_1^I[\Phi,g_{K=0}]} \  \frac{\delta S_1}{\delta \Phi(x)}.
\eea
The imaginary part of the above equation can be written as:
\bea
\int D\Phi \ e^{- 2 S_1^I[\Phi,g_{K=0}]} \  \frac{\delta S^I_1}{\delta \Phi(x)} = -\frac{1}{2} \int D\Phi \ \frac{\delta}{\delta \Phi(x)} \  e^{- 2 S_1^I[\Phi,g_{K=0}]} =0,
\eea
which vanishes, assuming $|\psi_{\mathrm{bulk}}[\Phi]|^2 \to 0$ as $\Phi(x) \to |\infty|$. This condition is required for normalizable QFT states. Following the same approach as before, we introduce an auxiliary functional $S_{\mathrm{aux}}[\Phi,g_{\mathrm{aux}}]$ and proceed with a Taylor expansion of $\left( e^{- 2 S_1^I + S_{\mathrm{aux}}} \ \frac{\delta S^R_1}{\delta \Phi(x)} \right)$ about $\Phi=0$ to obtain:
\bea
\left\langle \Pi_{\Phi}(x) \right\rangle_{\psi_{\mathrm{bulk}}} &=& \mathcal{N}^{2} \int D\Phi \ e^{- 2 S_1^I} \  \frac{\delta S^R_1}{\delta \Phi(x)}, \\
&=& \mathcal{N}^{2} \int D\Phi \ e^{-S_{\mathrm{aux}}} \ e^{- 2 S_1^I+S_{\mathrm{aux}}} \  \frac{\delta S^R_1}{\delta \Phi(x)}, \\
&=& \mathcal{N}^{2} \sum_{n=0}^{\infty} \frac{1}{n!}  \int d^dy_1 ... d^dy_n \ \mathcal{\tilde{G}}_n(\mathbf{y_n}) \ \mathcal{P}_{n}(\mathbf{y_n}),
\eea
where $\mathcal{P}_{n}$ is defined in the same way as in Equation \ref{auxcorrelation}, but now $\mathcal{\tilde{G}}_n$ is defined as:
\bea
\mathcal{\tilde{G}}_{1,n}(x,\mathbf{y_n})&:=& \left[\frac{\delta^n }{\delta \Phi(y_1)...\delta \Phi(y_n)} \left(e^{- 2 S_1^I+S_{\mathrm{aux}}} \ \frac{\delta S^R_1}{\delta \Phi(x)} \right)\right]_{\Phi=0}.
\eea
Now, the expansion of $\mathcal{\tilde{G}}_{1,n}$ involves not only the functional derivatives of $S_1^I$, but also the functional derivatives of $S_1^R$. These can be expressed in a similar manner as in the derivation of Equation \ref{realpart}:
\bea
 \frac{\delta^n S^R_1}{\delta \Phi(y_1)...\delta \Phi(y_n)}\Bigg|_{\Phi=0} = \sqrt{g_{K=0}(y_1)}...\sqrt{g_{K=0}(y_n)} \lim_{N \to \infty} \ \mathfrak{Im} \left\langle \mathcal{O}(y_1)...\mathcal{O}(y_n)\right\rangle_{\mathrm{c}}^{T^2}.
\eea
Therefore, $\tilde{\mathcal{G}}_n$ can be obtained by computing both the real and imaginary part of the connected correlation functions of dual operators $\mathcal{O}$'s in the $T^2$ theory at large $N$. 

Using a similar methodology, the expectation value of any bulk QFT operator in the bulk QFT state can be computed by calculating the connected correlation functions of the dual operators $\mathcal{O}$'s in the $T^2$ theory at large $N$.

\section{Applications
in dS/CFT} \label{dSCFTclosed}

In a closed universe, which lacks a time-like boundary, the holographic principle and the dual description present significant challenges, as it is unclear where the dual description should reside. One approach to dS/CFT posits that the dual description is provided by a non-reflection positive, Euclidean CFT residing on the future or past asymptotic spacelike boundary (referred to as $\mathcal{I}^+$ or $\mathcal{I}^-$). This differs from the unitary CFTs found in the AdS case, which reside on an asymptotic timelike boundary. However, an interpretation of states in terms of the Euclidean CFT corresponding to bulk quantum gravity states is still uncertain. Furthermore, the relationship between the quantities of the bulk spacetime and those of the dual CFT on the future or past asymptotic boundary remains unclear, raising further questions about the functioning of the holographic principle in this context.

An alternative approach involves introducing a hard wall in de Sitter spacetime and situating the dual field theory description on the wall. This idea seeks to provide a concrete location for the dual description, akin to the boundary in the traditional holographic principle. However, the nature of this wall, its origin, applicability, and how it encodes information about the bulk are all subjects of ongoing investigation. Some researchers argue that this wall might be an artifact of a specific construction or approximation, while others propose it as a fundamental aspect of the holographic principle in de Sitter space.

On the other hand, the canonical quantum gravity community has been exploring a distinct set of questions related to closed universe scenarios. Within the WDW formalism, the state of the entire universe is represented by a single wave functional, known as the wave function of the universe: $\Psi[g, \Phi]$ of the metric $g$ and matter field $\Phi$ on $\Sigma$ (a $d$-dimensional closed manifold, such as $S^d$). This wave function encompasses everything within the universe, including all its contents, observers, and physical systems.

The WDW formalism raises several intriguing questions and challenges, as the wave function of the universe lacks any notion of external time and external observers. Consequently, traditional concepts of quantum mechanics, which rely on an external time or external observer, need to be re-evaluated within the context of a closed universe. If we adopt Bohr's view that quantum mechanics is a theory describing a quantum system as observed by a classical system, then such a wave function of the universe cannot be considered a quantum mechanical state, as it lacks external observers. Furthermore, since the entire universe exists in this single state, questions arise regarding how we can perform physics by choosing different states and making observations within the framework of this single state.

Nonetheless, we must view any such wave function as an abstract object from which we can extract physics through appropriate approximations. Any such approximation is likely to involve a semiclassical approximation, as it is only in this case that a classical description emerges (at least for a subsystem in the universe), which is required to have observers who describe the rest of the universe quantum mechanically. The validity of the semiclassical approximation depends on the choice of $\Psi[g,\Phi]$ and may only be valid in a subregion of the superspace, referred to as the WKB region. Only within this WKB region can we obtain a good notion of time, classical spacetime, and observers.

In summary, understanding the holographic principle and the dual description in a closed universe presents a challenging area of research, with several competing ideas and numerous open questions. Developing a coherent framework that addresses these questions will necessitate a deeper understanding of the holographic principle, its relationship with bulk spacetime information, and the nature of quantum states in this unique setting. As the field progresses, new insights and innovative approaches will be essential for unraveling the complexities of holography in a closed universe.

This section aims to delve into these intricate questions and explore the framework provided by Cauchy slice holography for a closed universe (which we will explain in later sections), as well as the implications of the semiclassical approximation. By examining these aspects in detail, we hope to gain a better understanding of the challenges and potential solutions associated with holography and quantum states in closed universes.

\subsection{A minisuperspace model}\label{minisuperspacemodel}

To elucidate the concepts of quantum gravity as they pertain to closed universes, we turn to the minisuperspace model, as initially introduced in \cite{Hartle:1983ai}. We consider a scalar field, $\phi$, conformally coupled to gravity within a 3+1-dimensional closed universe, characterized by a positive cosmological constant, $\Lambda$.
\footnote{For explicit solutions to the WDW equation for pure gravity in $2+1$-dimensions under the minisuperspace approximation, refer to \cite{Araujo-Regado:2022jpj}.} In the minisuperspace model, all the degrees of freedom associated with the metric and matter fields are constrained, except for the overall size of the universe, denoted by $R$.\footnote{A cautionary note on the notation used in this paper is warranted. Elsewhere, the symbol $R$ refers to the Ricci scalar associated with the $d$-dimensional metric. However, in this specific subsection \ref{minisuperspacemodel}, $R$ denotes a numerical value corresponding to the scale factor of the universe.} Additionally, the matter field in this model is permitted to obtain only a spatially constant value: 
\bea
ds^2 = -N(t)^2 dt^2 + R(t)^2  d\Omega^2_3, \quad \quad  \quad \quad
\phi(x) = \phi =: \frac{\chi}{\sqrt{2} \pi R},
\eea
This drastically simplifies the infinite-dimensional configuration space of general relativity and quantum field theory, reducing it to a finite-dimensional minisuperspace. The Lorentzian Lagrangian simplifies to:
\bea\label{minisuperspacelagrangian}
L = \frac{1}{2} \frac{N}{R} \left(- \frac{1}{\sigma}\left(\frac{R}{N} \frac{dR}{dt}\right)^2 +\frac{R^2}{\sigma} -\frac{\Lambda}{3} \frac{R^4}{\sigma} +  \left(\frac{R}{N} \frac{d \chi}{dt}\right)^2 -\chi^2 \right),  
\eea
where $\sigma=\frac{2}{3\pi}G_N$. The corresponding Hamiltonian is given by:
\bea
H= \frac{N}{2 R} \left(-\sigma \Pi^2_R - \frac{R^2}{\sigma} +\frac{\Lambda}{3} \frac{R^4}{\sigma} + \Pi^2_\chi + \chi^2 \right).
\eea
Upon quantization, with the selection of normal ordering, the WDW equation simplifies to:
\bea
\frac{1}{2} \left(\sigma \frac{\partial^2}{\partial R^2} - \frac{R^2}{\sigma} +\frac{\Lambda}{3} \frac{R^4}{\sigma} -\frac{\partial^2}{\partial \chi^2} + \chi^2 + 2 \epsilon_0 \right) \Psi(R,\chi) = 0,
\eea
where $\epsilon_0$ denotes the matter-energy renormalization constant. Utilizing a method of separation of variables, the WDW equation becomes:
\bea
\Psi(R,\chi) &=& \sum_{n \geq 0} c_n(R) \  u_n(\chi), \\
\frac{1}{2} \left\{-\sigma \frac{\partial^2}{\partial R^2} + \frac{R^2}{\sigma} -\frac{\Lambda}{3} \frac{R^4}{\sigma} \right\}  c_n(R)  &=& \left(n+\frac{1}{2}+\epsilon_0\right)c_n(R)=: E_n c_n(R), \label{hampuregravityminisuperspace} \\
\frac{1}{2} \left\{ -\frac{d^2}{d \chi^2} +\chi^2  \right\} u_n(\chi) &=& \left(n+\frac{1}{2}\right) u_n(\chi),
\eea
where the final equation represents the eigenvalue equation of a simple harmonic oscillator. It yields the following unique solution when the boundary conditions $\lim_{|\chi| \to \infty} u_n(\chi) = 0$ is imposed:
\bea
u_n(\chi) = \frac{1}{\sqrt{2^n n! \sqrt{\pi}}} e^{-\frac{\chi^2}{2}} H_n(\chi),
\eea
where $H_n(\chi)$ are Hermite polynomials. It's important to note that the boundary condition for $u_n$ is necessary to ensure that the emergent states satisfying the emergent Schrödinger equation will be normalizable according to the standard square integrable norm in quantum mechanics. However, at this juncture, we lack a motivation for normalizing $c_n$. Equation \eqref{hampuregravityminisuperspace} yields two linearly independent solutions, denoted as $c^{+}_n(R)$ and $c^{-}_n(R)$: 
\begin{align}\label{HeunTsolutions}
c_n(R) &= \alpha_n^{+} \ c^{+}_n(R) + \alpha_n^{-} \ c^{-}_n(R), \notag \\
&= \alpha_n^{+} \ e^{\frac{i R \left(2 \Lambda R^2 -9\right)}{6 \sigma \sqrt{3 \Lambda} }} \ \text{HeunT}\left(\frac{3}{4 \Lambda \sigma^2}-\frac{2 E_n}{\sigma},\frac{2i}{\sigma} \sqrt{\frac{\Lambda}{3}}, -\frac{i}{\sigma} \sqrt{\frac{3}{\Lambda}}, 0,\frac{2i}{\sigma} \sqrt{\frac{\Lambda}{3}}, R\right) \notag\\
&+ \alpha_n^{-} \ e^{-\frac{i R \left(2\Lambda  R^2-9\right)}{6 \sigma \sqrt{3 \Lambda} }}  \ \text{HeunT}\left(\frac{3}{4 \Lambda \sigma^2}- \frac{2 E_n}{\sigma}, -\frac{2i}{\sigma} \sqrt{\frac{\Lambda}{3}}, \frac{i}{\sigma} \sqrt{\frac{3}{\Lambda}}, 0, -\frac{2i}{\sigma} \sqrt{\frac{\Lambda}{3}}, R\right),
\end{align}
where $\alpha_n^{+}$ and $\alpha_n^{-}$ represent arbitrary complex coefficients and $\text{HeunT}$ stands for the tri-confluent Heun function. The most general solution to the minisuperspace WDW equation is given as follows:
\bea
\Psi(R,\chi) &=& \sum_{n \geq 0} u_n(\chi) \left(\alpha^{+}_n c^{+}_n(R) +\alpha^{-}_n c^{-}_n(R) \right).
\eea

The WKB approximation's application is contingent on whether the states are heavily excited or not. For lighter states, where $E_n\ll\frac{3}{4 \Lambda \sigma}$, the Einstein-Hamilton principal function remains largely insensitive to the excitations denoted by $E_n$. Conversely, for heavier states, where $E_n\gg\frac{3}{4 \Lambda \sigma}$, the Einstein-Hamilton principal function exhibits sensitivity to these excitations. Let's examine each scenario individually.

\subsubsection{Light states ($E_n\ll\frac{3}{4 \Lambda \sigma}$)}

\begin{figure}[H]
    \centering
    \begin{minipage}{0.45\textwidth}
        \centering
        \includegraphics[width=\linewidth]{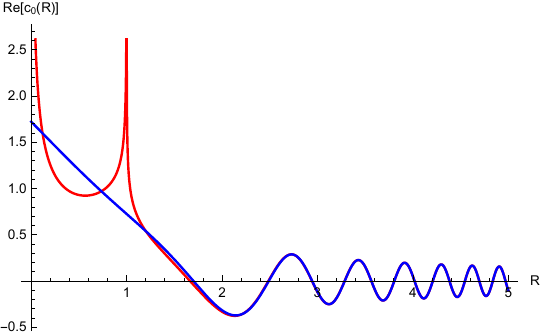}
        \includegraphics[width=\linewidth]{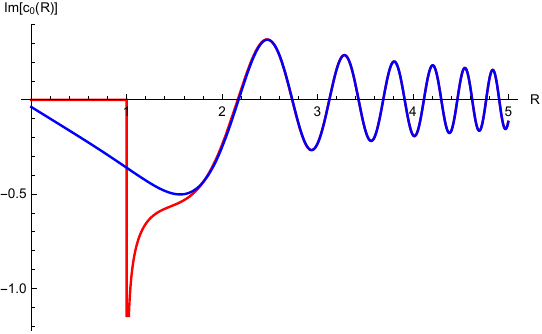}
    \end{minipage}
    \hfill
    \begin{minipage}{0.45\textwidth}
        \centering
        \includegraphics[width=\linewidth]{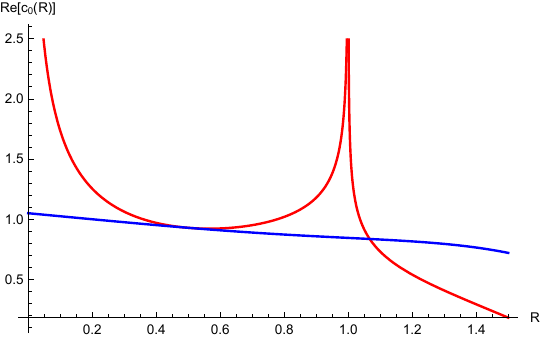}
        \includegraphics[width=\linewidth]{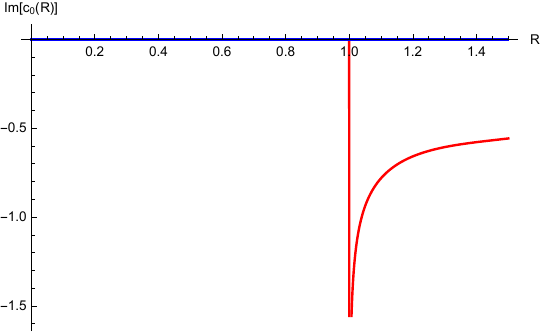}
    \end{minipage}
    \caption{\small The parameters chosen here are $\Lambda=3$, $\sigma=1$, $\epsilon_0=-1/2$, and $n=0$. The top two figures illustrate the plots of the real components of the functions, while the bottom two figures present the plots of the imaginary components of the functions. The red line symbolizes the WKB solution corresponding to the $\uparrow$ branch, $c_{0(\uparrow)}^{\mathrm{WKB}}(R)$. The blue lines in the figures on the left represent a specific linear combination of the exact solutions, $c_0^\uparrow(R)=(1.17311 - 1.07475 \ i) c_0^+(R)+ (0.548655 + 1.03709 \ i) c_0^-(R)$. These parameters are determined by requiring the linear combination of the exact solutions and its first derivative to coincide with those of $c_{0(\uparrow)}^{\mathrm{WKB}}(R)$ at $R=5$. The WKB approximation performs adequately in the classical region, however, it demonstrates limitations near the turning point at $R=1$. Within the classically forbidden region, it becomes necessary to select a different linear combination of exact solutions for comparison with the WKB solution. Accordingly, the blue lines in the figures on the right now signify $c_0^\uparrow(R)=(0.525487 - 0.251296 \ i) c_0^+(R)+ (0.525487 + 0.251296 \ i) c_0^-(R)$. These parameters are ascertained by necessitating that the linear combination of the exact solutions and its first derivative coincide with those of $c_{0(\uparrow)}^{\mathrm{WKB}}(R)$ at $R=0.5$. The WKB approximation is insufficient for the chosen parameters in the classically forbidden region. This demonstrates a scenario where the validity condition is satisfied when $R\gg 1$ for a finite value of $\sigma$.}
    \label{lightplot1}
\end{figure}

\begin{figure}[H]
    \centering
    \begin{minipage}{0.45\textwidth}
        \centering
        \includegraphics[width=\linewidth]{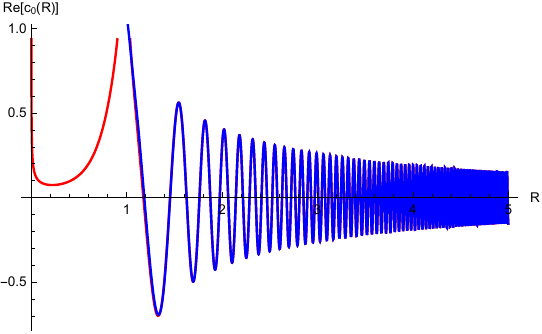}
        \includegraphics[width=\linewidth]{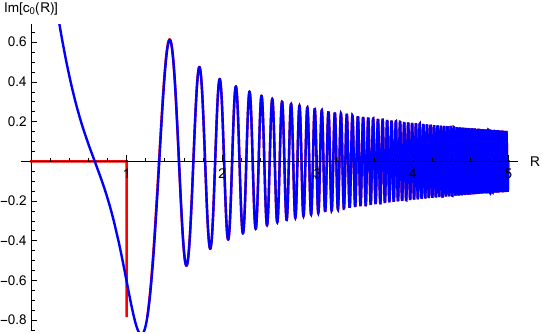}
    \end{minipage}
    \hfill
    \begin{minipage}{0.45\textwidth}
        \centering
        \includegraphics[width=\linewidth]{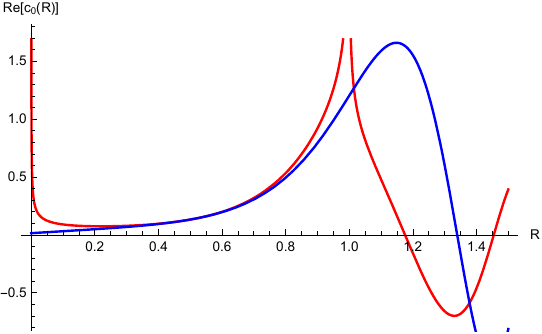}
        \includegraphics[width=\linewidth]{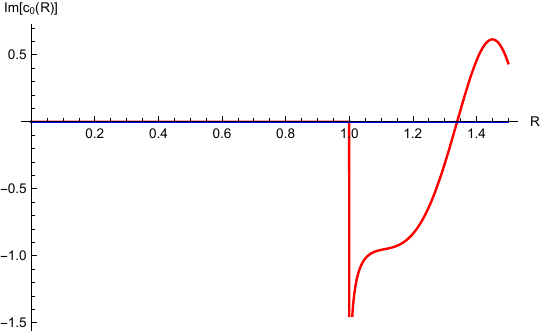}
    \end{minipage}
    \caption{\small The parameters chosen here are $\Lambda=3$, $\sigma=0.1$, $\epsilon_0=-1/2$, and $n=0$. The top two figures illustrate the plots of the real components of the functions, while the bottom two figures present the plots of the imaginary components of the functions. The red line symbolizes the WKB solution corresponding to the $\uparrow$ branch, $c_{0(\uparrow)}^{\mathrm{WKB}}(R)$. The blue lines in the figures on the left represent a specific linear combination of the exact solutions, $c_0^\uparrow(R)=(27.5553 - 10.3621 \ i) c_0^+(R)+ (26.8436 + 12.0388 \ i) c_0^-(R)$. These parameters are determined by requiring the linear combination of the exact solutions and its first derivative to coincide with those of $c_{0(\uparrow)}^{\mathrm{WKB}}(R)$ at $R=5$. The WKB approximation performs adequately in the classical region, however, it demonstrates limitations near the turning point at $R=1$. Notably, the smaller value of $\sigma$—as compared to that in Figure \ref{lightplot1}—has ameliorated the approximation closer to this turning point. Within the classically forbidden region, it becomes necessary to select a different linear combination of exact solutions for comparison with the WKB solution. Accordingly, the blue lines in the figures on the right now signify $c_0^\uparrow(R)=(0.00744697 + 0.0169472 \ i) c_0^+(R)+ (0.00744697 - 0.0169472 \ i) c_0^-(R)$. These parameters are ascertained by necessitating that the linear combination of the exact solutions and its first derivative coincide with those of $c_{0(\uparrow)}^{\mathrm{WKB}}(R)$ at $R=0.5$. Once again, the smaller value of $\sigma$ has improved the accuracy of the approximation within the classically forbidden region. This demonstrates a scenario where the validity condition is satisfied by maintaining $\sigma \ll 1$ for any value of $R$ where $p(R) \neq 0$  (i.e., in regions distant from turning points).}
    \label{lightplot2}
\end{figure}

Upon applying the WKB approximation to Equation \ref{hampuregravityminisuperspace} for light states, we derive up to $O(\sigma^0)$:
\bea
c^{\mathrm{WKB}}_{n(\uparrow\downarrow)}(R) &=& \mathrm{exp}\left[\frac{i}{\sigma} S^{(0)}_{(\uparrow\downarrow)}(R) + i S^{(1)}_{n(\uparrow\downarrow)}(R) \right]   \\
&=& \frac{\mathrm{exp}\left[(\mp) \frac{i}{\sigma \Lambda}\left(\frac{\Lambda}{3} R^2 -1\right)^{3/2} (\mp) i E_n \ \mathrm{ArcTan}\left(\sqrt{\frac{\Lambda}{3}R^2-1}\right)\right]}{\left(3 R^2 - \Lambda R^4\right)^{1/4}},
\eea
where the plus and minus signs are correspondingly associated with $\downarrow$ and $\uparrow$ respectively. The physical significance of these arrows will shortly become apparent. The validity regime for this approximation is characterized by the following condition:
\bea
\Bigg|\sigma \frac{p'}{p} \Bigg| \ll |p|, 
\eea
where $p(R) := \sqrt{-R^2 +\frac{\Lambda}{3} R^4}$. This condition is met under two potential scenarios: one scenario is when $\sigma \ll 1$ for any given value of $R$ where $p(R) \neq 0$, and the other scenario is when $R \gg 1$ for a finite value of $\sigma$. Within these regimes of validity, the exact solutions $c_n^{+}(R)$ and $c_n^{-}(R)$ given in equation \eqref{HeunTsolutions} may not necessarily represent WKB states but could instead embody a superposition of WKB states. Nevertheless, a specific linear combination of the $c_n^{+}(R)$ and $c_n^{-}(R)$ can be identified that would correspond to a WKB state:
\bea
c_{n}^{\uparrow}(R) = \alpha^+_\uparrow c_n^{+}(R) + \alpha^-_\uparrow c_n^{-}(R), \\
c_{n}^{\downarrow}(R) = \alpha^+_\downarrow c_n^{+}(R) + \alpha^-_\downarrow c_n^{-}(R).
\eea
The superposition coefficients can be determined by insisting that the exact solutions $c_{n}^{\uparrow\downarrow}(R)$ and their first derivatives coincide with those of the WKB state $c_{n(\uparrow\downarrow)}^{\mathrm{WKB}}(R)$ at a chosen point $R_0$, where the WKB approximation holds strong. This approach is valid since these are solutions to a second-order differential equation. For visual reference, refer to Figures \ref{lightplot1} and \ref{lightplot2}, where the WKB solutions have been plotted alongside the exact solutions for comparison, using Mathematica.

The term $S^{(0)}_{(\uparrow\downarrow)}(R)$ is real in the region where $R > \sqrt{\frac{3}{\Lambda}}$. According to equation \eqref{semiclassicalsuperspace}, this corresponds to the Lorentzian classical region. Conversely, the region where $R < \sqrt{\frac{3}{\Lambda}}$ corresponds to the Euclidean region or the classically forbidden region. The back-reacted momentum $\Pi_R$, which is conjugate to $R$ and derived from equation \eqref{backreactedmomentum}, is expressed as follows:
\bea
\Pi_R &=& \frac{1}{\sigma} \frac{\partial S^{(0)}_{(\uparrow\downarrow)}}{\partial R}  + \left\langle \frac{\partial }{\partial R} \mathfrak{Re}\left\{S^{(1)}_{n(\uparrow\downarrow)}\right\}\right\rangle ,\\
&=& (\mp) \left[ \frac{p(R)}{\sigma} + \frac{E_n}{p(R)} \right]. \label{backreactedmomentumminisuperspace}
\eea
The classical momentum is related to the time derivative of $R$ as derived from the Lagrangian in equation \eqref{minisuperspacelagrangian}: 
\bea
\Pi_R = -\frac{R \dot{R}}{\sigma N}, 
\eea
which implies that the WKB branch, $c^{\mathrm{WKB}}_{n(\uparrow)}(R)$, corresponds to an expanding universe ($\dot{R}>0$), while the WKB branch, $c^{\mathrm{WKB}}_{n(\downarrow)}(R)$, 
corresponds to a contracting one ($\dot{R}<0$). The first term in equation \eqref{backreactedmomentumminisuperspace} is also the momentum originating from an exact de Sitter spacetime. When $\sigma \ll 1$, this term is dominant, causing the WKB branches, $c^{\mathrm{WKB}}_{n(\uparrow\downarrow)}(R)$, to closely approximate a de Sitter configuration. Furthermore, when $R \gg 1$ (for any finite value of $\sigma$), this term also dominates, leading these WKB branches to asymptotically approach a de Sitter configuration.

At this point, $R$ can be used to define a notion of proper time, denoted as $\tau$:
\bea
 \tau = \int \frac{dR}{\dot{R}}= - \int dR \frac{R}{p(R)} = \begin{cases}
+\sqrt{\frac{{3}}{{\Lambda}}} \ \mathrm{{ArcTanh}}\left(\frac{{R }}{{\sqrt{R^2-\frac{3}{\Lambda}  }}}\right) + \mathrm{C}, \quad \text{for} \quad \uparrow\mathrm{branch}, \\
-\sqrt{\frac{{3}}{{\Lambda}}} \ \mathrm{{ArcTanh}}\left(\frac{{R }}{{\sqrt{R^2-\frac{3}{\Lambda}  }}}\right) - \mathrm{C}
, \quad \text{for}  \quad \downarrow\mathrm{branch},
\end{cases}
\eea
where $\mathrm{C}$ is the integration constant. For $R>\sqrt{\frac{3}{\Lambda}}$, the function $\mathrm{{ArcTanh}}\left(\frac{{R }}{{\sqrt{R^2-\frac{3}{\Lambda} }}}\right)$ carries a constant imaginary component equivalent to $-i \frac{\pi}{2}$. However, one may set the integration constant to $\mathrm{C}=i \frac{\pi}{2}\sqrt{\frac{3}{\Lambda}}$, to cancel this imaginary component. This is tantamount to setting $\tau=0$ at $R=\sqrt{\frac{3}{\Lambda}}$. Then, by inverting the above equation, one gets $R(\tau) =\sqrt{\frac{{3}}{{\Lambda}}} \mathrm{Cosh}\left(\sqrt{\frac{\Lambda}{3}} \tau\right)$.

\newpage

A general light state can be approximated as follows:
\bea
\Psi(R,\chi) &=&\sum_{n \ll \frac{1}{\Lambda \sigma}} u_n(\chi) \left(\alpha_n^{\uparrow} c_{n}^{\uparrow}(R) +\alpha_n^{\downarrow} c_{n}^{\downarrow}(R) \right) , \label{coefficientsinWDW}\\
&\approx& \Psi_{\mathrm{WKB}}^{\uparrow}(R,\chi)+\Psi_{\mathrm{WKB}}^{\downarrow}(R,\chi)\\
&=& \frac{1}{\left(3 R^2 - \Lambda R^4\right)^{1/4}}  \left[ e^{\frac{i}{\sigma} S^{(0)}_{(\uparrow)}(R)} \psi^{\uparrow}(\chi,R) + e^{\frac{i}{\sigma} S^{(0)}_{(\downarrow)}(R)} \psi^{\downarrow}(\chi,R) \right],
\eea
where $\psi^{\uparrow}(\chi,R)$ and $\psi^{\downarrow}(\chi,R)$ denote the emergent quantum mechanical states of $\chi$ that reside on the emergent de Sitter background. These states are represented as:
\bea\label{minisuperspaceqmstates}
\psi^{\uparrow\downarrow}(\chi,R) = \sum_{n \ll \frac{1}{\Lambda \sigma}} \alpha_n^{\uparrow\downarrow} u_n(\chi) \mathrm{exp}\left[(\mp) i E_n \ \mathrm{ArcTan}\left(\sqrt{\frac{\Lambda}{3}R^2-1}\right)\right].
\eea
This satisfies the emergent Schrödinger equation:
\bea
i \frac{\partial}{\partial \tau} \psi^{\uparrow\downarrow} &=& \mathrm{H}_{\mathrm{matter}} \psi^{\uparrow\downarrow},
\eea
where $\text{H}_{\text{matter}}$ is the time-dependent matter Hamiltonian:
\bea
\mathrm{H}_{\mathrm{matter}}=\frac{1}{2}\sqrt{\frac{\Lambda}{3}}\frac{1}{\mathrm{Cosh}\left(\sqrt{\frac{\Lambda}{3}}\tau\right)} \left\{ -\frac{\partial^2}{\partial \chi^2} +\chi^2 + 2\epsilon_0 \right\}.
\eea

\subsubsection{Heavy states ($E_n\gg\frac{3}{4 \Lambda \sigma}$)}

For heavily excited states, let's represent $E_n$ as:
\bea
E_n=: \frac{\zeta}{\sigma} + \eta.
\eea
In the context of the WKB approximation, the $\frac{\zeta}{\sigma}$ term will contribute to the Einstein-Hamilton's principal function:
\bea
\left(\frac{\partial\tilde{S}^{(0)}}{\partial R}\right)^2 = \tilde{p}(R)^2 = \left(-R^2+\frac{\Lambda}{3}R^4+2\zeta\right).
\eea
At the next order, we obtain:
\bea
\left(\frac{\partial\tilde{S}^{(1)}}{\partial R}\right) = \frac{\eta + \frac{i}{2}\tilde{p}'(R)}{\tilde{p}(R)}.
\eea
By solving the aforementioned equations, we can express the WKB states as follows:
\bea
\tilde{c}^{\mathrm{WKB}}_{n(\uparrow\downarrow)}(R) &=& \mathrm{exp}\left[\frac{i}{\sigma} \tilde{S}^{(0)}_{(\uparrow\downarrow)}(R) + i \tilde{S}^{(1)}_{n(\uparrow\downarrow)}(R) \right],
\eea
where
\begin{align}
&\tilde{S}^{(0)}_{(\uparrow\downarrow)}(R) = (\mp) \frac{1}{9 \Lambda \sqrt{-3 R^2 + 6 \zeta + R^4 \Lambda}} \Bigg\{\sqrt{3}
R \Lambda (-3 R^2 + 6 \zeta + R^4 \Lambda) \notag \\
&+ 
18 i \sqrt{2} \zeta \sqrt{\frac{\Lambda}{-3 + \sqrt{9 - 24 \zeta \Lambda}}} \sqrt{\frac{
3 - 2 R^2 \Lambda + \sqrt{9 - 24 \zeta \Lambda}}{
3 + \sqrt{9 - 24 \zeta \Lambda}}}
\sqrt{\frac{-3 + 2 R^2 \Lambda + \sqrt{9 - 24 \zeta \Lambda}}{-3 + \sqrt{
9 - 24 \zeta \Lambda}}} \notag \\
&\Bigg(\sqrt{3}
\text{EllipticE}\left[
i \text{ArcSinh}\left[
\sqrt{2} R \sqrt{\frac{\Lambda}{-3 + \sqrt{9 - 24 \zeta \Lambda}}}\right], \frac{
3 - \sqrt{9 - 24 \zeta \Lambda}}{
3 + \sqrt{9 - 24 \zeta \Lambda}}\right] \notag \\
&- 
\sqrt{3 - 8 \zeta \Lambda}
\text{EllipticF}\left[
i \text{ArcSinh}\left[
\sqrt{2} R \sqrt{\frac{\Lambda}{-3 + \sqrt{9 - 24 \zeta \Lambda}}}\right], \frac{
3 - \sqrt{9 - 24 \zeta \Lambda}}{
3 + \sqrt{9 - 24 \zeta \Lambda}}\right]\Bigg)\Bigg\},
\end{align}
where $\text{EllipticF}$ and $\text{EllipticE}$ are elliptic integrals of the first and second kinds respectively, and 
\begin{align}
 \tilde{S}^{(1)}_{n(\uparrow\downarrow)}(R)   &=+ \frac{1}{4} i \log\left[-3 R^2 + 6 \zeta + R^4 \Lambda\right] \notag \\
&-(\mp)\frac{i \sqrt{\frac{3}{2}} \eta \sqrt{1 - \frac{2 R^2 \Lambda}{3 \left(1 - \frac{\sqrt{3 - 8 \zeta \Lambda}}{\sqrt{3}}\right)}} \sqrt{1 - \frac{2 R^2 \Lambda}{3 \left(1 + \frac{\sqrt{3 - 8 \zeta \Lambda}}{\sqrt{3}}\right)}}
}{\sqrt{-R^2 + 2 \zeta + \frac{R^4 \Lambda}{3}} \sqrt{-\frac{\Lambda}{1 - \frac{\sqrt{3 - 8 \zeta \Lambda}}{\sqrt{3}}}}} \notag \\
& \text{EllipticF}\left[i \text{ArcSinh}\left[\sqrt{\frac{2}{3}} R \sqrt{-\frac{\Lambda}{1 - \frac{\sqrt{3 - 8 \zeta \Lambda}}{\sqrt{3}}}}\right], \frac{1 - \frac{\sqrt{3 - 8 \zeta \Lambda}}{\sqrt{3}}}{1 + \frac{\sqrt{3 - 8 \zeta \Lambda}}{\sqrt{3}}}\right] .
\end{align}

\begin{figure}[H]
    \centering
    \begin{minipage}{0.45\textwidth}
        \centering
        \includegraphics[width=\linewidth]{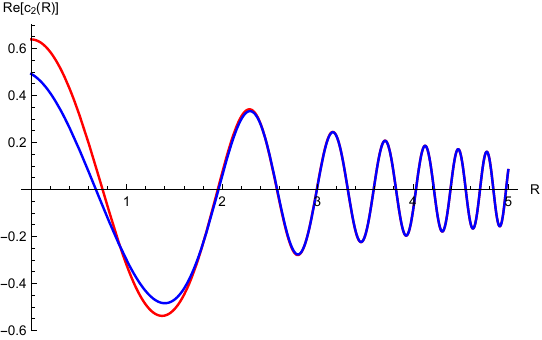}
        \caption*{(a) \small The parameters chosen here are $\Lambda=3$, $\sigma=1$, $\epsilon_0=-1/2$, $\zeta=1$, and $\eta=1$. Blue line represents $c_2^\uparrow(R)=(1.35861 - 0.291269 \ i) c_2^+(R)+ (-0.866516 + 0.152247 \ i) c_2^-(R)$. The red line represents $c_{2(\uparrow)}^{\mathrm{WKB}}(R)$.}
    \end{minipage}
    \hfill
    \begin{minipage}{0.45\textwidth}
        \centering
        \includegraphics[width=\linewidth]{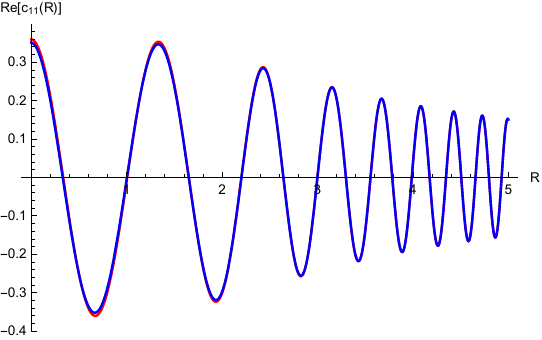}
        \caption*{(b) \small The parameters chosen here are $\Lambda=3$, $\sigma=1$, $\epsilon_0=-1/2$, $\zeta=10$, and $\eta=1$. Blue line represents $c_{11}^\uparrow(R)=(1.8223 - 0.0455046 \ i) c_{11}^+(R)+ (-1.47137 + 0.0367493 \ i) c_{11}^-(R)$. The red line represents $c_{{11}(\uparrow)}^{\mathrm{WKB}}(R)$.}
    \end{minipage}
     \hfill
    \begin{minipage}{0.5\textwidth}
        \centering
        \includegraphics[width=\linewidth]{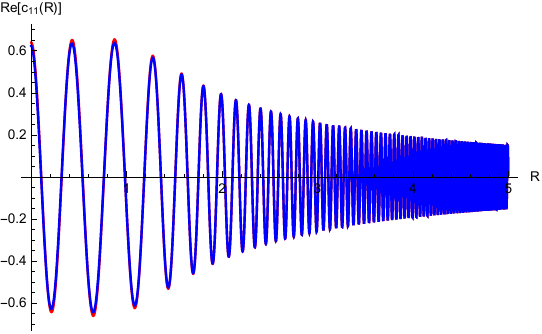}
        \caption*{(c) \small The parameters chosen here are $\Lambda=3$, $\sigma=0.1$, $\epsilon_0=-1/2$, $\zeta=1$, and $\eta=1$. Blue line represents $c_{11}^\uparrow(R)=(1.23694 - 0.0406383 \ i) c_{11}^+(R)+ (-0.613484 + 0.0201552 \ i) c_{11}^-(R)$. The red line represents $c_{{11}(\uparrow)}^{\mathrm{WKB}}(R)$.}
    \end{minipage}
    \caption{\small  All three plots display the real part of the functions. The plots for the imaginary part demonstrate similar agreement. The parameters for the linear combinations of exact solutions are obtained by requiring the linear combination and its first derivative to match with those of the WKB states at $R=5$. These three plots illustrate the three different scenarios wherein the validity condition is fulfilled by holding: $R\gg 1$, $\zeta \gg 1$, and $\sigma \ll 1$ respectively.}
    \label{heavyplots}
\end{figure}

The validity regime for this approximation is characterized by the following condition:
\bea
\Bigg|\sigma \frac{\tilde{p}'}{\tilde{p}} \Bigg|  \ll |\tilde{p}|, 
\eea
Note that for heavy states, $\tilde{p}(R)> 0$ for all values of $R$. Hence, the aforementioned condition can be fulfilled under three potential scenarios: when $\sigma \ll 1$, when $\zeta \gg 1$, or when $R \gg 1$. Refer to Figure \ref{heavyplots}, where the WKB solutions are plotted alongside the exact solutions for comparison. 

From this point, one could proceed to identify the emergent quantum mechanical states, as previously done for light states. However, we will not pursue this path, as the narrative remains similar. The notable difference lies in the fact that the quantum mechanical states for $\chi$ would also become WKB states in certain regions of $\chi$. Further, classical behavior for $\chi$ also emerges in the context of heavy states.

\newpage

\subsubsection{WKB branch merging and spreading}

\begin{figure}[h]
    
\begin{center}
\begin{tikzpicture}[x=0.75pt,y=0.75pt,yscale=-1,xscale=0.9]
    \newcommand{\symcurve}[2]{%
        \draw (#1,130+#2) .. controls (#1+40,130+#2/2) and (199,130) .. (480,130);
        \draw (#1,130-#2) .. controls (#1+40,130-#2/2) and (199,130) .. (480,130);
    }

    \symcurve{30}{110}
    \symcurve{30}{60}
    \symcurve{30}{30}
    \symcurve{30}{0}

    \draw (480,130) -- (580,130);

    \node at (-20, 50) {Heavy states};

    \node at (-20, 210) {Light states};

    \node at (485, 110) {Asymptotically de Sitter};
    
\end{tikzpicture}
\end{center}
    \caption{\small The different WKB branches are represented by curved lines in the figure. The value of $R$ increases from the left to the right in this figure. For the expanding branch, time runs along increasing in $R$, whereas for the contracting branch, time runs along decreasing in $R$. The lower lines pertain to light states, which adopt a Lorentzian, WKB character after a certain point in time. The upper lines relate to heavy states, which exhibit a Lorentzian and WKB nature right from the outset at $R=0$. All these varied WKB branches eventually converge, forming a single WKB branch that corresponds to an asymptotically de Sitter spacetime.}
    \label{branchmerging}
\end{figure}
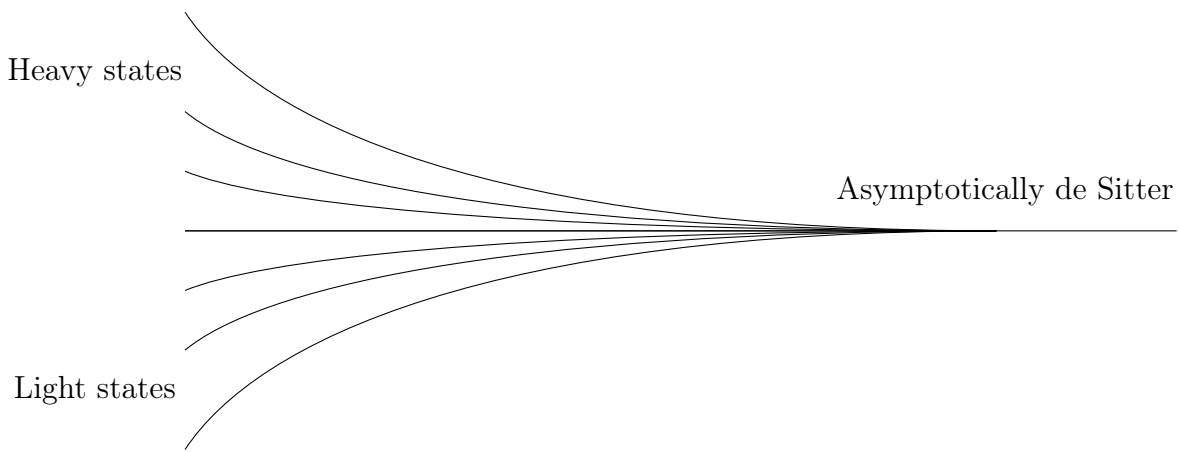

Let us summarise a pivotal observation. Through our analysis, we've learned that not all states exhibit WKB behavior across all values of $R$; however, they all transform into WKB states for $R \gg 1$. This is pertinent to the late-time limit for the expanding branch, denoted as $(\uparrow)$, and the early-time limit for the contracting branch, denoted as $(\downarrow)$. In the context of light states, spacetime becomes classical and Lorentzian only beyond a certain value of $R$. Conversely, for heavy states, the spacetime remains classical and Lorentzian across the entire range, extending to $R=0$. While these WKB branches differ at early times, they eventually converge into a single WKB branch at a late time, seamlessly transitioning into an asymptotically de Sitter spacetime, as depicted in Figure \ref{branchmerging}.

Another crucial insight can be drawn from equation \eqref{minisuperspaceqmstates}. The superposition coefficients present in these emergent quantum mechanical states are derived from the superposition coefficients of the complete WDW state given in Equation \eqref{coefficientsinWDW}. This suggests that when two WKB branches, each hosting different quantum mechanical states, merge into a single branch, the superposition coefficients from the two WKB states are integrated into the quantum mechanical state on the merged branch.

In other words, the information embedded in the quantum mechanical states of the matter component on different WKB branches amalgamates into the information incorporated in the quantum mechanical states of the matter component on the merged branch. Conversely, when a solitary WKB branch bifurcates into two or more branches, the information from the quantum mechanical states of the matter component on the initial branch is distributed among all the resultant branches.

The emergent quantum mechanical state, which satisfies an emergent Schrödinger equation, undergoes unitary evolution solely within individual WKB branches. When these branches merge, bifurcate, or enter a regime where the WKB approximation ceases to be valid, unitary evolution is no longer applicable. However, this doesn't imply the loss of information; instead, it simply signifies that the information is represented in a more intricate manner in the full WDW state. This phenomenon is an inherent part of the quantum mechanical nature of gravity. We believe that this phenomenon has significant implications for the black hole information paradox.

\subsection{Full superspace}

Within the all-encompassing superspace, the degrees of freedom for both the metric and matter fields remain unrestricted. To solve the WDW equation in this complete superspace, the formalism developed in \cite{GRW} can be employed. Though originally designed for the AdS case, this methodology can be adapted to the dS case with slight modifications. Let us briefly sketch the logic presented in \cite{GRW}.

Freidel \cite{Freidel} demonstrated that any solution to the radial WDW equation in an asymptotically AdS quantum gravity theory assumes a universal form. This was subsequently pointed out to also apply to the dS case by \cite{GRW}, and a rigorous derivation in the dS context was later provided by \cite{Araujo-Regado:2022jpj} and \cite{Chakraborty:2023yed}. In the dS scenario, any solution to the WDW equation tends towards the following form in the superspace region that corresponds to an infinite uniform rescaling of the metric and an appropriate rescaling of the matter fields:
\bea
\Psi [g,\Phi] \longrightarrow A_{+} \ e^{+\text{CT}[g,\Phi] } Z_{\mathrm{CFT}(+)}[g,\Phi] + A_{-} \ e^{- \text{CT}[g,\Phi] } Z_{\mathrm{CFT}(-)}[g,\Phi].
\eea
Here, $g$ is the Euclidean metric, and $\Phi$ is the matter field on the abstract $d$-dimensional compact manifold $\Sigma$. This infinite rescaling is associated with the large volume limit, i.e., $\mathrm{Vol}[g] \to \infty$. $\text{CT}[g,\Phi]$ is a spatial integral of local relevant operators and is universal across all WDW states. $Z_{\mathrm{CFT}(\pm)}$ mimic CFTs with opposing anomalies:
\begin{equation} 
 \mathcal{W}(x) \, Z_{\mathrm{CFT}(\pm)}^{(\epsilon)} = (\pm) i\mathcal{A}(x) \, Z_{\mathrm{CFT}(\pm)}^{(\epsilon)} ,
\end{equation}
where $ \mathcal{W}(x)$ is the Weyl generator, $\mathcal{A}(x)$ is the Weyl anomaly, and $\epsilon$ is a regulator necessary to regulate the logarithmic divergences (which cannot be eliminated by counterterms) arising from the anomaly. The asymptotic analysis only indicates that $Z_{\mathrm{CFT}(\pm)}^{(\epsilon)}$ is a functional of the metric and matter fields that satisfy the Weyl anomaly equation. It does not assert whether it is a proper CFT, which would additionally need to fulfill the axioms of Euclidean QFT in curved spacetime.

In order to solve the WDW equation beyond this asymptotic limit, it is necessary to apply a $T^2$ deformation to each of the functionals $Z_{\mathrm{CFT}(\pm)}$, leading to the following form:
\bea
\Psi\left[g,\Phi \right] &=& A_{+} \ Z_{T^2(+)}[g,\Phi] + A_{-} \ Z_{T^2(-)}[g,\Phi].
\eea
For further details on the $T^2$ deformation of $Z_{\mathrm{CFT}(\pm)}$ to yield $Z_{T^2(\pm)}$, we refer the reader to \cite{GRW} and \cite{Araujo-Regado:2022jpj}. Choosing a WDW state essentially involves selecting $Z_{\mathrm{CFT}(\pm)}$ and the coefficients $A_{\pm}$. As the Weyl anomaly equation is linear, any linear combination of $Z_{\mathrm{CFT}(+)}$ will satisfy it, leading to the formation of a vector space from all functionals that satisfy this Weyl anomaly equation. The same principle applies to $Z_{\mathrm{CFT}(-)}$. Thus, making a linear combination of WDW states is equivalent to making a linear combination of $Z_{\mathrm{CFT}}$.

To discern the branch corresponding to the WKB branch of an expanding universe, it's sufficient to examine the behavior of $Z_{T^2(\pm)}$ in the large volume limit ($\mathrm{Vol}[g] \to \infty$). The counter terms $\text{CT}(\mu)$ yield the dominant contribution. For the case of a scalar field coupled to gravity in a $3+1$ dimensional spacetime, the counterterms identified in \cite{GRW} are as follows:
\bea 
\text{CT} &=& \frac{i}{16 \pi G_N}\int d^3y \sqrt{g} \left(4  \sqrt{\frac{\Lambda}{3}}   -  \sqrt{\frac{3}{\Lambda}} \ R \right) + i\frac{\Delta_{\phi}}{2}  \sqrt{\frac{\Lambda}{3}} \int d^3y \sqrt{g}   \ \Phi^2 . 
\eea
When $g$ is confined to the minisuperspace metric $g_{ab} dx^a dx^b = r^2 d\Omega^2_3$, the pure gravity component of $\mathrm{CT}$ becomes:
\bea\label{minisuperspacecountertermpuregravity}
\frac{i}{16 \pi G_N}\int d^3y \sqrt{g} \left(4  \sqrt{\frac{\Lambda}{3}}   -  \sqrt{\frac{3}{\Lambda}} \ R \right) =  \frac{i 2 \pi^2 }{16 \pi G_N}  \left(4 r^3 \sqrt{\frac{\Lambda}{3}}   - 6 r \sqrt{\frac{3}{\Lambda}}  \right). 
\eea
Please note the change in notation here. In subsection \ref{minisuperspacemodel}, $R$ was used to denote the scale factor, but in this subsection, it signifies the Ricci scalar. Thus, we use $r$ to represent the scale factor in this context. We can now compare this with the WKB states of the minisuperspace model. For $r \gg 1$, performing a binomial expansion of $\frac{i}{\sigma} S^{(0)}_{(\uparrow\downarrow)}(r)= (\mp) \frac{i}{\sigma \Lambda} \left(\frac{\Lambda}{3} r^2 - 1\right)^{3/2}$, we obtain:
\bea
\frac{i}{\sigma} S^{(0)}_{(\uparrow\downarrow)}(r) &=& (\mp) \frac{i 2 \pi^2}{ 16 \pi G_N }  \left( 4 r^3 \sqrt{\frac{\Lambda}{3}} - 6 r \sqrt{\frac{3}{\Lambda}} \right) + O(r^{-1}). 
\eea
The initial two terms in this expansion align with the counter terms in equation \ref{minisuperspacecountertermpuregravity}. Each of the $T^2$ branches, therefore, approaches WKB branches asymptotically as $r$ increases significantly and so $Z_{T^2(-)}$ corresponds to the expanding branch $(\uparrow)$, and $Z_{T^2(+)}$ corresponds to the contracting branch $(\downarrow)$:
\bea
\Psi^{\uparrow\downarrow}[g,\Phi] = Z_{T^2(\mp)}.
\eea
In this sense, we can interpret $Z_{\mathrm{CFT}(-)}$ as living on $\mathcal{I}^+$ (and similarly, $Z_{\mathrm{CFT}(+)}$ on $\mathcal{I}^-$), thereby encoding both the dynamics and state of the universe.

Focusing on any of the individual branches, we can extrapolate from the minisuperspace calculations, which suggest that a typical expanding branch, $\Psi^{\uparrow}$, is likely to be in a WKB state only in the large volume (or late-time) limit. Whether it exists in a WKB state at early times depends on the chosen state. In the regime where WKB is valid, the spacetime generated by the WDW state and the emergent QFT state in the curved spacetime it generates can be calculated using the same methods developed in subsections \ref{bulkclassicalduals} and \ref{bulkQFTduals}. The $d$-dimensional extrinsic curvature $K_{ab}$ can be extracted from the $T^2$ theory using the same relation as in equation \ref{KrelationtoT1ptfun}. Similarly, expectation values of the bulk field operators in the QFT state that emerges from the WDW state can be derived employing the formula in equation \ref{onepointexpectationvalue}. The notable distinction here is that it's unnecessary to evaluate these quantities on the maximal volume slice. When taking the late-time limit, these expectation values of the bulk field operators become the cosmological correlators.

\subsection{Operational observers} \label{operationalobservers}

In the context of our discussion, it is crucial to clarify our understanding of the term ``observer'', as its interpretation can vary greatly among researchers. Here, we employ the term in a specific sense, and to distinguish what we mean exactly, we refer to it as ``\textbf{operational observers}''. 

In the realm of quantum mechanics, a ``\textbf{Heisenberg cut}'' is conceptualized as the abstract boundary separating quantum events from the observer's information, knowledge, or conscious awareness. Below this cut, everything is governed by the wave function; however, above the cut, everything resorts to a classical description. A realistic observer always employs a classical object for conducting measurements, such as a particle detector, a photographic plate, or a classical computer. Thus, operational observers reside above the Heisenberg cut, relying on these classical tools to interpret quantum phenomena.

While operational observers are classical entities, it's essential to recognize that they are made up of entities that are fundamentally quantum mechanical in nature. Every physical apparatus employed in measurements or experiments is composed of the same material that also adheres to the laws of quantum mechanics. Indeed, even the human performing the measurements is made of material that obeys these quantum laws.

Consequently, it is plausible to assign a wave function to describe an operational observer. Since an operational observer is inherently a classical concept, its manifestation must occur through an approximation, specifically the WKB approximation. When we consider the wave function of a system that includes an observer and a quantum mechanical subsystem, it must be feasible only when the wave function is in a WKB state concerning the degrees of freedom that constitute the observer. This is akin to the classical-quantum split elucidated in subsection \ref{CQsplitinQM}, where the state was in a WKB state with respect to the heavy particle, but not necessarily with respect to the light particle.

To elucidate this concept further, let's take an example from conventional quantum physics. Imagine a system, $S$, which consists of an observer, $O$, and a qubit in the state $(\alpha |0\rangle + \beta |1\rangle)$. Here, $\alpha$ and $\beta$ are complex numbers that satisfy the condition $|\alpha|^2 + |\beta|^2 = 1$. 

We also introduce an external observer, $E$, who is not part of the system $S$ but is observing it. Since observer $O$ is now considered part of system $S$, the state of the entire system $S$ at the initial time $t=0$, from the perspective of external observer $E$, can be described as:
\bea
|\psi_{(0)}\rangle = |O_{(0)}\rangle (\alpha |0\rangle + \beta |1\rangle),
\eea
where $|O_{(0)}\rangle$ is the initial state of observer $O$. Importantly, this must be a WKB state to enable the observers to behave classically, allowing them to be classified as operational observers.

Assume that the initial state $|O_{(0)}\rangle$ corresponds to a situation where the observer has decided to measure the qubit before the time $t=1$. External observer $E$ can then use unitary dynamics to evolve the state of the entire system $S$. By time $t=1$, observer $O$ has measured the qubit and perceives it to have collapsed to either the state $|0\rangle$ or $|1\rangle$. 

From the perspective of observer $E$, however, the state of the entire system $S$ at time $t$ is $|\psi_{(t)}\rangle$, and no collapse has taken place. Instead, observer $O$, as a subsystem, has merely interacted with the qubit, another subsystem. Observer $E$ would then describe the system $S$ to be in the state:
\bea
|\psi_{(1)}\rangle = \alpha |O_{(1)}^0\rangle |0\rangle + \beta |O_{(1)}^1\rangle |1\rangle =: \alpha|\psi_{(1)}^{0}\rangle + \beta|\psi_{(1)}^{1}\rangle,
\eea
where $|O_{(1)}^0\rangle$ and $|O_{(1)}^1\rangle$ correspond to the state of observer $O$ who has registered a measurement output of $0$ and $1$, respectively. At this stage, the state is no longer a single WKB state. Instead, it's a summation of two WKB states, or two WKB branches, each accommodating the observer's perception of a specific measurement outcome.

Indeed, the system $S$ exists in the state $|\psi_{(1)}\rangle$ with respect to the external observer $E$. However, if one were to ask about the state of the qubit from the perspective of observer $O$, the answer would depend on the branch of $|\psi_{(1)}\rangle$ being considered. \emph{This example illustrates a foundational principle of quantum mechanics: the state of a system is defined relative to a specific operational observer. Moreover, different observers may have distinct understandings of the state of the system, emphasizing the necessity of explicitly stating both the system and the operational observer when discussing a wave function.}

It is crucial to highlight an ostensibly straightforward point: all experimental devices occupy a specific location in space, and measurements are carried out at some finitely-sized regions of space-time. Further, the outcome of any measurement is recorded in a physical object located at some region of space. An ``operational observer'' refers to this entire process, encompassing the beginning of the measurement through to the end, where the outcome is recorded. Such concepts lose their significance without a background in which to situate the operational observer; the notions of space and time are thus integral to a meaningful observer construct. 

In scenarios where gravity becomes intensely quantum mechanical, the breakdown of the concept of space and time—due to the superposition of background geometry—leads to a corresponding breakdown in the very notion of operational observers. The existence of a classical object, as traditionally described in classical mechanics, is contingent upon having a classical spacetime background. All classical notions, including those of measuring devices, only make sense when grounded in a background spacetime. Consequently, the existence of operational observers necessitates a classical spacetime background, and therefore requires the WDW state to be a semiclassical state—one that encompasses at least one region of spacetime that is classical, and within which observers can reside.

For a closed universe, the wave function of the universe is all-encompassing, including even the observers within the universe itself. Therefore, it becomes necessary to define which portion of the universe represents the observer and which part comprises the system that these observers are studying. This division then gives rise to a separation of arguments in the WDW state, whereby some degrees of freedom would correspond to the observers, others would correspond to the system under investigation, and the remaining ones would correspond to the background spacetime and other things.

Consequently, the WDW state leads to several emergent descriptions of subsystems within the universe, as observed by other subsystems, depending on the context of this division. For observers to even exist, the WDW state must initially allow for a semiclassical description. The WDW state could be composed of a multitude of different WKB branches, each of which could either host distinct observers or the same observer perceiving different outcomes of a specific measurement. 

Hence, despite having chosen a particular wave function of the universe, the induced states could be numerous, with different branches corresponding to different induced states. This concept—that subsystems observe various states while examining other subsystems, despite the entire universe existing in a single state—will henceforth be referred to as an ``\textbf{enumeration of states}''.

In the context of AdS/CFT correspondence, the boundary of AdS spacetime provides a natural laboratory. It is natural to interpret the WDW state as providing a description of the entire bulk (in the WDW patch) as observed by the ``boundary observers''. However, any observer inside the bulk—referred to as a ``bulk observer''—cannot be a fundamental entity in this picture because they exist within the spacetime that the WDW state is describing. Hence, they must be emergent from the WDW state.

\section{Subregion Classicalisation} \label{subregionclassicalisationscreens}

In subsection \ref{constructingaclassicalspacetime}, we constructed a classical spacetime starting from a chosen Cauchy slice, which relied on the applicability of the WKB approximation throughout this initial Cauchy slice. However, for WDW states describing quantum gravitational processes in specific regions of spacetime, the WKB approximation will not necessarily be valid across an entire initial Cauchy slice.\footnote{It is often the case in reality that a field, such as an electromagnetic field, is treated classically in certain regions of spacetime (for instance, in capacitors) and quantum mechanically in others (for instance, in particle colliders).} In this section, we delve into the situations when gravity can be treated quantum mechanically in certain regions of spacetime and classically in others.

\subsection{Emergent WDW screens}

\begin{figure}[h]
\centering
\tikzset{every picture/.style={line width=0.75pt}} 

\begin{tikzpicture}[x=0.75pt,y=0.75pt,yscale=-1,xscale=1]

\draw[dashed]    (210,217.38) .. controls (279.33,181.38) and (375.33,180.38) .. (450,217.38) ;
\draw    (70,277.38) .. controls (117.33,262.38) and (159.33,245.38) .. (210,217.38) ;
\draw    (450,217.38) .. controls (503.33,243.38) and (540.33,267.38) .. (590,277.38) ;

\draw    (68,264) .. controls (50,227) and (220,157) .. (233,198) ;
\draw    (432,198) .. controls (444,146) and (619,233) .. (589,263) ;
\draw    (233,198) -- (235.3,204.13) ;
\draw [shift={(236,206)}, rotate = 249.44] [color={rgb, 255:red, 0; green, 0; blue, 0 }  ][line width=0.75]    (10.93,-3.29) .. controls (6.95,-1.4) and (3.31,-0.3) .. (0,0) .. controls (3.31,0.3) and (6.95,1.4) .. (10.93,3.29)   ;
\draw    (432,198) -- (429.53,207.07) ;
\draw [shift={(429,209)}, rotate = 285.26] [color={rgb, 255:red, 0; green, 0; blue, 0 }  ][line width=0.75]    (10.93,-3.29) .. controls (6.95,-1.4) and (3.31,-0.3) .. (0,0) .. controls (3.31,0.3) and (6.95,1.4) .. (10.93,3.29)   ;

\draw (285,202.4) node [anchor=north west][inner sep=0.75pt]    {$( \Sigma^0_{\text{quant}} ,g^0_{\text{quant}})$};
\draw (121,262.4) node [anchor=north west][inner sep=0.75pt]    {$( \Sigma^0_{\text{class}} ,g^0_{\text{class}})$};
\draw (127,175) node [anchor=north west][inner sep=0.75pt]   [align=left] {$\Sigma^0_H$};
\draw (326,165) node [anchor=north west][inner sep=0.75pt]   [align=left] {$\Sigma^0_\slashed{H}$};

\end{tikzpicture}

\caption{\small This figure denotes various segments of the starting Cauchy slice. $\Sigma^0_H$ denotes the area where the Halliwell criterion is met, implying that this part can be handled classically. We refer to this as the partial Halliwell slice. Nevertheless, a minor region of the partial Halliwell slice is treated quantum mechanically by choice, and the remaining part, which is treated classically, is represented as $\Sigma^0_{\text{class}}$. The complement of $\Sigma^0_{\text{class}}$, treated quantum mechanically, is denoted as $\Sigma^0_{\text{quant}}$. The complement of $\Sigma^0_H$, where the Halliwell criterion is not met, is denoted as $\Sigma^0_\slashed{H}$.}
\label{subregionclassicalisationfig}
\end{figure}
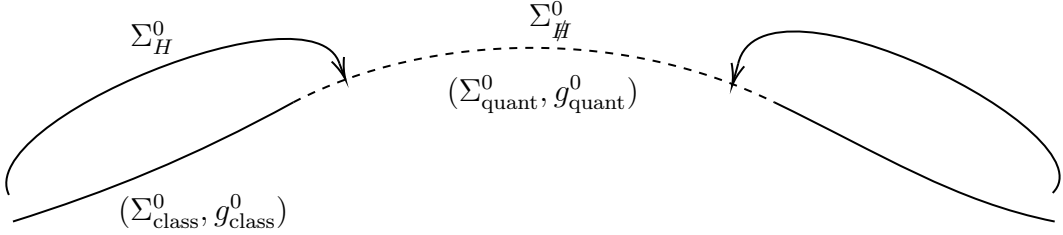

First, we will discuss the general framework for treating gravity classically outside a region of spacetime and quantum mechanically within that region. Consider an initial Cauchy slice, $\Sigma^0$, equipped with the metric $g^0$ (refer to Fig. \ref{subregionclassicalisationfig}). As detailed in subsection \ref{semiclassicaleinsteinequationfromWDW}, the Halliwell criterion, as applied to gravity, is a local criterion. Consequently, it can be sufficiently satisfied on certain subsets of the starting Cauchy slice, while failing to hold in others. \footnote{We kindly direct the reader's attention to the end of subsection \ref{semiclassicaleinsteinequationfromWDW} to remind themselves of what it means to satisfy the Halliwell criterion sufficiently.}
In the context of WDW states and an initial Cauchy slice, $\Sigma^0$, equipped with the metric $g^0$, let's explore scenarios where the Halliwell criterion is satisfied everywhere on $\Sigma^0$ except within a finite, bounded subregion.

Let us denote the portion of the starting Cauchy slice that satisfies the Halliwell criterion as $\Sigma^0_H$, or the partial Halliwell slice. We will refer to the complementary set of the partial Halliwell slice as the partial anti-Halliwell slice, symbolized as $\Sigma^0_{\slashed{H}}$. The partial anti-Halliwell slice cannot be treated classically, as this is the region where the WKB approximation ceases to apply. A classical spacetime can only emerge from the partial Halliwell slice, and thus, gravity can only be classically addressed here. Notwithstanding, in areas where gravity can be classically treated, it is not erroneous to handle it quantum mechanically, though this might be excessive\footnote{This is akin to treating the moon quantum mechanically by formulating a wavefunction for it. While this would indeed provide a correct description of the moon, it would be unnecessarily intricate when a classical description would suffice.}. 

Let us select a particular subset of the partial Halliwell slice, denoted as $\Sigma^0_{\text{class}} \subseteq \Sigma^0_\slashed{H}$, and exclusively construct a classical spacetime from it. This subset may represent the entire partial Halliwell slice or just a portion of it. We will signify $\Sigma^0_{\text{class}}$ with a solid line in Fig. \ref{subregionclassicalisationfig}. Let's denote the complement of $\Sigma^0_{\text{class}}$ as $\Sigma^0_{\text{quant}}$ and represent it by a dashed line in Fig. \ref{subregionclassicalisationfig}.

For WDW states satisfying the Halliwell criterion at least on $\Sigma^0_{\text{class}}$, the reduced Wigner functional of gravity would assume the following form:
\bea
W[g,\Pi]\Big|_{g=g^0} \sim  \mathcal{F}[g^0 ,\Pi|_{x \in \Sigma^0_{\text{quant}}}] \left( \prod_{x \in \Sigma^0_{\text{class}}} \delta\left(\Pi_{ab}(x) - f_{ab}(x) \right) \right),
\eea
where $\mathcal{F}[g^0 ,\Pi|_{x \in \Sigma^0_{\text{quant}}}]$ is any arbitrary functional and $f_{ab}$ satisfies the backreacted Einstein-Hamilton-Jacobi equation. Consequently, the classical canonical momentum conjugate to the metric on the classical region $\Sigma^0_{\text{class}}$ is:
\bea
\Pi^{ab}_{\mathrm{cl}}(x) =f_{ab}(x). \label{subregionclassicalmetricmomentum}
\eea

\begin{figure}[h]
\centering
\tikzset{every picture/.style={line width=0.75pt}} 
\begin{tikzpicture}[x=0.75pt,y=0.75pt,yscale=-1,xscale=1]

\draw[dashed]    (210,217.38) .. controls (279.33,181.38) and (375.33,180.38) .. (450,217.38) ;
\draw    (70,277.38) .. controls (117.33,262.38) and (159.33,245.38) .. (210,217.38) ;
\draw    (450,217.38) .. controls (503.33,243.38) and (540.33,267.38) .. (590,277.38) ;
\draw    (70,257.38) .. controls (117.33,242.38) and (159.33,225.38) .. (210,197.38) ;
\draw    (450,197.38) .. controls (503.33,223.38) and (540.33,247.38) .. (590,257.38) ;
\draw    (210,197.38) -- (210,217.38) ;
\draw    (450,197.38) -- (450,217.38) ;



\draw (121,262.4) node [anchor=north west][inner sep=0.75pt]    {$( \Sigma^0_{\text{class}} ,g^0_{\text{class}})$};
\draw (100,200) node [anchor=north west][inner sep=0.75pt]   [align=left] {$(\tilde{g},\tilde{\Sigma})$};
\draw (190,170) node [anchor=north west][inner sep=0.75pt]   [align=left] {$\partial\tilde{\Sigma}$};

\end{tikzpicture}
\caption{\small From the WDW state, we can determine the extrinsic curvature on $\Sigma^0_{\text{class}}$, which is used to compute the metric $\tilde{g}$ on the subsequent partial Cauchy slice $\tilde{\Sigma}$. However, $\tilde{\Sigma}$ may not necessarily be a partial Halliwell slice. To confirm this, one must resort to the analysis of the Wigner function.}
\label{subregionclassicalisationfigemergent}
\end{figure}
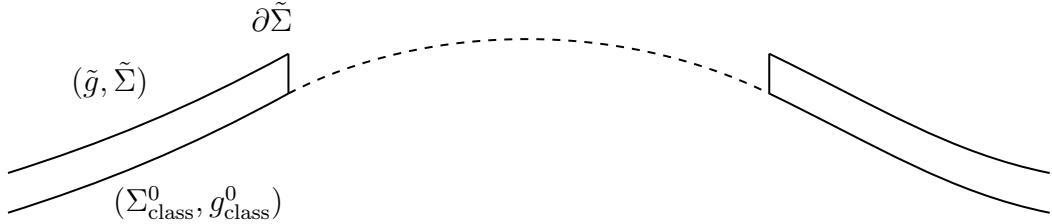

Next, we proceed similarly to the procedure detailed in subsection \ref{constructingaclassicalspacetime}, finding the metric on the ``infinitesimally next''  partial Cauchy slice (which we denote as $\tilde{\Sigma}$). We accomplish this by determining the extrinsic curvature at $x$ using $\Pi^{ab}_{\mathrm{cl}}(x)$ and choosing appropriate lapse and shifts:
\bea \label{nextpartialslicemetric}
 \tilde{g}_{ab} = \Delta t \left( 2 N K_{ab} + D_a N_b + D_b N_a \right) + (g^0_{\text{class}})_{ab},
\eea
where $\tilde{g}$ represents the metric on $\tilde{\Sigma}$. We denote the inner boundary of $\tilde{\Sigma}$ as $\partial\tilde{\Sigma}$ (refer to Fig \ref{subregionclassicalisationfigemergent}).

The difference between the procedure in this subsection and that of subsection \ref{constructingaclassicalspacetime} is that here, we construct a spacetime only from a subregion of the Cauchy slices. This is due to the state possibly not conforming to the WKB approximation across the entire slice.

The next step is to find the spatial completions of $\tilde{g}$. A metric $g^{c} \in \mathcal{S}$ on a manifold $\Sigma^c$ (where $\tilde{\Sigma}$ is a submanifold of $\Sigma^c$) is considered a spatial completion of $\tilde{g}$ (which is defined on $\tilde{\Sigma}$) if an embedding of $\tilde{\Sigma}$ in $\Sigma^c$ can be found such that the pullback of $g^{c}$ is $\tilde{g}$ within $\tilde{\Sigma}$. The spatial completion of a metric on a partial Cauchy slice is not unique, and we merely select any one. Different choices of such spatial completions would correspond to various methods of extending the partial Cauchy slice as an embedding in a classical spacetime interior to the constructed subregion spacetime, if it exists. Ideally, one must search through all the embeddings to identify the optimal or some preferred ones.

\begin{figure}[h]
\centering
\tikzset{every picture/.style={line width=0.75pt}} 
\begin{tikzpicture}[x=0.75pt,y=0.75pt,yscale=-1,xscale=1]

\draw[dashed]    (210,217.38) .. controls (279.33,181.38) and (375.33,180.38) .. (450,217.38) ;
\draw    (70,277.38) .. controls (117.33,262.38) and (159.33,245.38) .. (210,217.38) ;
\draw    (450,217.38) .. controls (503.33,243.38) and (540.33,267.38) .. (590,277.38) ;
\draw    (70,257.38) .. controls (117.33,242.38) and (159.33,225.38) .. (210,197.38) ;
\draw    (450,197.38) .. controls (503.33,223.38) and (540.33,247.38) .. (590,257.38) ;
\draw[dotted]    (210,197.38) .. controls (279.33,161.38) and (375.33,160.38) .. (450,197.38) ;
\draw    (210,197.38) -- (210,217.38) ;
\draw    (450,197.38) -- (450,217.38) ;

\draw (121,262.4) node [anchor=north west][inner sep=0.75pt]    {$( \Sigma^0_{\text{class}} ,g^0_{\text{class}})$};
\draw (318,152) node [anchor=north west][inner sep=0.75pt]   [align=left] {$g^c$};
\draw (190,170) node [anchor=north west][inner sep=0.75pt]   [align=left] {$\partial\tilde{\Sigma}$};
\draw (100,200) node [anchor=north west][inner sep=0.75pt]   [align=left] {$(\tilde{g},\tilde{\Sigma})$};

\end{tikzpicture}
\caption{\small The dotted curve, in conjunction with $\tilde{\Sigma}$, represents $\Sigma^c$. This is furnished with the metric $g^c$, which serves as a spatial completion of $\tilde{g}$.}
\label{subregionclassicalisationfigemergent2}
\end{figure}
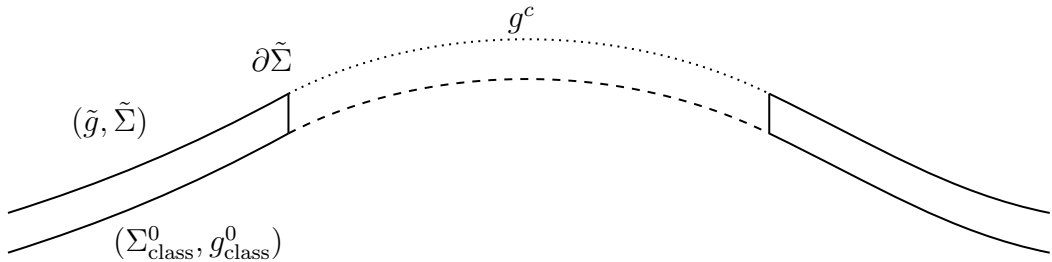

The next step involves analyzing the WDW state for the metric $g^{c}$ on the slice $\Sigma^c$ to check for the Halliwell criterion. This enables the identification of the partial Halliwell slice $\Sigma^c_{H}$ on this next slice $\Sigma^c$. By selecting an improper subset of it, denoted as $\Sigma^c_{\text{class}}$, one partitions $\Sigma^c$ into a classical part, $\Sigma^c_{\text{class}}$, and a quantum mechanical part, $\Sigma^c_{\text{quant}}$, as previously done. 

Additionally, one obtains the metric $g^c_{\text{class}}$ on $\Sigma^c_{\text{class}}$ as the pullback of $g^c$. The location of this division is entirely dependent on our selection of the improper subset of the partial Halliwell slice. The extent to which we can make this choice is dictated by the WDW state. Specifically, the division can be placed anywhere on the partial Halliwell slice based on our discretion, but it cannot extend beyond the boundary of the partial Halliwell slice. The location of  this boundary of the partial Halliwell slice is specified by the WDW state.

This split, denoted as $\partial\tilde{\Sigma}_{\text{quant}}$, could either lie inside $\partial\tilde{\Sigma}$ or outside it. If it lies outside $\partial\tilde{\Sigma}$, then the region between these two points must be discarded, as gravity is not treated classically there, even though the metric on this area was obtained from equation \eqref{nextpartialslicemetric}. If $\partial\tilde{\Sigma}_{\text{quant}}$ lies inside $\partial\tilde{\Sigma}$, then additional information about the classical spacetime is derived from the WDW state, exceeding the information about the next partial Cauchy slice obtained using equation \eqref{nextpartialslicemetric}.

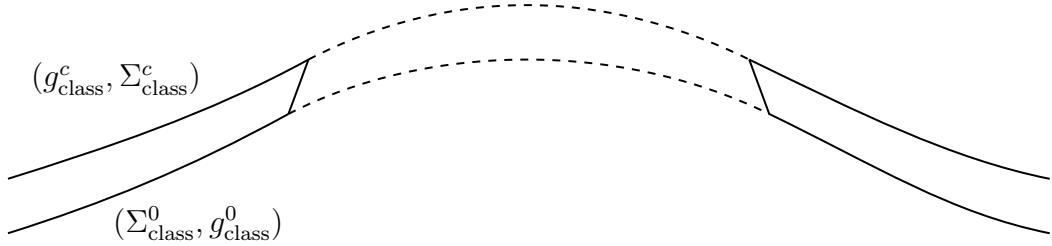
\begin{figure}[h]
\centering

\tikzset{every picture/.style={line width=0.75pt}} 

\begin{tikzpicture}[x=0.75pt,y=0.75pt,yscale=-1,xscale=1]

\draw[dashed]    (210,217.38) .. controls (279.33,181.38) and (375.33,180.38) .. (450,217.38) ;
\draw    (70,277.38) .. controls (117.33,262.38) and (159.33,245.38) .. (210,217.38) ;
\draw    (450,217.38) .. controls (503.33,243.38) and (540.33,267.38) .. (590,277.38) ;
\draw[dashed]    (220,190) .. controls (289.33,154) and (365.33,153) .. (440,190) ;
\draw    (70,250) .. controls (117.33,235) and (169.33,218) .. (220,190) ;
\draw    (440,190) .. controls (493.33,216) and (540.33,240) .. (590,250) ;

\draw    (220,190) -- (210,217.38) ;
\draw    (440,190) -- (450,217.38) ;

\draw (121,262.4) node [anchor=north west][inner sep=0.75pt]    {$( \Sigma^0_{\text{class}} ,g^0_{\text{class}})$};
\draw (80,190) node [anchor=north west][inner sep=0.75pt]   [align=left] {$(g^c_{\text{class}},\Sigma^c_{\text{class}})$};

\end{tikzpicture}

\caption{\small One checks the Halliwell criterion on the spatially completed slice to identify the subsequent partial Halliwell slice. From this, a choice is made regarding which subset will be treated classically, denoted as $\Sigma^c_{\text{class}}$.}
\label{subregionclassicalisationfigemergentnextclassical}
\end{figure}

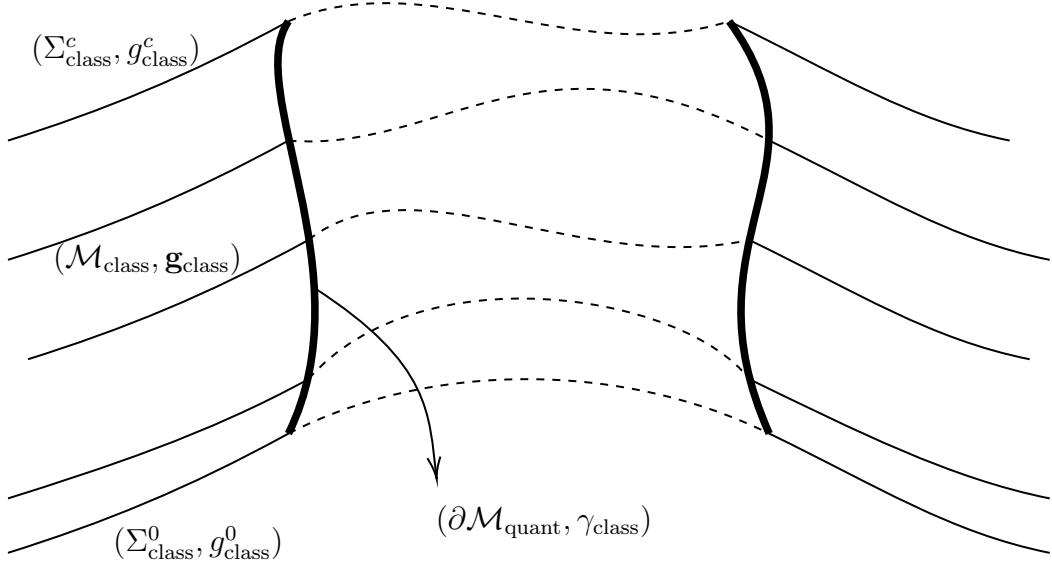
\begin{figure}[h]
\centering

\tikzset{every picture/.style={line width=0.75pt}} 

\begin{tikzpicture}[x=0.75pt,y=0.75pt,yscale=-1,xscale=1]

\draw[dashed]     (210,217.38) .. controls (279.33,181.38) and (375.33,180.38) .. (450,217.38) ;
\draw    (70,277.38) .. controls (117.33,262.38) and (159.33,245.38) .. (210,217.38) ;
\draw    (450,217.38) .. controls (503.33,243.38) and (540.33,267.38) .. (590,277.38) ;
\draw    (220,190) -- (210,217.38) ;
\draw    (440,190) -- (450,217.38) ;
\draw[dashed]     (220,190) .. controls (275,126.4) and (407,146.4) .. (440,190) ;
\draw    (70,250) .. controls (117.33,235) and (169.33,218) .. (220,190) ;
\draw    (440,190) .. controls (493.33,216) and (540.33,240) .. (590,250) ;
\draw[line width=1mm, black]    (210,10) .. controls (187,43.4) and (249,137.4) .. (210,217.38) ;
\draw[line width=1mm, black]    (430,10) .. controls (484,84.4) and (406,118.4) .. (450,217.38) ;
\draw    (70,70) .. controls (117.33,55) and (159.33,38) .. (210,10) ;
\draw    (430,10) .. controls (483.33,36) and (520.33,60) .. (570,70) ;
\draw    (80,180) .. controls (127.33,165) and (169.33,148) .. (220,120) ;
\draw    (70,130) .. controls (117.33,115) and (159.33,98) .. (210,70) ;
\draw    (440,120) .. controls (493.33,146) and (530.33,170) .. (580,180) ;
\draw     (450,70) .. controls (503.33,96) and (540.33,120) .. (590,130) ;
\draw[dashed]     (210,70) .. controls (287,76.4) and (333,8.4) .. (450,70) ;
\draw[dashed]     (220,120) .. controls (272,78.4) and (357,138.4) .. (440,120) ;
\draw[dashed]    (210,10) .. controls (280,-19.6) and (353,34.4) .. (430,10) ;

\draw    (223,144.2) .. controls (281.8,182.42) and (280.09,208.15) .. (283.77,238.35) ;
\draw [shift={(284,240.2)}, rotate = 262.65] [color={rgb, 255:red, 0; green, 0; blue, 0 }  ][line width=0.75]    (10.93,-3.29) .. controls (6.95,-1.4) and (3.31,-0.3) .. (0,0) .. controls (3.31,0.3) and (6.95,1.4) .. (10.93,3.29)   ;

\draw (121,262.4) node [anchor=north west][inner sep=0.75pt]    {$( \Sigma^0_{\text{class}} ,g^0_{\text{class}})$};
\draw (80,15) node [anchor=north west][inner sep=0.75pt]   [align=left] {$(\Sigma^c_{\text{class}},g^c_{\text{class}})$};
\draw (91,120) node [anchor=north west][inner sep=0.75pt]   [align=left] {$(\mathcal{M}_{\text{class}},\mathbf{g}_{\text{class}})$};
\draw (282,252) node [anchor=north west][inner sep=0.75pt]   [align=left] {$(\partial\mathcal{M}_{\text{quant}},\gamma_{\text{class}})$};

\end{tikzpicture}
\caption{\small By iteratively repeating the partial stitching process, one obtains a classical exterior $(\mathcal{M}_{\text{class}},\mathbf{g}_{\text{class}})$ and an emergent time-like boundary $(\partial\mathcal{M}_{\text{quant}},\gamma_{\text{class}})$, which we refer to as a WDW screen. The metrics of the exterior and on the WDW screen depend entirely on the WDW state.}
\label{subregionclassicalspacetime}
\end{figure}

Beginning with the newly obtained metric $g^c_{\text{class}}$ on the partial Cauchy slice $\Sigma^c_{\text{class}}$, the steps outlined in this subsection are repeated. This iterative process enables the construction of a sequence of partial Cauchy slices, each equipped with a metric (and suitable choices of shift and lapse). Consequently, a subregion of classical spacetime $(\mathcal{M}_{\text{class}},\mathbf{g}_{\text{class}})$ is constructed, accompanied by a time-like boundary $(\partial\mathcal{M}_{\text{quant}},\gamma_{\text{class}})$, as depicted in Figure \ref{subregionclassicalspacetime}. The region bounded by $\partial\mathcal{M}_{\text{quant}}$ is denoted as $\mathcal{M}_{\text{quant}}$, although there may not be a classical spacetime present in this region. The constructed spacetime $(\mathcal{M}_{\text{class}},\mathbf{g}_{\text{class}})$ adheres to the semiclassical Einstein equations. This adherence is due to its construction using equation \eqref{subregionclassicalmetricmomentum}, which satisfies the back-reacted Einstein-Jacobi equation, as assured by the WKB approximation.

Let's now remove the Einstein-Hamilton principal function part from $\Psi_{\text{WDW}}$ and define a resulting state $\psi_{\text{emergent}}$:
\bea 
\psi_{\text{emergent}}[\phi,g] &:= \mathcal{N}[g_{\text{class}}] \tilde{\psi}_{\text{emergent}}[\phi,g],\\
\tilde{\psi}_{\text{emergent}}[\phi,g] &:= e^{-i\tilde{S}[g_{\text{class}}]} \Psi_{\text{WDW}}[g,\phi], \label{stripaway} \\
\mathcal{N}^{-2}[g_{\text{class}}] &:= \int \frac{D\phi D g_{\text{quant}}}{\text{Diff}\Sigma_{\text{quant}}} \big|\tilde{\psi}_{\text{emergent}}[\phi,g] \big|^2.
\eea
In the above equations, $\tilde{S}[g_{\text{class}}]$ is any functional of the metric on a partial Cauchy slice that gives rise to $(\mathcal{M}_{\text{class}},\mathbf{g}_{\text{class}})$. The emergent state $\psi_{\text{emergent}}$ satisfies the WDW equation in $\mathcal{M}_{\text{quant}}$, as its functional dependency concerning the metric and matter fields inside $\mathcal{M}_{\text{quant}}$ aligns with that of the WDW state. Furthermore, $\psi_{\text{emergent}}$ would conform to the Tomonaga-Schwinger equation within $\mathcal{M}_{\text{class}}$. This is because $\Psi_{\text{WDW}}$ is WKB in $\mathcal{M}_{\text{class}}$, and the Einstein-Hamilton principal function part of $\Psi_{\text{WDW}}$ has been removed, leaving behind next order terms that satisfy the Tomonaga-Schwinger equation, as explained in the preceding sections.

One can now express the WDW state as:
\bea
\Psi_{\text{WDW}}[g,\phi] = \frac{1}{\mathcal{N}[g_{\text{class}}]} e^{i \tilde{S}[g_{\text{class}}]} \psi_{\text{emergent}}[\phi,g].
\eea
By establishing a coordinate system within $\mathcal{M}_{\text{class}} \cup \partial\mathcal{M}_{\text{quant}}$, one can consider $\psi_{\text{emergent}}$ as residing on $\Sigma^{\text{ext}}_t \cup \partial\Sigma^{\text{quant}}_t$, where $t$ represents a time parameter in $\mathcal{M}_{\text{class}} \cup \partial\mathcal{M}_{\text{quant}}$. Here, $\Sigma^{\text{ext}}_t$ is a co-dimension one spacelike slice in $\mathcal{M}_{\text{class}}$, $\partial\Sigma^{\text{quant}}_t$ is a co-dimension one spacelike slice in $\partial\mathcal{M}_{\text{quant}}$, and $\Sigma^{\text{ext}}_t$ is anchored to $\partial\Sigma^{\text{quant}}_t$. The location of these slices is specified by the metric $g_{\text{class}}$ on them. By associating $g_{\text{class}}$ with the proper time parameter $\tau$, one can rewrite the WDW state as follows:
\bea
\Psi_{\text{WDW}}\bigg|_{\text{classical exterior}} = \frac{1}{\mathcal{N}[\tau]} e^{i \tilde{S}[\tau]} \psi_{\text{emergent}}[\phi,g_{\text{quant}},\tau].
\eea
The notation $\Psi_{\text{WDW}}\bigg|_{\text{classical exterior}}$ signifies that the WDW state is being evaluated using a metric, a portion of which aligns with $g_{\text{class}}$ on any slice within $\mathcal{M}_{\text{class}}$.

In an effort to comprehend the manner in which $\psi_{\text{emergent}}$ evolves with the boundary time $\tau_{\text{bdy}}$ (referring to the $\tau$ at $\partial\mathcal{M}_{\text{quant}}$), it's beneficial to utilize path integrals. Let us assume that the WDW state can be represented via a particular gravitational path integral. In specific regions where the WKB approximation holds, the gravitational path integral would be dominated by the classical saddle. However, this applies solely to those regions where WKB is valid. Consequently, this would be reduced to a path integral in which contributions from all possible matter fields are considered, as well as a restricted range of spacetime geometries such that the metric in $\mathcal{M}_{\text{class}}$ is fixed to be $\mathbf{g}_{\text{class}}$. Meanwhile, the metric in $\mathcal{M}_{\text{quant}}$ could be arbitrary as long as it fulfills Dirichlet boundary conditions on $\partial\mathcal{M}_{\text{quant}}$ (given the metric cannot exhibit discontinuity on $\partial\mathcal{M}_{\text{quant}}$). As proven by Hayward and Wong in \cite{Hayward:1992ix}, imposing such a restriction on the gravitational path integral would make it satisfy the boundary Schr\"{o}dinger equation. Consequently, we have:
\bea
i \frac{\delta}{\delta \tau_{\text{bdy}}(y)}\psi_{\text{emergent}}[\phi,g_{\text{quant}},\tau] = \mathcal{H}_{\text{ADM}}(y)\psi_{\text{emergent}}[\phi,g_{\text{quant}},\tau],
\eea
where $y \in \partial\mathcal{M}_{\text{quant}}$. In this sense, one must think of $\psi_{\text{emergent}}$ as a QFT state living on partial Cauchy slices outside the quantum region, tensor-producted with an induced WDW state living on the Cauchy slices of the emergent time-like boundary or a superposition of them. Henceforth, we will refer to this emergent time-like boundary as a \textbf{WDW screen}.

\textbf{Summary:} Let us summarize the results of this subsection. Starting from a WDW state, it is possible to obtain a classical spacetime (obeying the semiclassical Einstein equation) in a subregion while treating a finitely bounded region quantum mechanically. Additionally, a time-like boundary emerges, housing an induced WDW state induced by the initial WDW state. This induced WDW state encodes the quantum mechanical information of both the matter and gravitational field in $\mathcal{M}_{\text{quant}}$. We refer to this time-like boundary as a WDW screen. Furthermore, a QFT state emerges within the classical subregion, entangled with the induced WDW state. As long as the WKB approximation holds outside the WDW screen, the QFT state on the partial Cauchy slices outside the screen and the induced WDW state on the Cauchy slices of the WDW screen exist together in a pure state, evolving unitarily through the Tomonaga-Schwinger equation and a boundary Schrödinger equation. The location of the WDW screen can be chosen arbitrarily in any region where the WKB approximation is valid, and its extent is determined by the WDW state, as it indicates where precisely the breakdown of the WKB approximation occurs. However, if the Halliwell criterion fails on or outside the WDW screen, this description will break down. We will analyze the validity conditions for this in the next subsection.

\subsection{Spherical validity condition for subregion classicalization}

In this subsection, we develop a set of validity conditions that must be adhered to in order to represent gravity classically outside a region denoted as $\mathcal{M}_{\text{quant}}$ (inside this region, gravity could potentially behave non-classically). For gravity to be perfectly classical outside $\mathcal{M}_{\text{quant}}$, it is required that the Halliwell criterion be fulfilled outside $\mathcal{M}_{\text{quant}}$. This implies that the reduced Wigner functional of the metric on any partial Cauchy slice, denoted as $\Sigma_{\text{out}}$, outside $\mathcal{M}_{\text{quant}}$, must exhibit the following form:
\bea
W^{\text{classical}}_{\text{out}}[g^{\text{out}},\Pi^{\text{out}}] &=& \mathcal{P}[g^{\text{out}}] \delta^\infty\left[\frac{1}{\sqrt{g^{\text{out}}}}\left(\Pi_{\text{out}}^{ab}-F^{ab}\right)\right].
\eea
Beginning with the Wigner functional provided in equation \eqref{WDWWignerfunctional}, we perform integration over the matter variables ($\phi$, $\Pi_\phi$) as well as the gravitational variables ($g_{ab}^{\text{in}}$ , $\Pi^{ab}_{\text{in}}$) inside $\Sigma_{\text{in}}$. Here, $\Sigma_{\text{in}}$ represents the complement of $\Sigma_{\text{out}}$. Consequently, we obtain:
\begin{align}
W_{\text{out}}[g^{\text{out}},\Pi^{\text{out}}] &= \int \frac{Dg^{\text{in}} D\Pi_{\text{in}}}{\text{Diff}(\Sigma_{\text{in}})} \int D\phi D\Pi_\phi \ W[g,\Pi,\phi,\Pi_\phi],\\
&= \int \frac{Dg^{\text{in}} D\Pi_{\text{in}}}{\text{Diff}(\Sigma_{\text{in}})} \int D\phi  \int D(32\pi G_N h) e^{\left(2 i (32\pi G_N) \int d^dx h_{ab} \Pi^{ab}  \right)} \\
&\; \quad \quad \Psi^*[g+(32\pi G_N) h,\phi]\Psi[g-(32\pi G_N) h,\phi],\\
&=   \int D(32\pi G_N h^{\text{out}}) e^{\left(2 i (32\pi G_N) \int_{\Sigma_{\text{out}}} d^dx h_{ab} \Pi^{ab}  \right)}  \\
&\; \quad \quad \int \frac{Dg^{\text{in}} D\phi }{\text{Diff}(\Sigma_{\text{in}})} \Psi^*[g+(32\pi G_N) h^{\text{out}},\phi]\Psi[g-(32\pi G_N) h^{\text{out}},\phi].
\end{align}
By Taylor expanding about $h^{\text{out}}_{ab}=0$ we get:
\bea
W_{\text{out}}[g^{\text{out}},\Pi^{\text{out}}] &=&  \int D(32\pi G_N h^{\text{out}}) e^{\left(2 i (32\pi G_N) \int_{\Sigma_{\text{out}}} d^dx h_{ab} \Pi^{ab}  \right)}  \\
&\;& \int \frac{Dg^{\text{in}} D\phi }{\text{Diff}(\Sigma_{\text{in}})} e^{-2 S^I[g,\phi]}
e^{-2i(32\pi G_N) \int_{\Sigma_{\text{out}}} d^dx h^{\text{out}}_{ab} \frac{\delta S^R}{\delta g_{ab}}}.
\eea
By defining:
\bea
\mathcal{P}[g^{\text{out}}] := &\;& \int \frac{Dg^{\text{in}} D\phi }{\text{Diff}(\Sigma_{\text{in}})} e^{-2 S^I[g,\phi]}, \\
\left\langle \mathcal{A} \right\rangle := &\;& \frac{1}{\mathcal{P}[g^{\text{out}}]} \int D\mathcal{C} \ e^{-2 S^I[g,\phi]} \mathcal{A}, \\
D\mathcal{C} := &\;& \frac{Dg^{\text{in}} D\phi}{\text{Diff}(\Sigma_{\text{in}})}, \label{subregionmetricmeasure}\\
\psi[\phi,g] := &\;& \frac{e^{i S_1[g,\phi]}}{\sqrt{\mathcal{P}[g^{\text{out}}]}} =: e^{i \mathbf{s}}, \label{emergentminiWDWstatesubregion}
\eea
we get:
\bea
W_{\text{out}}[g^{\text{out}},\Pi^{\text{out}}] &=& \mathcal{P}[g^{\text{out}}]  \int D(32\pi G_N h^{\text{out}}) e^{\left(2 i (32\pi G_N) \int_{\Sigma_{\text{out}}} d^dx h^{\text{out}}_{ab} \Pi^{ab}  \right)}  \\
&\;& \left\langle e^{-2i(32\pi G_N) \int_{\Sigma_{\text{out}}} d^dx h^{\text{out}}_{ab} \frac{\delta S^R}{\delta g_{ab}}} \right\rangle.
\eea
If the following condition is satisfied (which we will take as a necessary condition for the subsequent analysis to hold):
\bea\label{subregionWDWvalidityconditions}
\left\langle e^{-2i(32\pi G_N) \int_{\Sigma_{\text{out}}} d^dx h^{\text{out}}_{ab} \frac{\delta S^R}{\delta g_{ab}}} \right\rangle =  e^{-2i(32\pi G_N) \int_{\Sigma_{\text{out}}} d^dx h^{\text{out}}_{ab} \left\langle\frac{\delta S^R}{\delta g_{ab}} \right\rangle},
\eea
then we get:
\begin{align}
W_{\text{out}}[g^{\text{out}},\Pi^{\text{out}}] &= \mathcal{P}[g^{\text{out}}]  \int D(32\pi G_N h^{\text{out}}) e^{2 i (32\pi G_N) \int_{\Sigma_{\text{out}}} d^dx h^{\text{out}}_{ab} \left(\Pi^{ab} -\left\langle\frac{\delta S^R}{\delta g_{ab}} \right\rangle\right)},\\
&= \mathcal{P}[g^{\text{out}}] \ \delta^{\infty}_{\text{out}}\left[\frac{1}{\sqrt{g}}\left(\Pi^{ab} -\left\langle\frac{\delta S^R}{\delta g_{ab}} \right\rangle\right)\right].
\end{align}
Therefore, the Halliwell criterion is upheld on $\Sigma_{\text{out}}$.

The validity conditions presented in \eqref{subregionWDWvalidityconditions} yield an infinite set of conditions for describing the metric classically outside $\mathcal{M}_{\text{quant}}$. Specifically, they are the following for each value of $n$:
\bea \label{towervcquantumball}
\left\langle \prod_{i=1}^n \frac{1}{\sqrt{g_i}(x_i)}\frac{\delta S^R}{\delta g_{a_ib_i}(x_i)}\right\rangle - \prod_{i=1}^n \left\langle \frac{1}{\sqrt{g_i}(x_i)}\frac{\delta S^R}{\delta g_{a_ib_i}(x_i)} \right\rangle  = 0.
\eea
In this expression, we should replace $\sqrt{g_i}$ with $\sqrt{g}$ when $x_i$ is outside $\mathcal{M}_{\text{quant}}$ and use the determinant of the induced metric $\sqrt{\gamma}$ if $x_i$ is on $\partial\mathcal{M}_{\text{quant}}$. Considering the case $n=2$, where $x$ lies on $\partial\Sigma_{\text{quant}}$ (a codimension one Cauchy slice of $\partial\mathcal{M}_{\text{quant}}$) and $y$ is outside, we obtain:
\bea 
\left\langle \frac{1}{\sqrt{\gamma(x)}}\frac{\delta S^R}{\delta g_{ab}(x)} \frac{1}{\sqrt{g(y)}}\frac{\delta S^R}{\delta g_{cd}(y)} \right\rangle - \left\langle \frac{1}{\sqrt{\gamma(x)}}\frac{\delta S^R}{\delta g_{ab}(x)} \right\rangle \left\langle\frac{1}{\sqrt{g(y)}}\frac{\delta S^R}{\delta g_{cd}(y)} \right\rangle = 0.
\eea
Upon expansion of $S$ in a power series in $G_N$, the leading order term $S_0$ does not depend on the quantum variables $(g^{\text{in}},\phi)$. Consequently, it comes out of the expectation values and reduces to zero. Therefore, the $O(G_N^0)$ term of aforementioned condition simplifies to:
\bea 
\left\langle \frac{1}{\sqrt{\gamma(x)}}\frac{\delta S_1^R}{\delta g_{ab}(x)} \frac{1}{\sqrt{g(y)}}\frac{\delta S_1^R}{\delta g_{cd}(y)} \right\rangle - \left\langle \frac{1}{\sqrt{\gamma(x)}}\frac{\delta S_1^R}{\delta g_{ab}(x)} \right\rangle \left\langle\frac{1}{\sqrt{g(y)}}\frac{\delta S_1^R}{\delta g_{cd}(y)} \right\rangle = 0.
\eea
The emergent state satisfies the Tomonaga-Schwinger equation at the point $y$ (which is situated outside $\mathcal{M}_{\text{quant}}$ and away from the boundary $\partial\mathcal{M}_{\text{quant}}$). Moreover, it satisfies the boundary WDW equation at the point $x$ (which lies on $\partial\Sigma_{\text{quant}}$). Consequently, the previously mentioned equation simplifies to:
\bea 
\mathcal{F}(x,y):=&\;&\left\langle \frac{\mathcal{H}_{\text{ADM}}(x)}{\sqrt{\gamma(x)}} \frac{\mathcal{H}_{\text{matter}}(y)}{\sqrt{g(y)}}\right\rangle - \left\langle \frac{\mathcal{H}_{\text{ADM}}(x)}{\sqrt{\gamma(x)}}\right\rangle \left\langle\frac{\mathcal{H}_{\text{matter}}(y)}{\sqrt{g(y)}}\right\rangle \\
&\;&- \frac{1}{\sqrt{\gamma(x)}\sqrt{g(y)}} \left\langle \frac{\delta \mathbf{s}^I}{\delta \tau(x)} \frac{\delta \mathbf{s}^I}{\delta \tau(y)}\right\rangle=0.
\eea
The preceding condition is indeed restrictive, stemming from our requirement for the metric outside $\mathcal{M}_{\text{quant}}$ to be perfectly classical. Instead, let us necessitate the reduced Wigner functional of the metric outside $\mathcal{M}_{\text{quant}}$ to embody a quasiclassical Wigner functional (analogous to \eqref{quasiclassicalwignerfunction} and also explained at the end of subsection \ref{semiclassicaleinsteinequationfromWDW}). This implies, we only demand that the variance of the Wigner functional does not diverge in the $G_N \to 0$ limit. The reason for this condition is that, if divergence occurs, the metric will not exhibit classical behavior since a strong correlation between the metric and its canonically conjugate momentum will be absent: the situation would then be far away from satisfying the Halliwell criterion. 

When $\mathcal{F}(x,y)$ deviates from $0$, it affects the classicality of the metric at points $x$ and $y$. In simpler terms, the variance of the Wigner functional at $x$ (which signifies the strength of correlations between the metric and its canonically conjugate momentum at $x$) receives contributions from $\mathcal{F}(x,y)$ for each $y$. Thus, to describe the metric classically at $x$, we need this variance to be of order $O(G_N^0)$, after taking contributions from all such points $y$. This implies the following condition:
\bea
\mathcal{F}(x):= \int_{\Sigma_{\text{out}}}d^dy \sqrt{g} \mathcal{F}(x,y) \sim O(G_N^0). \label{quasiclassical2ptvc}
\eea
In the case of spherical symmetry, we integrate the above equation over $\partial\Sigma_{\text{quant}}$ to yield:

\bea \label{sphericalvalidity}
\boxed{\left\langle \mathbf{H}^{\partial\Sigma_{\text{quant}}}_{\text{ADM}} \mathbf{H}^{\Sigma_{\text{out}}}_{\text{matter}}\right\rangle - \left\langle \mathbf{H}^{\partial\Sigma_{\text{quant}}}_{\text{ADM}}\right\rangle \left\langle\mathbf{H}^{\Sigma_{\text{out}}}_{\text{matter}}\right\rangle - \left\langle \frac{\partial \mathbf{s}^I}{\partial t_{\text{bdy}}} \frac{\partial \mathbf{s}^I}{\partial t_{\text{out}}}\right\rangle \sim A \ O(G_
N^0) },
\eea
where, 
\bea
\mathbf{H}^{\partial\Sigma_{\text{quant}}}_{\text{ADM}} &=&\int_{\partial\Sigma_{\text{quant}}} d^{d-1}x \  \mathcal{H}_{\text{ADM}}(x),\\
\mathbf{H}^{\Sigma_{\text{out}}}_{\text{matter}}&=& \int_{\Sigma_{\text{out}}} d^dy \ \mathcal{H}_{\text{matter}}(y) ,\\
\frac{\partial }{\partial t_{\text{bdy}}}&=& \int_{\partial\Sigma_{\text{quant}}} d^{d-1}x \frac{\delta}{\delta \tau(x)} ,\\
\frac{\partial }{\partial t_{\text{out}}}&=&  \int_{\Sigma_{\text{out}}} d^dy \frac{\delta}{\delta \tau(y)} ,\\
A &=& \int_{\partial\Sigma_{\text{quant}}} d^{d-1}x \sqrt{\gamma},
\eea
where $\gamma$ represents the determinant of the induced metric on $\partial\Sigma_{\text{quant}}$ derived from $g_{\text{out}}$. Hereafter, we will refer to equation \eqref{sphericalvalidity} as the Spherical Validity Condition (SVC). For spherically symmetric scenarios, this condition must be satisfied to appropriately describe gravity in a classical manner in the vicinity of the WDW screen.

\subsection{Beyond $n=2$}
The infinite tower of validity conditions \eqref{towervcquantumball} are complicated and it would be interesting to put them in a more understandable form. Doing this in generality is left as an open problem. Here let us simplify them under certain assumptions. Consider the following:
\bea 
\left\langle \prod_{i=1}^n \mathcal{H}(x_i)\right\rangle &=& \int D\mathcal{C} \ \psi^* \ \prod_{i=1}^n \mathcal{H}(x_i) \ \psi,\\
&=& (-1)^n \int D\mathcal{C} \ \psi^* \psi \   \frac{(-i)^n}{\psi[\phi,g]} \frac{\delta^n \psi[\phi,g]}{\delta\tau(x_1) \dots \delta\tau(x_n)}, \\
\\
&=& (-1)^n \left\langle   \frac{(-i)^n}{\psi[\phi,g]} \frac{\delta^n \psi[\phi,g]}{\delta\tau(x_1) \dots \delta\tau(x_n)} \right\rangle,
\eea
where $\mathcal{H}(x_i)$ is $\mathcal{H}_{\text{ADM}}(x_i)$ or $\mathcal{H}_{\text{matter}}(x_i)$ depending on where $x_i$ is located and $D\mathcal{C}$, $\psi$ were defined in equations \eqref{subregionmetricmeasure} and \eqref{emergentminiWDWstatesubregion} respectively. Let us define:
\bea
\begin{tikzpicture}[baseline=(o.base)]
            \setlength{\feynhandblobsize}{3.5mm}
            \begin{feynhand}
                \vertex [ringblob] (o) at (0,0) {}; 
                \vertex [dot] (i1) at (0,0.5) {};
                \vertex [dot] (i2) at (0.354,-0.354) {};
                \propag [plain] (o) to [] (i1);
                \propag [plain] (o) to [] (i2);
                \vertex at (0.49,-0.098) {.};
                \vertex at (0.278,0.416) {.};
                \vertex at (0.462,0.191) {.};
            \end{feynhand}
        \end{tikzpicture} &\;&:= \quad \frac{(-i)^n}{\psi[\phi,g]} \frac{\delta^n \psi[\phi,g]}{\delta\tau(x_1) \dots \delta\tau(x_n)}, \\
\begin{tikzpicture}[baseline=(o.base)]
            \setlength{\feynhandblobsize}{3.5mm}
            \begin{feynhand}
                \vertex [] (o) at (0,0); 
                \vertex [dot] (i1) at (0,0.5) {};
                \vertex [dot] (i2) at (0.354,-0.354) {};
                \propag [plain] (o) to [] (i1);
                \propag [plain] (o) to [] (i2);
                \vertex at (0.49,-0.098) {.};
                \vertex at (0.278,0.416) {.};
                \vertex at (0.462,0.191) {.};
            \end{feynhand}
        \end{tikzpicture}&\;& := \quad (-i)^{(n-1)} \frac{\delta^n \mathbf{s}[\phi,g]}{\delta\tau(x_1) \dots \delta\tau(x_n)},        
\eea
where $\mathbf{s}[\phi,g]=-i \ln{\psi[\phi,g]}$. So we can write:
\bea
\left\langle \prod_{i=1}^n \mathcal{H}(x_i)\right\rangle &=& (-1)^n \left\langle \begin{tikzpicture}[baseline=(o.base)]
\setlength{\feynhandblobsize}{3.5mm}
\begin{feynhand}
                \vertex [ringblob] (o) at (0,0) {}; 
                \vertex [dot] (i1) at (0,0.5) {};
                \vertex [dot] (i2) at (0.354,-0.354) {};
                \propag [plain] (o) to [] (i1);
                \propag [plain] (o) to [] (i2);
                \vertex at (0.49,-0.098) {.};
                \vertex at (0.278,0.416) {.};
                \vertex at (0.462,0.191) {.};
            \end{feynhand}
\end{tikzpicture} \right\rangle.
\eea

One can show the following:
\begin{align}
\left\langle(-i)^{(n-1)} \frac{\delta^n S^R[\phi,g]}{\delta\tau(x_1) \dots \delta\tau(x_n)} \right\rangle=:\left\langle \begin{tikzpicture}[baseline=(o.base)]
            \setlength{\feynhandblobsize}{3.5mm}
            \begin{feynhand}
                \vertex [] (o) at (0,0); \vertex at (-0.15,-0.15) {\smaller\smaller{$R$}};
                \vertex [dot] (i1) at (0,0.5) {};
                \vertex at (0,0.7) {\smaller\smaller{$n$}};
                \vertex [dot] (i2) at (0.354,-0.354) {};
                \vertex at (0.55,-0.4) {\smaller\smaller{$1$}};
                \propag [plain] (o) to [] (i1);
                \propag [plain] (o) to [] (i2);
                \vertex at (0.49,-0.098) {.};
                \vertex at (0.278,0.416) {.};
                \vertex at (0.462,0.191) {.};
            \end{feynhand}
        \end{tikzpicture} \right\rangle  = \left(\prod_{a=1}^{n-1} (-i)\frac{\delta}{\delta \tau(x_a)}\left\langle \mathcal{H}(x_n) \right\rangle  \right) + \left\{ \partial s^I  \right\},
\end{align}
where $\left\{ \partial s^I  \right\}$ is a set of terms involving atleast one time derivative of $s^I$ along with any other term. By taking $N$ derivatives of $n^{\text{th}}$ condition in \eqref{towervcquantumball}, one can show:
\bea
\mathlarger{\mathlarger{\mathlarger{\sum}}} \left\langle \begin{tikzpicture}[baseline=(o.base)]
            \setlength{\feynhandblobsize}{3.5mm}
            \begin{feynhand}
                \vertex [] (o) at (0,0); \vertex at (-0.15,-0.15) {\smaller\smaller{$R$}};
                \vertex [dot] (i1) at (0,0.5) {};
                \vertex at (0,0.7) {\smaller\smaller{$N_1$}};
                \vertex [dot] (i2) at (0.354,-0.354) {};
                \vertex at (0.55,-0.4) {\smaller\smaller{$1$}};
                \propag [plain] (o) to [] (i1);
                \propag [plain] (o) to [] (i2);
                \vertex at (0.49,-0.098) {.};
                \vertex at (0.278,0.416) {.};
                \vertex at (0.462,0.191) {.};
            \end{feynhand}
        \end{tikzpicture} \quad \dots \quad \begin{tikzpicture}[baseline=(o.base)]
            \setlength{\feynhandblobsize}{3.5mm}
            \begin{feynhand}
                \vertex [] (o) at (0,0); \vertex at (-0.15,-0.15) {\smaller\smaller{$R$}};
                \vertex [dot] (i1) at (0,0.5) {};
                \vertex at (0,0.7) {\smaller\smaller{$N_n$}};
                \vertex [dot] (i2) at (0.354,-0.354) {};
                \vertex at (0.55,-0.4) {\smaller\smaller{$1$}};
                \propag [plain] (o) to [] (i1);
                \propag [plain] (o) to [] (i2);
                \vertex at (0.49,-0.098) {.};
                \vertex at (0.278,0.416) {.};
                \vertex at (0.462,0.191) {.};
            \end{feynhand}
        \end{tikzpicture} \right\rangle 
        =
        \mathlarger{\mathlarger{\sum}} \left\langle \begin{tikzpicture}[baseline=(o.base)]
            \setlength{\feynhandblobsize}{3.5mm}
            \begin{feynhand}
                \vertex [] (o) at (0,0); \vertex at (-0.15,-0.15) {\smaller\smaller{$R$}};
                \vertex [dot] (i1) at (0,0.5) {};
                \vertex at (0,0.7) {\smaller\smaller{$N_1$}};
                \vertex [dot] (i2) at (0.354,-0.354) {};
                \vertex at (0.55,-0.4) {\smaller\smaller{$1$}};
                \propag [plain] (o) to [] (i1);
                \propag [plain] (o) to [] (i2);
                \vertex at (0.49,-0.098) {.};
                \vertex at (0.278,0.416) {.};
                \vertex at (0.462,0.191) {.};
            \end{feynhand}
        \end{tikzpicture} \right\rangle \quad \dots \quad \left\langle \begin{tikzpicture}[baseline=(o.base)]
            \setlength{\feynhandblobsize}{3.5mm}
            \begin{feynhand}
                \vertex [] (o) at (0,0); \vertex at (-0.15,-0.15) {\smaller\smaller{$R$}};
                \vertex [dot] (i1) at (0,0.5) {};
                \vertex at (0,0.7) {\smaller\smaller{$N_n$}};
                \vertex [dot] (i2) at (0.354,-0.354) {};
                \vertex at (0.55,-0.4) {\smaller\smaller{$1$}};
                \propag [plain] (o) to [] (i1);
                \propag [plain] (o) to [] (i2);
                \vertex at (0.49,-0.098) {.};
                \vertex at (0.278,0.416) {.};
                \vertex at (0.462,0.191) {.};
            \end{feynhand}
        \end{tikzpicture} \right\rangle + \left\{ \partial s^I  \right\},
\eea
where the summation is over all permutations of space points and also over partitions of $(N+n)$ into $N_1,\dots,N_n$. We also know that:
\bea
\begin{tikzpicture}[baseline=(o.base)]
            \setlength{\feynhandblobsize}{3.5mm}
            \begin{feynhand}
                \vertex [ringblob] (o) at (0,0) {};
                \vertex [dot] (i1) at (0,0.5) {};
                \vertex at (0,0.7) {\smaller\smaller{$n$}};
                \vertex [dot] (i2) at (0.354,-0.354) {};
                \vertex at (0.55,-0.4) {\smaller\smaller{$1$}};
                \propag [plain] (o) to [] (i1);
                \propag [plain] (o) to [] (i2);
                \vertex at (0.49,-0.098) {.};
                \vertex at (0.278,0.416) {.};
                \vertex at (0.462,0.191) {.};
            \end{feynhand}
        \end{tikzpicture} = \mathlarger{\sum} \begin{tikzpicture}[baseline=(o.base)]
            \setlength{\feynhandblobsize}{3.5mm}
            \begin{feynhand}
                \vertex [] (o) at (0,0);
                \vertex [dot] (i1) at (0,0.5) {};
                \vertex at (0,0.7) {\smaller\smaller{$N_1$}};
                \vertex [dot] (i2) at (0.354,-0.354) {};
                \vertex at (0.55,-0.4) {\smaller\smaller{$1$}};
                \propag [plain] (o) to [] (i1);
                \propag [plain] (o) to [] (i2);
                \vertex at (0.49,-0.098) {.};
                \vertex at (0.278,0.416) {.};
                \vertex at (0.462,0.191) {.};
            \end{feynhand}
        \end{tikzpicture} \dots \begin{tikzpicture}[baseline=(o.base)]
            \setlength{\feynhandblobsize}{3.5mm}
            \begin{feynhand}
                \vertex [] (o) at (0,0);
                \vertex [dot] (i1) at (0,0.5) {};
                \vertex at (0,0.7) {\smaller\smaller{$N_m$}};
                \vertex [dot] (i2) at (0.354,-0.354) {};
                \vertex at (0.55,-0.4) {\smaller\smaller{$1$}};
                \propag [plain] (o) to [] (i1);
                \propag [plain] (o) to [] (i2);
                \vertex at (0.49,-0.098) {.};
                \vertex at (0.278,0.416) {.};
                \vertex at (0.462,0.191) {.};
            \end{feynhand}
        \end{tikzpicture},
\eea
where the summation performs a partition of $n$ space points into $m$ pieces (along with along summing over $m$ too) and permutes all of the space points too. 

So we get:
\begin{align}
\left\langle \prod_{i=1}^n \mathcal{H}(x_i)\right\rangle &= (-1)^n \left\langle \begin{tikzpicture}[baseline=(o.base)]
\setlength{\feynhandblobsize}{3.5mm}
\begin{feynhand}
                \vertex [ringblob] (o) at (0,0) {}; 
                \vertex [dot] (i1) at (0,0.5) {};
                \vertex [dot] (i2) at (0.354,-0.354) {};
                \propag [plain] (o) to [] (i1);
                \propag [plain] (o) to [] (i2);
                \vertex at (0.49,-0.098) {.};
                \vertex at (0.278,0.416) {.};
                \vertex at (0.462,0.191) {.};
            \end{feynhand}
\end{tikzpicture} \right\rangle,\\
&= (-1)^n \left\langle \mathlarger{\sum} \begin{tikzpicture}[baseline=(o.base)]
            \setlength{\feynhandblobsize}{3.5mm}
            \begin{feynhand}
                \vertex [] (o) at (0,0);
                \vertex [dot] (i1) at (0,0.5) {};
                \vertex at (0,0.7) {\smaller\smaller{$N_1$}};
                \vertex [dot] (i2) at (0.354,-0.354) {};
                \vertex at (0.55,-0.4) {\smaller\smaller{$1$}};
                \propag [plain] (o) to [] (i1);
                \propag [plain] (o) to [] (i2);
                \vertex at (0.49,-0.098) {.};
                \vertex at (0.278,0.416) {.};
                \vertex at (0.462,0.191) {.};
            \end{feynhand}
        \end{tikzpicture} \dots \begin{tikzpicture}[baseline=(o.base)]
            \setlength{\feynhandblobsize}{3.5mm}
            \begin{feynhand}
                \vertex [] (o) at (0,0);
                \vertex [dot] (i1) at (0,0.5) {};
                \vertex at (0,0.7) {\smaller\smaller{$N_m$}};
                \vertex [dot] (i2) at (0.354,-0.354) {};
                \vertex at (0.55,-0.4) {\smaller\smaller{$1$}};
                \propag [plain] (o) to [] (i1);
                \propag [plain] (o) to [] (i2);
                \vertex at (0.49,-0.098) {.};
                \vertex at (0.278,0.416) {.};
                \vertex at (0.462,0.191) {.};
            \end{feynhand}
        \end{tikzpicture} \right\rangle,\\
        &= (-1)^n \mathlarger{\sum}\left\langle  \begin{tikzpicture}[baseline=(o.base)]
            \setlength{\feynhandblobsize}{3.5mm}
            \begin{feynhand}
                \vertex [] (o) at (0,0);
                \vertex [dot] (i1) at (0,0.5) {};
                \vertex at (0,0.7) {\smaller\smaller{$N_1$}};
                \vertex [dot] (i2) at (0.354,-0.354) {};
                \vertex at (0.55,-0.4) {\smaller\smaller{$1$}};
                \propag [plain] (o) to [] (i1);
                \propag [plain] (o) to [] (i2);
                \vertex at (0.49,-0.098) {.};
                \vertex at (0.278,0.416) {.};
                \vertex at (0.462,0.191) {.};
            \end{feynhand}
        \end{tikzpicture} \dots \begin{tikzpicture}[baseline=(o.base)]
            \setlength{\feynhandblobsize}{3.5mm}
            \begin{feynhand}
                \vertex [] (o) at (0,0);
                \vertex [dot] (i1) at (0,0.5) {};
                \vertex at (0,0.7) {\smaller\smaller{$N_m$}};
                \vertex [dot] (i2) at (0.354,-0.354) {};
                \vertex at (0.55,-0.4) {\smaller\smaller{$1$}};
                \propag [plain] (o) to [] (i1);
                \propag [plain] (o) to [] (i2);
                \vertex at (0.49,-0.098) {.};
                \vertex at (0.278,0.416) {.};
                \vertex at (0.462,0.191) {.};
            \end{feynhand}
        \end{tikzpicture} \right\rangle,\\
        &= (-1)^n \mathlarger{\sum}  \left\langle  \begin{tikzpicture}[baseline=(o.base)]
            \setlength{\feynhandblobsize}{3.5mm}
            \begin{feynhand}
                \vertex [] (o) at (0,0);
                \vertex at (-0.15,-0.15) {\smaller\smaller{$R$}};
                \vertex [dot] (i1) at (0,0.5) {};
                \vertex at (0,0.7) {\smaller\smaller{$N_1$}};
                \vertex [dot] (i2) at (0.354,-0.354) {};
                \vertex at (0.55,-0.4) {\smaller\smaller{$1$}};
                \propag [plain] (o) to [] (i1);
                \propag [plain] (o) to [] (i2);
                \vertex at (0.49,-0.098) {.};
                \vertex at (0.278,0.416) {.};
                \vertex at (0.462,0.191) {.};
            \end{feynhand}
        \end{tikzpicture} \dots \begin{tikzpicture}[baseline=(o.base)]
            \setlength{\feynhandblobsize}{3.5mm}
            \begin{feynhand}
                \vertex [] (o) at (0,0);
                \vertex at (-0.15,-0.15) {\smaller\smaller{$R$}};
                \vertex [dot] (i1) at (0,0.5) {};
                \vertex at (0,0.7) {\smaller\smaller{$N_m$}};
                \vertex [dot] (i2) at (0.354,-0.354) {};
                \vertex at (0.55,-0.4) {\smaller\smaller{$1$}};
                \propag [plain] (o) to [] (i1);
                \propag [plain] (o) to [] (i2);
                \vertex at (0.49,-0.098) {.};
                \vertex at (0.278,0.416) {.};
                \vertex at (0.462,0.191) {.};
            \end{feynhand}
        \end{tikzpicture} \right\rangle +  \left\{ \partial s^I  \right\},\\
        &= (-1)^n \mathlarger{\sum}  \left\langle  \begin{tikzpicture}[baseline=(o.base)]
            \setlength{\feynhandblobsize}{3.5mm}
            \begin{feynhand}
                \vertex [] (o) at (0,0);
                \vertex at (-0.15,-0.15) {\smaller\smaller{$R$}};
                \vertex [dot] (i1) at (0,0.5) {};
                \vertex at (0,0.7) {\smaller\smaller{$N_1$}};
                \vertex [dot] (i2) at (0.354,-0.354) {};
                \vertex at (0.55,-0.4) {\smaller\smaller{$1$}};
                \propag [plain] (o) to [] (i1);
                \propag [plain] (o) to [] (i2);
                \vertex at (0.49,-0.098) {.};
                \vertex at (0.278,0.416) {.};
                \vertex at (0.462,0.191) {.};
            \end{feynhand}
        \end{tikzpicture} \right\rangle \dots \left\langle \begin{tikzpicture}[baseline=(o.base)]
            \setlength{\feynhandblobsize}{3.5mm}
            \begin{feynhand}
                \vertex [] (o) at (0,0);
                \vertex at (-0.15,-0.15) {\smaller\smaller{$R$}};
                \vertex [dot] (i1) at (0,0.5) {};
                \vertex at (0,0.7) {\smaller\smaller{$N_m$}};
                \vertex [dot] (i2) at (0.354,-0.354) {};
                \vertex at (0.55,-0.4) {\smaller\smaller{$1$}};
                \propag [plain] (o) to [] (i1);
                \propag [plain] (o) to [] (i2);
                \vertex at (0.49,-0.098) {.};
                \vertex at (0.278,0.416) {.};
                \vertex at (0.462,0.191) {.};
            \end{feynhand}
        \end{tikzpicture} \right\rangle +  \left\{ \partial s^I  \right\},\\
      \left\langle \prod_{i=1}^n \mathcal{H}(x_i)\right\rangle - \prod_{i=1}^n \left\langle \mathcal{H}(x_i)\right\rangle  &=    \left\{ \partial \langle\mathcal{H}\rangle  \right\} +  \left\{ \partial s^I  \right\} :=\left\{ \partial \langle\mathcal{H}\rangle, \partial s^I  \right\} , \label{npthamvc}
\end{align}
where $\left\{ \partial \langle\mathcal{H}\rangle  \right\}$ is the set of terms involving at least one time derivative of $\langle \mathcal{H}(x)\rangle$ along with any other term except (except derivatives of $s^I$ because they are already included in $\left\{ \partial s^I  \right\}$). Similarly, $\left\{ \partial \langle\mathcal{H}\rangle, \partial s^I  \right\}$ is defined to be the set of terms that include at least one time derivative of $\langle \mathcal{H}(x)\rangle$ or at least one time derivative of $s^I $.

Equation \eqref{npthamvc} is a necessary condition to have a perfectly classical metric, and to have a quasiclassical spacetime, we have the relaxed integrated version of the above condition for a spherically symmetric system (similar to equation \eqref{sphericalvalidity}):
\begin{equation} \label{nSVC}
\boxed{\left\langle \left(\mathbf{H}^{\partial\Sigma_{\text{quant}}}_{\text{ADM}}\right)^n \left( \mathbf{H}^{\Sigma_{\text{out}}}_{\text{matter}}\right)^m\right\rangle - \left\langle \mathbf{H}^{\partial\Sigma_{\text{quant}}}_{\text{ADM}}\right\rangle ^n  \left\langle\mathbf{H}^{\Sigma_{\text{out}}}_{\text{matter}}\right\rangle^m - \int\left\{ \partial \langle \mathcal{H} \rangle, \partial s^I \right\} \sim A^{p_{nm}} \  O(G_
N^0)},
\end{equation}
where $\int\left\{ \partial \langle \mathcal{H} \rangle, \partial s^I \right\}$ represents the set of integrated terms in $\left\{ \partial \langle \mathcal{H} \rangle, \partial s^I \right\}$, and $p_{nm}$ is a certain power of $A$, the explicit form of which we do not consider necessary to ascertain.

\subsection{Comments on finite region holography in AdS and dS}

\subsection*{AdS}

\begin{figure}[h]
\centering
\begin{tikzpicture}[scale=1.5]
\draw (0,0) ellipse (1.25 and 0.3);
\draw (-1.25,0) -- (-1.25,-3.5);
\draw (1.25,-3.5) -- (1.25,0) node [midway] (N) {}; 
\draw (0,-3.5) ellipse (1.25 and 0.3);
\draw (0,0) ellipse (1.25*1.5 and 0.3*1.5);
\draw (-1.25*1.5,0) -- (-1.25*1.5,-3.5);
\draw (1.25*1.5,-3.5) -- (1.25*1.5,0) node [midway] (M) {}; 
\draw (0,-3.5) ellipse (1.25*1.5 and 0.3*1.5);
\fill [gray,opacity=0.2] (-1.25,0) -- (-1.25,-3.5) arc (180:360:1.25 and 0.3) -- (1.25,0) arc (0:180:1.25 and 0.3);
\draw (M) to  (N)[->,>=stealth'];
\node at (M) [label={[xshift=-4.5mm, yshift=-1mm]$\mu$}] {};
\node at (M) [label={[xshift=4mm, yshift=7mm]$\partial\mathcal{B}$}] {};
\node at (N) [label={[xshift=4.5
mm, yshift=7mm]$\partial\mathcal{B}^{(\mu)}$}] {};
\end{tikzpicture}
\caption{\small Illustration of the usual finite-cutoff AdS/CFT duality. The $T^2$ deformed theory lives on a finite radius brick wall, labeled by $\partial\mathcal{B}^{(\mu)}$. Here, $\mu$ serves as the deformation parameter, quantifying the extent of the deformation.}
\label{usual}
\end{figure}
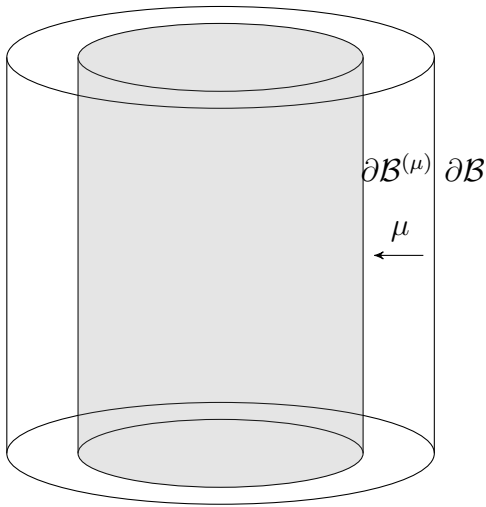

In the framework of finite cutoff holography \cite{Hartman:2018tkw}, the initial starting point is the dual CFT situated on the asymptotic AdS boundary. One then deforms this CFT by a radial $T^2$ deformation, resulting in a theory thought to inhabit a cylinder with a finite bulk radius. The states in this deformed theory are understood to holographically encode the bulk spacetime encapsulated within that cylinder, as depicted in figure \ref{usual}.

In essence, one should be able to derive this finite cutoff holographic framework from the CFT situated on the asymptotic boundary, as it encompasses the quantum gravity information pertaining to the entire bulk. To accomplish this, one would begin with a CFT state on the asymptotic boundary and evaluate its dual WDW state, as specified by the CSH dictionary in equation \ref{dualWDWstate}. It would be required for this state to be semiclassical beyond the boundary of the cylinder. This is a necessity to ensure the existence of operational observers outside the cylinder, which would then permit an interpretation of the states of the theory on the cylinder's surface, describing the cylinder's interior from the perspective of these external operational observers. This requirement aligns with the concept of subregion classicalization, making the validity conditions outlined in equation \eqref{nSVC} (or their generalized version beyond spherical symmetry) essential for the existence of finite cutoff holography in the bulk.


\subsection*{dS}

In the opening of Section \ref{dSCFTclosed}, we touched upon an alternative approach for dS/CFT, which involves the introduction of a hard wall in de Sitter spacetime and positing the dual field theory description on this hard wall. This hard wall can now be interpreted as the WDW screen.

Our expectation is that the scenario may be more akin to the following: the universal wavefunction could comprise a multitude of branches. Among these, we should confine ourselves to branches that are at least partially WKB in certain areas, as they may permit the presence of operational observers. 

Following this, we should further condition the wavefunction based on the data gathered by these operational observers up to the present. This would result in an induced state of the remainder of the universe from the perspective of the operational observers. If the wavefunction of the geometry on a certain segment of the universe is not in a WKB state, then the resulting induced state would also incorporate an induced WDW state located on the WDW screen.

\section{Black Hole Information Paradox} \label{BHinfoparadox}

In this subsection, we will leverage insights gleaned from the semiclassical approximation of the WDW state to gain a deeper understanding of the black hole information paradox \cite{Hawking:1976ra}. We'll begin by assessing the validity conditions in the context of an evaporating black hole.

\subsection{Breakdown of semiclassical gravity near the Page time for an evaporating black hole}\label{demiseofsemiclassicalgravity}
\begin{figure}[h]
\centering
\tikzset{every picture/.style={line width=1.5pt}} 

\tikzset{every picture/.style={line width=0.75pt}} 

\begin{tikzpicture}[x=0.75pt,y=0.75pt,yscale=-1,xscale=1]

\draw    (490,20) -- (490,460) ;
\filldraw   (245,150) .. controls (245,147.24) and (247.24,145) .. (250,145) .. controls (252.76,145) and (255,147.24) .. (255,150) .. controls (255,152.76) and (252.76,155) .. (250,155) .. controls (247.24,155) and (245,152.76) .. (245,150) -- cycle ;
\draw[dashed]    (250,20) -- (250,150) ;
\draw    (250,150) -- (90,310) ;
\draw[dashed]   (90,310) -- (90,460) ;
\draw[dashed]    (90,150) -- (90,310) ;
\draw[snake=zigzag]    (90,150) -- (250,150) ;
\filldraw   (180,215) .. controls (180,212.24) and (182.24,210) .. (185,210) .. controls (187.76,210) and (190,212.24) .. (190,215) .. controls (190,217.76) and (187.76,220) .. (185,220) .. controls (182.24,220) and (180,217.76) .. (180,215) -- cycle ;
\draw    (100,150) .. controls (131.4,171) and (148.4,306) .. (160,460) ;
\draw    (90,420) .. controls (186.4,348) and (362.4,445) .. (490,410) ;
\draw    (90,290) .. controls (245.4,272) and (365.4,301) .. (490,230) ;
\draw    (90,240) .. controls (192.4,174) and (362.4,265) .. (490,230) ;
\draw    (90,160) .. controls (325.4,160) and (396.4,144) .. (490,230) ;
\draw    (250,50) .. controls (290,20) and (425.4,68) .. (490,40) ;

\draw (281,382) node [anchor=north west][inner sep=0.75pt]   [align=left] {$\displaystyle \Sigma_{\text{past}} $};
\draw (251,262) node [anchor=north west][inner sep=0.75pt]   [align=left] {$\displaystyle \Sigma_{\text{early}} $};
\draw (311,210) node [anchor=north west][inner sep=0.75pt]   [align=left] {$\displaystyle \Sigma_{\mathcal{P}} $};
\draw (351,142) node [anchor=north west][inner sep=0.75pt]   [align=left] {$\displaystyle \Sigma_{\mathcal{E}} $};
\draw (255,132) node [anchor=north west][inner sep=0.75pt]   [align=left] {$\displaystyle \mathcal{E} $};
\draw (175,192) node [anchor=north west][inner sep=0.75pt]   [align=left] {$\displaystyle \mathcal{P} $};
\draw (155,242) node [anchor=north west][inner sep=0.75pt]   [align=left] {$\displaystyle H $};
\draw (361,22) node [anchor=north west][inner sep=0.75pt]   [align=left] {$\displaystyle \Sigma_{\text{future}} $};

\end{tikzpicture}

\caption{\small This figure illustrates a conventional Penrose diagram of evaporating black holes within the framework of semiclassical gravity. The event horizon is denoted by $H$, the Page time point is represented by $\mathcal{P}$, and the final point of evaporation is symbolized by $\mathcal{E}$. This point $\mathcal{E}$ is a naked singularity. Various Cauchy slices, which traverse through different stages of black hole collapse and subsequent evaporation, are also depicted. It should be noted that the depicted Penrose diagram represents a spacetime that is not globally generalized hyperbolic; therefore, these Cauchy slices are not generalized Cauchy surfaces.}
\label{hawkingpenrosediagram}
\end{figure}
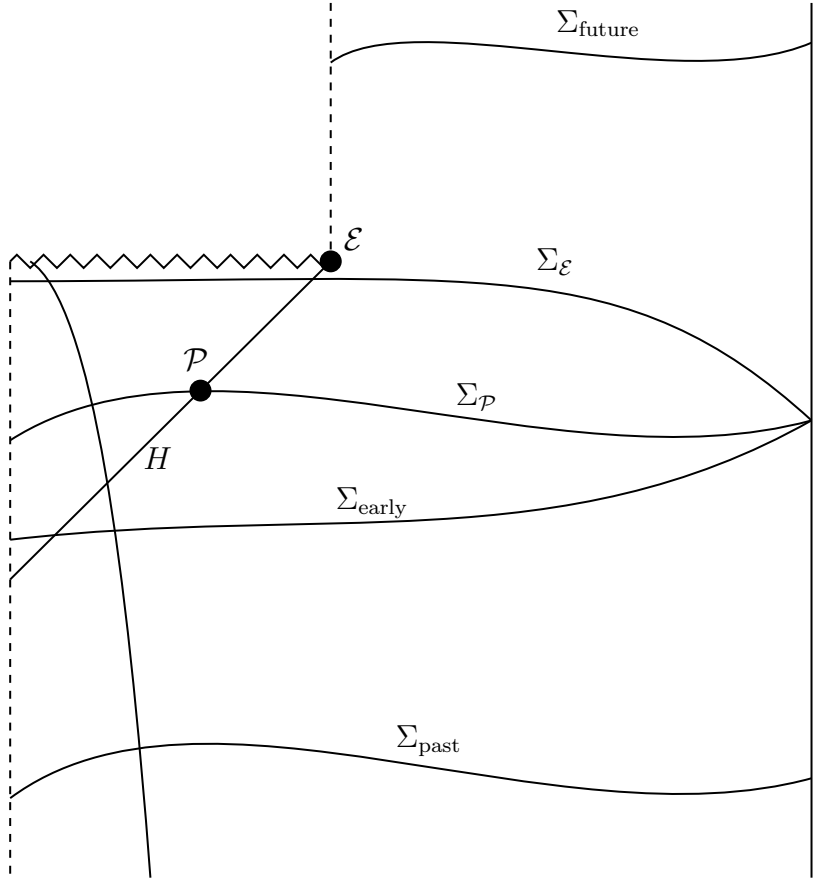

Consider the Penrose diagram for a spherically symmetric, collapsing, and evaporating black hole in asymptotically AdS spacetime (refer to Fig \ref{hawkingpenrosediagram}). Similar arguments are also applicable in the dS scenario. In Hawking's original formulation of the black hole information paradox, it was assumed that semiclassical gravity was universally applicable, with the exception of areas in close proximity to the singularity. 

Let us now examine the implications of the validity conditions for semiclassical gravity in the context of the near-horizon region. In this case, we take the horizon to be the WDW screen, and therefore, the validity conditions \eqref{nSVC} that are necessary for space-time to be classical in this near-horizon region can be expressed as follows:
\bea 
\left\langle \mathbf{H}_{\text{BH}}^n  \mathbf{H}_{\text{HR}}^m\right\rangle - \left\langle \mathbf{H}_{\text{BH}}\right\rangle ^n  \left\langle\mathbf{H}_{\text{HR}}\right\rangle^m - \int\left\{ \partial \langle \mathcal{H} \rangle, \partial s^I \right\} \sim A^{p_{nm}} \ O(G_
N^0).\label{sphericalbhvc}
\eea
Here, $\mathbf{H}_{\text{BH}}$ represents the ADM energy of the black hole, while $\mathbf{H}_{\text{HR}}$ represents the energy of the Hawking radiation. For a black hole resulting from the collapse of a large mass, the evaporation rate is slow, before and around the Page time. Consequently, the terms in $\int\left\{ \partial \langle \mathcal{H} \rangle, \partial s^I \right\}$ are expected to be of order $O(G^0_N)$. \footnote{We assume that the QFT in this background exhibits stable behavior, in that the expectation value of the stress tensor doesn't diverge. If this wasn't the case, the backreaction on the metric could be so potent that it could induce singularities near the horizon at the Page time itself, resulting in the breakdown of semiclassical gravity due to these singularities.}

Therefore, this suggests that the $n+m$ point correlator of energies (the first two terms in equation \eqref{sphericalbhvc}) must also be of order $O(G^0_N)$. Given that correlation functions are linked to the entanglement entropy between the black hole and the Hawking radiation, this indicates that the entanglement entropy between the black hole and Hawking radiation must be of order $O(G_N^0)$.

For Cauchy slices, $\Sigma_{\text{early}}$, that pass through the black hole during its early stages of evaporation, the amount of Hawking radiation present is minimal. As a result, the entanglement entropy (the Von Neumann entropy between the Hawking radiation and the black hole) is relatively small, at least on the order of $O(G_N^0)$. Therefore, during the early stages of black hole evaporation, reliance on semiclassical gravity can be justified.

If semiclassical gravity remained valid up until the Cauchy slice $\Sigma_{\mathcal{P}}$ (the slice passing through the Page time), then the QFT state on $\Sigma_{\mathcal{P}}$ must also meet the validity conditions to further evolve with semiclassical gravity. However, the entanglement entropy of the black hole at the Page time is $\frac{A_P}{4 G_N} \sim  O(G_N^{-1})$, with $A_P$ denoting the area of the black hole at the Page time. This suggests that the validity conditions are violated around the Page time, causing a breakdown in semiclassical gravity at and beyond this point. More accurately, it fails whenever the Cauchy slice intersects the Page time point, $\mathcal{P}$. \emph{Therefore, one should not use semiclassical gravity to evolve the state beyond these particular Cauchy slices!}\footnote{See \cite{Juarez-Aubry:2023kvl} for a discussion on how semiclassical gravity breaks down in a different context.}

\subsection{A calculational framework for predicting the future state from the past state}


The bulk's breakdown of semiclassical gravity around the Page time precludes its use in predicting future states based on past ones. To progress the state, we need to apply the complete quantum theory of gravity, i.e., the holographic CFT. If the objective is to predict the states of the far future, the following steps must be taken:

Begin with a holographic CFT state $\psi_i$ in the distant past, which is dual to a bulk semiclassical WDW state $\Psi_i$ (as given by the CSH dictionary in Equation \ref{dualWDWstate}).  From this state, we can derive the metric $g_{ab}^{(i)}$ and classical momenta $\Pi^{ab}_{(i)}$ on $\Sigma_i$ using the procedure outlined in subsection \ref{bulkclassicalduals}. Additionally, the effective QFT state $\psi^{\text{bulk}}_{i}$ on $\Sigma_i$ can be obtained from the semiclassical WDW state $\Psi_i$ through the procedure described in subsection \ref{bulkQFTduals}. If semiclassical gravity holds, the initial conditions are given by $\left(g_{ab}^{(i)},\Pi^{ab}_{(i)},\psi^{\text{bulk}}_{i} \right)$ and one can evolve it using semiclassical gravity to find the future state. However, if semiclassical gravity may potentially fail at some point during the intermediate stages, then one must proceed to evolve the holographic CFT state $\psi_i$ using the CFT Hamiltonian to obtain the future holographic CFT state $\psi_f$. 

Subsequently, it must be verified whether the bulk WDW state $\Psi_f$, dual to $\psi_f$, is a WKB state. If it is, one can derive the future metric, momenta, and the effective bulk QFT state $\left(g_{ab}^{(f)},\Pi^{ab}_{(f)},\psi^{\text{bulk}}_{f} \right)$ from $\Psi_f$. The future state is dependent on the past state due to the unitary nature of the boundary evolution.

This framework should be applied to predict the outcomes of a black hole. The future WDW state may not be a WKB state, in which case the description can only be provided holographically. Alternatively, the future WDW state might be a superposition of WKB states. In such a situation, bulk observers residing in different branches wouldn't be able to ascertain what the initial state was, although boundary observers (or those near the boundary) could.\footnote{In this scenario, information about the initial state could seep into the matter-gravity entanglement even when there are multiple WKB branches. See \cite{Kay:2022wpn,Kay:2018mxr,Kay:2019ukr} for the matter-gravity entanglement hypothesis.}

\begin{figure}[ht!]
\centering
\tikzset{every picture/.style={line width=0.75pt}} 

\begin{tikzpicture}[x=0.75pt,y=0.75pt,yscale=-1,xscale=1]

\draw    (320,20) -- (320,460) ;
\draw    (90,20) -- (90,460) ;
\draw    (90,440) .. controls (186.4,368) and (192.4,475) .. (320,440) ;
\draw    (90,50) .. controls (130,20) and (255.4,68) .. (320,40) ;
\draw    (90,60) .. controls (130,30) and (255.4,78) .. (320,50) ;
\draw    (90,70) .. controls (130,40) and (255.4,88) .. (320,60) ;
\draw    (90,80) .. controls (130,50) and (255.4,98) .. (320,70) ;
\draw    (90,90) .. controls (130,60) and (255.4,108) .. (320,80) ;
\draw    (90,100) .. controls (130,70) and (255.4,118) .. (320,90) ;
\draw    (90,110) .. controls (130,80) and (255.4,128) .. (320,100) ;
\draw    (90,120) .. controls (130,90) and (255.4,138) .. (320,110) ;
\draw    (90,430) .. controls (186.4,358) and (192.4,465) .. (320,430) ;
\draw    (90,420) .. controls (186.4,348) and (192.4,455) .. (320,420) ;
\draw    (90,410) .. controls (186.4,338) and (192.4,445) .. (320,410) ;
\draw    (90,400) .. controls (186.4,328) and (192.4,435) .. (320,400) ;
\draw    (90,385) .. controls (186.4,313) and (192.4,420) .. (320,385) ;
\draw   (315,440) .. controls (315,437.24) and (317.24,435) .. (320,435) .. controls (322.76,435) and (325,437.24) .. (325,440) .. controls (325,442.76) and (322.76,445) .. (320,445) .. controls (317.24,445) and (315,442.76) .. (315,440) -- cycle ;
\draw   (315,40) .. controls (315,37.24) and (317.24,35) .. (320,35) .. controls (322.76,35) and (325,37.24) .. (325,40) .. controls (325,42.76) and (322.76,45) .. (320,45) .. controls (317.24,45) and (315,42.76) .. (315,40) -- cycle ;
\draw   (330,80) -- (340,60) -- (350,80) ;
\draw    (340,60) -- (340,430) ;
\fill [gray,opacity=0.5]    (90,180) .. controls (290,179.4) and (290,299.4) .. (90,300) ;

\draw (151,420) node [anchor=north west][inner sep=0.75pt]   [align=left] {$\displaystyle \Sigma_i $};
\draw (255,22) node [anchor=north west][inner sep=0.75pt]   [align=left] {$\displaystyle \Sigma_f $};
\draw (331,430) node [anchor=north west][inner sep=0.75pt]   [align=left] {$\displaystyle \psi _{i}$};
\draw (331,32) node [anchor=north west][inner sep=0.75pt]   [align=left] {$\displaystyle \psi _{f}$};
\draw (91,212) node [anchor=north west][inner sep=0.75pt]   [align=left] {Semiclassical \\ gravity invalid};

\end{tikzpicture}

\caption{\small To investigate scenarios in which spacetime exhibits strong quantum mechanical characteristics during intermediate times (indicating the invalidity of semiclassical gravity), one must rely on boundary evolution, which covertly encodes the non-perturbative information of quantum gravity in the bulk. In order to predict future semiclassical variables (such as the metric, extrinsic curvature, and QFT state) from their past equivalents, one must evolve the boundary CFT state beyond the point where all quantum gravitational fluctuations wash out. Subsequently, the CSH dictionary can be used to derive the bulk WDW state. This bulk state can then be used to determine semiclassical data through an eikonal approximation.}
\label{evolutionframework}
\end{figure}
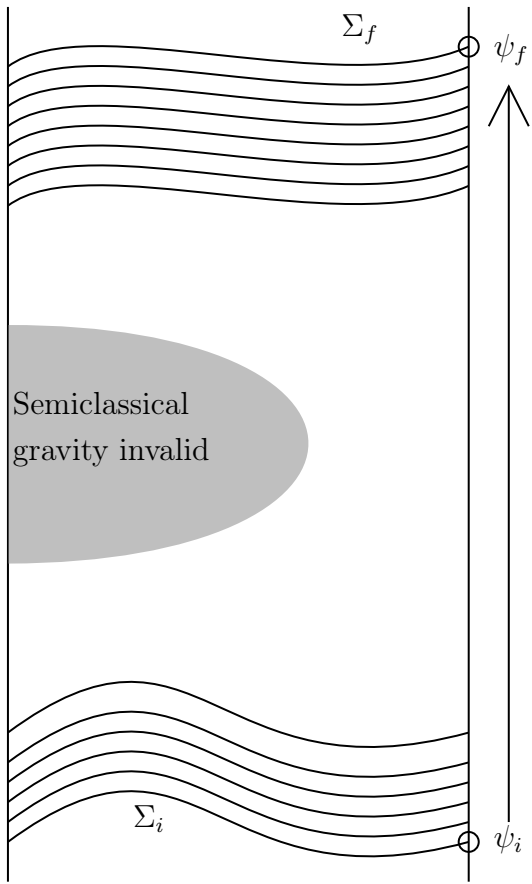

\subsection{Perspective of observers outside an evaporating black hole}

To analyze the evaporation of a black hole during intermediate times, it's necessary to evolve the holographic CFT state to an intermediate point, and then map it to its corresponding dual WDW state to study the bulk with these intermediate WDW states (refer to Fig \ref{evolutionframework}). This process should, in theory, provide precise information regarding the duration of semiclassical gravity's validity and pinpoint its exact breakdown point. It should also shed light on the processes occurring within the black hole during its evaporation. We will endeavor to gain a qualitative understanding of these phenomena in this subsection.

If we operate under the assumption that semiclassical gravity holds until $\Sigma_{\mathcal{P}}$, we find that it gives rise to its own demise beyond this point. Hence, we must conclude that spacetime does not maintain a classical nature in the near-horizon region beyond $\Sigma_{\mathcal{P}}$. This is because the validity conditions \eqref{sphericalbhvc} were essential for the metric to exhibit classical properties near the horizon of the black hole.

\begin{figure}[h]
\centering
\tikzset{every picture/.style={line width=1.5pt}} 

\tikzset{every picture/.style={line width=0.75pt}} 

\begin{tikzpicture}[x=0.75pt,y=0.75pt,yscale=-1,xscale=1]

\draw    (490,20) -- (490,460) ;
\draw[dashed]   (90,310) -- (90,460) ;
\draw[dashed]    (90,150) -- (90,310) ;
\draw    (90,150) -- (174,234) ;
\draw    (90,310) -- (170,230) ;
\draw[dashed]    (250,20) -- (250,120) ;

\draw    (90,420) .. controls (186.4,348) and (362.4,445) .. (490,410) ;
\draw    (250,50) .. controls (290,20) and (425.4,68) .. (490,40) ;
\draw[line width=0.5mm]    (90,310) .. controls (232,170.25) and (284.83,173.17) .. (250,120) ;
\draw    (174,234) .. controls (291.83,350.67) and (320.83,190.67) .. (490,230);
\draw    (260,150) .. controls (313.83,123.17) and (436.83,181.17) .. (490,150) ;
\draw    (100,160) .. controls (134.83,210.83) and (145.83,369.83) .. (150,460) ;
\draw    (90,150) .. controls (128.83,116.33) and (182.83,182.33) .. (250,120) ;
\draw[line width=0.5mm]   (250,120) .. controls (281.83,164.33) and (229.83,181.33) .. (174,234) ;
\draw    (110,290) .. controls (230,379.6) and (446,337.6) .. (490,300) ;

\fill [gray,opacity=0.2] (90,150) .. controls (128.83,116.33) and (182.83,182.33) .. (250,120) .. controls (281.83,164.33) and (229.83,181.33) .. (174,234) ;
\filldraw   (169,234) .. controls (169,231.24) and (171.24,229) .. (174,229) .. controls (176.76,229) and (179,231.24) .. (179,234) .. controls (179,236.76) and (176.76,239) .. (174,239) .. controls (171.24,239) and (169,236.76) .. (169,234) -- cycle ;
\filldraw   (105,290) .. controls (105,287.24) and (107.24,285) .. (110,285) .. controls (112.76,285) and (115,287.24) .. (115,290) .. controls (115,292.76) and (112.76,295) .. (110,295) .. controls (107.24,295) and (105,292.76) .. (105,290) -- cycle ;
\filldraw   (245,120) .. controls (245,117.24) and (247.24,115) .. (250,115) .. controls (252.76,115) and (255,117.24) .. (255,120) .. controls (255,122.76) and (252.76,125) .. (250,125) .. controls (247.24,125) and (245,122.76) .. (245,120) -- cycle ;

\draw (101,300) node [anchor=north west][inner sep=0.75pt]   [align=left] {$\displaystyle \partial \mathcal{Q}_{t_0} $};
\draw (281,382) node [anchor=north west][inner sep=0.75pt]   [align=left] {$\displaystyle \Sigma_{\text{past}} $};
\draw (361,22) node [anchor=north west][inner sep=0.75pt]   [align=left] {$\displaystyle \Sigma_{\text{future}} $};
\draw (236,181) node [anchor=north west][inner sep=0.75pt]   [align=left] {$\partial \mathcal{Q}$};
\draw (240,260) node [anchor=north west][inner sep=0.75pt]   [align=left] {$\displaystyle \Sigma_{\text{last}} $};
\draw (165,242) node [anchor=north west][inner sep=0.75pt]   [align=left] {$\displaystyle \mathcal{C} $};
\draw (275,325) node [anchor=north west][inner sep=0.75pt]   [align=left] {$\displaystyle \Sigma_{\text{out}} $};
\draw (231,102) node [anchor=north west][inner sep=0.75pt]   [align=left] {$\displaystyle \mathcal{E} $};

\end{tikzpicture}

\caption{\small Quantum Penrose Diagram: The shaded region represents the part of spacetime where semiclassical gravity is no longer valid. }
\label{quantumhawkingpenrosediagram}
\end{figure}
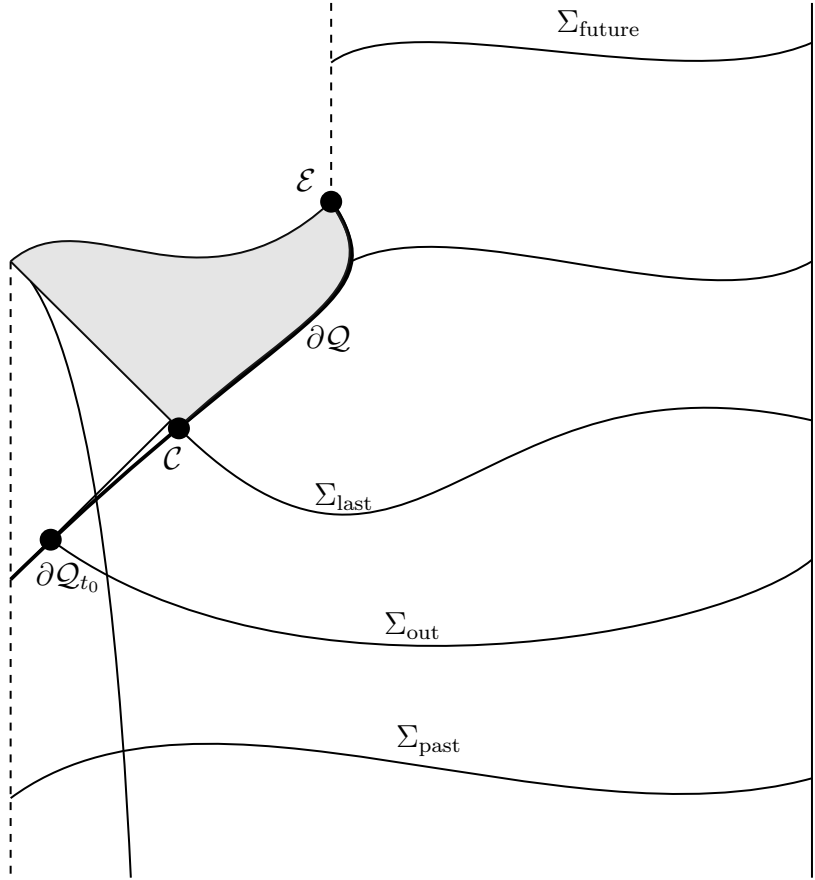

Penrose diagrams are suitable for usage when the metric is classical everywhere, but they're not applicable when a significant portion of spacetime is highly quantum mechanical. However, given that the metric remains classical far from the black hole, we can use a portion of the Penrose diagram for the classical region and describe the quantum region holographically, as outlined in section \ref{subregionclassicalisationscreens}. An apt diagram to portray this scenario is presented in Fig \ref{quantumhawkingpenrosediagram}.

The optimal approach to handle a region of spacetime that is quantum mechanical (henceforth denoted as $\mathcal{Q}$) involves using the WDW screen (now referred to as $\partial\mathcal{Q}$). We should position it outside the event horizon (as close as conditions allow, which depends on the classical nature of spacetime near the event horizon), as depicted in Fig \ref{quantumhawkingpenrosediagram}. 

The induced WDW states reside on a co-dimension 2 surface, $\partial\mathcal{Q}_{t_0}$, which is a slice of the WDW screen, $\partial\mathcal{Q}$. These states are entangled with the QFT state existing on a partial Cauchy slice, $\Sigma_{\text{out}}$. This slice connects the WDW screen to the asymptotic boundary. 

The evolution process must employ semiclassical gravity in regions where gravity maintains its classical nature (i.e., outside the WDW screen). Meanwhile, within $\partial\mathcal{Q}$, we should use the boundary Schrödinger equation for the evolution.

As we progress along the WDW screen, we note that the induced WDW state on the screen, although initially a WKB state, inevitably transitions into a non-WKB state close to the Page time. This transition is attributable to the breakdown of validity conditions, as elucidated in section \ref{demiseofsemiclassicalgravity}. We'll designate this point as $\mathcal{C}$.

Given that the induced WDW states on the WDW screen remain a WKB state prior to point $\mathcal{C}$, the WDW patch corresponding to point $\mathcal{C}$ will also maintain a classical character. Consequently, the description of the black hole interior remains consistent with that provided by semiclassical gravity. It is only beyond this point that the description changes, and no longer aligns with the principles of semiclassical gravity.

For this region, the optimal course of action involves removing the Penrose diagram corresponding to this region and incorporating a shaded area in the diagram to denote its quantum mechanical nature. Subsequently, the evolution, as dictated by the boundary Schrödinger equation on the WDW screen and semiclassical gravity outside the screen, will uphold unitarity, provided the spacetime remains classical outside the screen.

As the size of the co-dimension 1 slices of the WDW screen (which is a co-dimension 2 surface in the bulk) decreases with future progression, the induced WDW state continues to exist within a Hilbert space of progressively diminishing dimensions. This shift consequently forces information to leave the black hole. It is only in this specific sense that information can be understood to emerge from the black hole from the perspective of semiclassical observers outside the black hole.

Ultimately, by the time the Cauchy slice of the WDW screen becomes zero, the induced WDW state approaches a state that closely resembles that of a closed universe. This state is to be represented by the partition function of the $T^2$ theory on a manifold with a minuscule boundary. At this point, all the information resides on $\Sigma_{\text{out}}$, which transforms into a Cauchy surface upon the complete evaporation of the black hole.

\subsection{Perspective of observers falling into an evaporating black hole}

The black hole's interior is classical within the WDW patch of $\mathcal{C}$ (which corresponds to the unshaded region inside the black hole in Fig \ref{quantumhawkingpenrosediagram}). Observers falling into the black hole would experience physics consistent with that of semiclassical gravity within this region. However, if such observers exit the WDW patch of $\mathcal{C}$, they would cease to exist due to the breakdown of their concept of time and space.


The WDW state induced on the WDW screen, $\partial\mathcal{Q}$, during the very late stages (near $\mathcal{E}$) is quite restricted—it approximates the WDW state of a closed universe. Nonetheless, it is supposed to encapsulate all the information within its WDW patch, which includes the entire interior of the black hole. So how does this highly constrained WDW state encode the black hole's interior, given that it appears to be a single state?

The answer lies in the understanding that the induced WDW state at $\mathcal{E}$ encompasses many branches, not all of which remain classical near $\mathcal{E}$, but do exhibit classical behavior at points further away and within the interior from $\mathcal{E}$. Therefore, observers situated in the black hole's interior exist within a single partial WKB branch of the induced WDW state at $\mathcal{E}$. This idea is closely related to the enumeration of states concept, which was discussed in subsection \ref{operationalobservers}.

\section{Discussion}

\subsection{Summary}

Our research delved into the investigation of the semiclassical approximation to the WDW equation. We elucidated how a classical background, satisfying the semiclassical Einstein equations, emerges in concert with an emergent QFT state. We applied this understanding in conjunction with the CSH dictionary to unravel how the semiclassical bulk is encoded in the correlation functions of the dual QFT living on Cauchy slices, in both the context of AdS/CFT and dS/CFT correspondence.

We then explored the scenario where a subregion of spacetime can be treated classically, while its complement is handled quantum mechanically. This exploration led to the emergence of a time-like boundary, which we termed the WDW screen. This screen became host to induced WDW states, encoding the quantum gravitational information of the interior of the screen. We further derived the conditions for the validity of such an approximation and applied them to the study of the black hole information paradox. The application suggested a potential breakdown of semiclassical gravity near the Page time in the near-horizon region. We also proposed a computational framework to determine the future semiclassical state even when semiclassical gravity may break down at intermediate times. 

Our investigation also ventured into the realm of dS spacetimes, providing an understanding of subsystem holography in these spaces through the process of subregion classicalization. We discussed the phenomenon of WKB branch merging and analyzed how different subsystems of the universe could be described by distinct states depending on their specific WKB branch. This understanding offered deeper insights into the fate of information trapped in an evaporating black hole.

\subsection{Holographic Cauchy slice correlation functions}

For the past several decades, our primary engagement in QFT has revolved around the computation of correlation functions. In our quest to unravel the mysteries of quantum gravity, we've predominantly approached the subject through perturbative quantum gravity, examining it against a fixed background. In these scenarios, our computations are again centered on correlation functions set on this fixed background.

However, through the lens of Cauchy slice holography, the WDW states encapsulate the entirety of the quantum gravitational data within the WDW patch. These states manifest as partition functions of the $T^2$ theory. Consequently, quantum gravity information finds its representation in the correlation functions of the $T^2$ theory on $\Sigma$. By evaluating these correlation functions on the $d$-dimensional manifold $\Sigma$, we indirectly compute the QG data within the WDW patch of a $(d+1)$-dimensional quantum gravity theory.

The salient advantage here is the elimination of the necessity to lock in a $(d+1)$-dimensional spacetime background for correlation function computations. Instead, we can fix a $d$-dimensional spatial geometry, compute correlation functions based on it, and thereby decipher all attributes of the WDW state proximate to this spatial geometry in superspace. 

To determine which correlation functions encapsulate specific QG data, further investigation is imperative. In the context of the semiclassical regime, the approximation presented in this paper suggests that the large N limit of certain one-point functions conveys the classical data of the bulk background. Concurrently, the subleading terms in the higher-point correlation functions within the large N expansion capture details about the bulk QFT states. Drawing from this, one could now embark on a large N expansion to discern corrections to the semiclassical framework and pinpoint where the perturbative expansion might falter.

Perhaps one could develop a diagrammatic technology to compute these correlation functions as a perturbative expansion in $1/N$. These diagrams would be analogous to Feynman diagrams but situated on the abstract Cauchy slices $\Sigma$. This approach would allow us to strategically focus our efforts on computing specific diagrams on the abstract Cauchy slice, addressing a particular query, rather than attempting to compute the entire partition function or WDW state.

\subsection{Making $\text{ER} = \text{EPR}$ precise}

The TFD state, represented as $|\psi\rangle_{\text{TFD}} = \sum_n e^{-\beta E_n} |n\rangle_\text{L} |n\rangle_{\text{R}}$, where $|n\rangle_\text{L}$ and $|n\rangle_\text{R}$ denote the energy eigenstates of the left and right CFTs respectively, is dual to an eternal Schwarzschild black hole. If one assumes that each of these energy eigenstates is dual to a semiclassical bulk geometry that remains disconnected, their superposition results in a state dual to a connected geometry. This observation then morphs into a motivation for the ER=EPR conjecture.

In order to rigorously scrutinize and refine the ER=EPR conjecture, an initial, essential step would be to map the energy eigenstates to a bulk WDW state and subsequently evaluate its semiclassical attributes. If these are WKB states, it becomes apparent that within the bulk, the TFD state is a linear combination of multiple WDW WKB states.

It's crucial to note that an arbitrary linear combination of WKB states is not necessarily a WKB state. However, it could be possible that some specific linear combination is. This might be the case with the TFD state. In this situation, with the coefficients of the linear combination in the TFD state, perhaps the entire linear combination of WKB WDW states is another WKB WDW state, but this time with the Einstein-Hamilton-Jacobi function of a connected geometry. It would be interesting to demonstrate this using the CSH dictionary and Equation \eqref{KrelationtoT1ptfun}. Furthermore, if one considers a different state in which the left and the right CFTs are entangled, it might not necessarily be a WKB state. Thus, one can test the setting in which ER=EPR is valid and determine how broadly it is true or applicable.

\subsection{Some Questions}

\subsubsection*{Origin of entropy bounds?}

What is the origin of entropy bounds? All of the entropy bounds usually relate some entropy to a geometric quantity, such as the area. To have a good notion of the geometric quantity, the WDW state must at least be a partial WKB state from which such geometric quantities can be extracted. In the formalism of QFT in curved spacetime alone, there indeed exist states in which the entanglement entropy can be ramped up arbitrarily high. However, that doesn't mean one can do this in the real world because the formalism of QFT in curved spacetime is only approximately true in reality. But, at least in an effective theory of quantum gravity, if WDW states are reliable, and one is free to choose any WDW state, then whatever QFT state it admits in the semiclassical approximation would be applicable in the real world. Therefore, perhaps the very thing that becomes problematic when trying to violate the entropy bound is the semiclassical approximation of WDW states. Could it be that the entropy bounds stem from the semiclassical validity conditions?

\subsubsection*{Flat Space-time Holography?}

The utilization of the $T^2$ deformation technique successfully depicts solutions to the WDW equation for both positive and negative cosmological constants. Unfortunately, an analogous approach for generating solutions to the WDW equation when the cosmological constant is zero remains undiscovered. Resolving this would naturally pave the way to understanding flat space-time QG dynamics, which would, in turn, perhaps hinge on the dynamics of the dual field theory.

A noteworthy observation is that all Cauchy surfaces in flat spacetime anchor on $i^0$. This raises an intriguing query: should we postulate that the dual field theory resides on $i^0$? Though $i^0$ is fundamentally a co-dimension one surface, its representation as a co-dimension two surface in the Penrose diagram is a characteristic artifact of conformal compactification. If we're to effectively implement Cauchy slice holography in flat spacetimes, it becomes imperative to discern the precise nature and structure of the field theory that lives on $i^0$.

A further point of intrigue is the realization that, upon anchoring a Cauchy surface on a specific co-dimension one surface of $i^0$, it can access nearly every point within the spacetime, barring the future and past null infinity. This unique attribute can potentially have significant ramifications.

In the field of celestial holography, the scattering matrix in flat spacetimes is recasted as correlators on the Celestial sphere. This raises another question: how would this formulation connect with a field theory situated on $i^0$ that steers the boundary time evolution? Also, in celestial holography, the bulk space-time is fixed, and perturbative quantum gravity in this spacetime is studied. To go beyond this, and to incorporate manifest background independence, a well-defined Hilbert space, and dynamics confined to finite time in celestial holography is challenging yet necessary.

An avenue worth exploring could involve co-dimension one holography within flat spacetimes. By formulating dictionaries capable of accommodating bulk superpositions, insights might be gleaned to forge a computational structure adept at addressing black hole collapse and evaporation in flat spacetimes.

\subsubsection*{Connection to experiments?}

The validity conditions in \eqref{nSVC} essentially suggest that when there's significant entanglement between the interior and exterior of a finitely sized region, quantum gravitational effects become prominent. This could cause semiclassical gravity to break down on the surface of that region due to pronounced metric fluctuations, particularly if the Wigner function of quantum gravity starts to breach the Halliwell criteria. This potential breakdown can occur even if the region in question is large.

Consider a scenario where one sets up an interferometry experiment using QFT states that exhibit these traits. In this experiment, a particle exists in a superposition of two Gaussian wavepackets, with only one wavepacket passing near the region's surface. Given the potential breakdown of semiclassical gravity in such a large region, one might wonder if the interference patterns would deviate from expectations under a fixed spacetime assumption.

It's plausible to believe they would, given that interference patterns are inherently influenced by the optical distance, which in turn depends on the spacetime metric. If the metric undergoes fluctuations, it could indeed alter the interference pattern. The subsequent pressing question is: if this effect does manifest, is its magnitude significant enough to be detected with our current or upcoming experimental technologies?

\subsubsection*{Overlapping WDW Patches?}

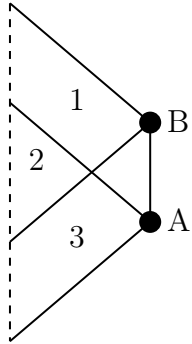
\begin{figure}[h]
\centering

\tikzset{every picture/.style={line width=0.75pt}} 

\begin{tikzpicture}[x=0.75pt,y=0.75pt,yscale=-1,xscale=1]

\draw    (320,80) -- (320,130) ;
\draw    (250,20) -- (320,80) ;
\draw    (250,140) -- (320,80) ;
\draw    (250,70) -- (320,130) ;
\draw    (250,190) -- (320,130) ;
\draw[dashed]    (250,20) -- (250,190) ;

\filldraw   (315,130) .. controls (315,127.24) and (317.24,125) .. (320,125) .. controls (322.76,125) and (325,127.24) .. (325,130) .. controls (325,132.76) and (322.76,135) .. (320,135) .. controls (317.24,135) and (315,132.76) .. (315,130) -- cycle ;
\filldraw   (315,80) .. controls (315,77.24) and (317.24,75) .. (320,75) .. controls (322.76,75) and (325,77.24) .. (325,80) .. controls (325,82.76) and (322.76,85) .. (320,85) .. controls (317.24,85) and (315,82.76) .. (315,80) -- cycle ;

\draw (278,62) node [anchor=north west][inner sep=0.75pt]   [align=left] {1};
\draw (258,92) node [anchor=north west][inner sep=0.75pt]   [align=left] {2};
\draw (278,131) node [anchor=north west][inner sep=0.75pt]   [align=left] {3};
\draw (327,122) node [anchor=north west][inner sep=0.75pt]   [align=left] {A};
\draw (327,72) node [anchor=north west][inner sep=0.75pt]   [align=left] {B};

\end{tikzpicture}

\caption{\small The vertical line AB represents a segment of the asymptotic boundary. WDW state $\Psi_A$ lives on point A and it boundary time evolves to the WDW state $\Psi_B$ on point B. Each of these states encode bulk information in their respective WDW patches. The WDW patch of A is region $2 \cup 3$ and the WDW patch of B is region $1\cup 2$. There is an overlapping region $2$ due to which there is some mixing between the dynamics and kinematics of QG.}
\label{overlappingWDWpatches}
\end{figure}

The WDW states (or equivalently their dual CFT states) encode bulk QG information in the entire WDW patch. For states situated close to eachother along the boundary time (see figure \ref{overlappingWDWpatches}), the WDW patches intersect. In this figure, the line AB represents a segment of the boundary. The WDW patch of A covers regions $2 \cup 3$, while the WDW patch of B covers regions $1 \cup 2$. Region $2$ is the overlapping region. If the WDW state $\Psi_B$ on B is the boundary time-evolved state of WDW state $\Psi_A$ on A, and since both WDW states must encode the same information in the overlapping region $2$, there exists some interplay between the kinematics and dynamics in the theory of quantum gravity.
Does this impose any constraints on the dynamics of the dual CFT?

\subsubsection*{Future-Past mixing in QG}
As elucidated in section \eqref{bulkclassicalduals}, there is potential branch mixing in the WDW state, even in the large $N$ limit. If one could segregate these branches within a particular foliation or gauge, each branch would offer insights about Cauchy slices on one side of the maximal volume slice. However, when the semiclassical limit isn't stringent, discerning whether a Cauchy slice is situated above or below the maximal slice becomes ambiguous. This implies that even when introducing small perturbations around a given background, higher order terms in the WKB expansion could introduce processes correlating the QFT state on Cauchy slices above the maximal volume slice with the QFT state on a Cauchy slice below the maximal volume slice, provided both possess identical induced metrics. This ambiguity arises because the WDW state, using the metric as input, can no longer distinguish the time ordering solely from the metric input. It would be intriguing to explore if this can be elucidated further through some expansion around a background.

In a similar vein, the radial WDW state might begin blending interior and exterior regions if the background results in the volumes of the radial slices of the Cauchy slice being non-monotonic. For instance, a Cauchy slice entering evaporating black holes could exhibit non-monotonic volumes for the codimension 2 radial slices in the bulk.

\subsubsection*{Algebraic Holographic QG}

It would be wonderful to formulate holography as an algebraic theory, similar to algebraic QFT. The formalism would go along the lines of first starting with the dual CFT in the algebraic QFT framework. Then recasting the boundary algebra of observables in some bulk form preserving the bulk background independence. Then only when certain CFT states that have a semiclassical dual are considered, the bulk observables rearrange in a net form as that of bulk algebraic QFT, but it wouldn't be exactly that because there might be a radial netting of this algebra of observables (similar to the time slice axiom of algebraic QFT). 

\subsubsection*{What goes wrong with the wrong deformation in the IR?}

So far in Cauchy slice holography, one needs to specify the bulk action to derive the bulk Hamiltonian constraint from which the required $T^2$ deformation operator can be obtained. But in principle, if one is handed a CFT, one should also be able to directly derive the bulk's low-energy effective action. Unfortunately at this stage, we do not know how to do this. One can perhaps ask a variant of this question. Let's say one is given a CFT for which the dual bulk action is known from which one obtains the correct $T^2$ deformation. What happens if one does a different $T^2$ deformation associated with a different bulk action? Does anything go wrong in the IR? Perhaps the wrongly deformed theory is not UV completable? If that is the case, then perhaps one should do the following: when given a CFT, do all possible deformations and ask which of those deformed theories are UV completable as field theories. This might give a way to derive the bulk action from the CFT. Of course, this is a hard task to do. And if one finds more than one $T^2$ deformations of a CFT that are UV completable, then the different seeming bulk theories are all dual to each other, as previously pointed out in \cite{GRW}.

\subsection{Three key elements of quantum gravity}

We desire any theory of quantum gravity to have the following three pivotal elements:

\begin{itemize}
\item Manifest background independence,
\item A notion of effective theory of quantum gravity and a concept of UV completion,
\item A feasible semiclassical limit.
\end{itemize}

Our goal is to use the boundary CFT to formulate the bulk theory of quantum gravity that incorporates all the above elements. Therefore, the ingredients at our disposal are the manifold $\partial \mathcal{B}$, on which the CFT resides, the CFT Hilbert space, and the CFT dynamics as provided by its Hamiltonian. The CFT states are situated on $\partial\Sigma$, which serve as Cauchy surfaces of the boundary $\partial\mathcal{B}$. Moreover, the CFT Hamiltonian evolves these states with respect to boundary time. Cauchy slice holography \cite{GRW} precisely does this.

\textbf{Manifest Background Independence:} The CSH dictionary maps the CFT states situated on $\partial\Sigma$ to bulk WDW states residing on an abstract $d$-dimensional Riemannian manifold $\Sigma$, which is anchored to $\partial\Sigma$. These WDW states encode all the bulk QG information within the WDW patch of $\partial\Sigma$. In the WDW states, the spatial metric serves as the argument for the wave function, and no spacetime background is presumed a priori. Consequently, background independence is manifest. This stands in contrast to perturbative quantum gravity, where a background spacetime is pre-assumed, and the metric perturbations and other fields around this background spacetime are quantized.

\textbf{A Notion of Effective Theory of Quantum Gravity and a Concept of UV Completion:} In the CSH dictionary, the WDW state is expressed as a partition function of the $T^2$ theory on $\Sigma$, with boundary conditions provided by the CFT state on $\partial\Sigma$. The $T^2$ theory is obtained by first placing an Euclidean CFT on $\Sigma$ and then deforming it with the $T^2$ operator. This deformation satisfies the Callan-Symanzik equation, and hence, the flow incited by this deformation is an RG flow. Coupled with the fact that the deformation operator is irrelevant, it is implied that the $T^2$ partition function describes an effective field theory. According to the dictionary, this suggests interpreting the WDW states in a Wilsonian manner. This implies that the canonical theory of quantum gravity is an effective theory of quantum gravity. By UV completing this $T^2$ field theory, one might then UV complete quantum gravity while preserving manifest background independence.

\textbf{A Feasible Semiclassical Limit:} By employing the semiclassical approximation to the WDW states outlined in this paper, we were able to discern the emergence of the classical bulk from the WDW states and, consequently, from the boundary CFT states. This also provided us with validity conditions to better discern when to trust and when not to trust semiclassical gravity. When applied to an evaporating black hole, we discovered that the validity conditions indicated a breakdown of the semiclassical approximation close to the Page time, assuming that semiclassical gravity held up to the Page time.

\section*{Acknowledgments}

This work was supported in part by AFOSR grant FA9550-19-1-0260 “Tensor Networks and Holographic Spacetime”, and Trinity Henry Barlow scholarship. We are grateful for helpful conversations with Aron Wall, Ronak Soni, Prahar Mitra, Bernard Kay, Fernando Quevedo, Henrique Gomes, Jeremy Butterfield, Ted Jacobson, Jorge Santos, Filipe Miguel, Goncalo Araujo-Regado, Amr Ahmadain, Chethan Krishnan, Suvrat Raju, Vasudev Shyam, Zihan Yan, and Santiago Agüi Salcedo.

\vspace{1cm}
\color{black}
\noindent\rule[0.25\baselineskip]{\textwidth}{1pt}

\bibliographystyle{ieeetr}
\bibliography{references}

\end{document}